\documentclass[manuscrip]{aastex62}

\usepackage{amsmath,amsfonts,amssymb}
\usepackage{graphicx}
\usepackage{multirow}
\usepackage{ulem}
\usepackage{tablefootnote}
\usepackage{threeparttable}
\accepted{January 12, 2020}
\submitjournal{ApJS}

\shorttitle{ALMA high frequency long baseline campaign in 2017}
\shortauthors{Asaki et al.}

\begin{document}

\title{
    ALMA High-frequency Long Baseline Campaign in 2017: \\
    Band-to-band Phase Referencing in Submillimeter Waves
}

\correspondingauthor{Yoshiharu Asaki}
\email{yoshiharu.asaki@nao.ac.jp}

\author[0000-0002-0976-4010]{Yoshiharu Asaki}
\affil{Joint ALMA Observatory, 
        Alonso de C\'{o}rdova 3107, Vitacura, Santiago, 763 0355, Chile}
\affil{National Astronomical Observatory of Japan, \\
        Alonso de C\'{o}rdova 3788, Office 61B, Vitacura, Santiago, Chile}
\affil{Department of Astronomical Science, School of Physical Sciences, \\
        The Graduate University for Advanced Studies (SOKENDAI), 
        2-21-1 Osawa, Mitaka, Tokyo 181-8588, Japan}
\nocollaboration        

\author{Luke T. Maud}
\affil{ESO Headquarters, 
        Karl-Schwarzchild-Str 2 85748 Garching, Germany}
\affil{Allegro, Leiden Observatory, Leiden University, 
        PO Box 9513, 2300 RA Leiden, The Netherlands} 
\nocollaboration

\author{Edward B. Fomalont}
\affil{Joint ALMA Observatory, 
        Alonso de C\'{o}rdova 3107, Vitacura, Santiago, 763 0355, Chile}
\affil{National Radio Astronomy Observatory, 
        520 Edgemont Rd. Charlottesville, VA 22903, USA}
\nocollaboration

\author{Neil M. Phillips}
\affil{ESO Headquarters, 
        Karl-Schwarzchild-Str 2 85748 Garching, Germany}
\nocollaboration

\author{Akihiko Hirota}
\affil{Joint ALMA Observatory, 
        Alonso de C\'{o}rdova 3107, Vitacura, Santiago, 763 0355, Chile}
\affil{National Astronomical Observatory of Japan, \\
        Alonso de C\'{o}rdova 3788, Office 61B, Vitacura, Santiago, Chile}
\nocollaboration

\author{Tsuyoshi Sawada}
\affil{Joint ALMA Observatory, 
        Alonso de C\'{o}rdova 3107, Vitacura, Santiago, 763 0355, Chile}
\affil{National Astronomical Observatory of Japan, \\
        Alonso de C\'{o}rdova 3788, Office 61B, Vitacura, Santiago, Chile}
\nocollaboration

\author{Loreto Barcos-Mu\~{n}oz}
\affil{Joint ALMA Observatory, 
        Alonso de C\'{o}rdova 3107, Vitacura, Santiago, 763 0355, Chile}
\affil{National Radio Astronomy Observatory, 
        520 Edgemont Rd. Charlottesville, VA 22903, USA}
\nocollaboration

\author{Anita M. S. Richards}
\affil{Jodrell Bank Centre for Astrophysics, Department of Physics and Astronomy, \\
        University of Manchester, Manchester M13 9PL, UK}
\nocollaboration

\author{William R. F. Dent}
\affil{Joint ALMA Observatory, 
        Alonso de C\'{o}rdova 3107, Vitacura, Santiago, 763 0355, Chile}
\nocollaboration

\author{Satoko Takahashi}
\affil{Joint ALMA Observatory, 
        Alonso de C\'{o}rdova 3107, Vitacura, Santiago, 763 0355, Chile}
\affil{National Astronomical Observatory of Japan, \\
        Alonso de C\'{o}rdova 3788, Office 61B, Vitacura, Santiago, Chile}
\affil{Department of Astronomical Science, School of Physical Sciences, \\
        The Graduate University for Advanced Studies (SOKENDAI), 
        2-21-1 Osawa, Mitaka, Tokyo 181-8588, Japan}
\nocollaboration

\author{Stuartt Corder}
\affil{Joint ALMA Observatory, 
        Alonso de C\'{o}rdova 3107, Vitacura, Santiago, 763 0355, Chile}
\nocollaboration

\author{John M. Carpenter}
\affil{Joint ALMA Observatory, 
        Alonso de C\'{o}rdova 3107, Vitacura, Santiago, 763 0355, Chile}
\nocollaboration

\author{Eric Villard}
\affil{Joint ALMA Observatory, 
        Alonso de C\'{o}rdova 3107, Vitacura, Santiago, 763 0355, Chile}
\nocollaboration

\author{Elizabeth M. Humphreys}
\affil{ESO Headquarters, 
        Karl-Schwarzchild-Str 2 85748 Garching, Germany}
\affil{Joint ALMA Observatory, 
        Alonso de C\'{o}rdova 3107, Vitacura, Santiago, 763 0355, Chile}
\nocollaboration




\begin{abstract}
In 2017, an Atacama Large Millimeter/submillimeter Array (ALMA) high-frequency 
long baseline campaign was organized to test image capabilities with 
baselines up to 16~km at submillimeter (submm) wavelengths. 
We investigated image qualities using ALMA receiver Bands~7, 8, 9, and 10 
(285--875~GHz) by adopting band-to-band (B2B) phase 
referencing in which a phase calibrator is tracked at a lower frequency. 
For B2B phase referencing, it is expected that a closer phase calibrator to a target 
can be used, comparing to standard in-band phase referencing. 
In the first step, it is ensured that an instrumental phase offset difference 
between low- and high-frequency Bands can be removed 
using a differential gain calibration in which a 
phase calibrator is certainly detected while frequency switching. 
In the next step, 
comparative experiments 
are arranged 
to investigate the image quality between B2B and 
in-band 
phase referencing with phase calibrators at various 
separation angles. 
In the final step, we conducted long baseline imaging tests 
for a quasar 
at 289~GHz in Band~7 and 405~GHz in Band~8 
and complex structure sources of HL~Tau and VY~CMa 
at $\sim$670~GHz in Band~9.  
The B2B phase referencing was successfully applied, 
allowing us to achieve an angular resolution of 
$14 \times 11$ 
and 
$10 \times 8$~mas 
for HL~Tau and VY~CMa, 
respectively. 
There is a high probability of finding a low-frequency calibrator within 
$5^{\circ}.4$ in B2B phase referencing, bright enough to use 
an 8~s scan length combined with a 7.5~GHz bandwidth.
\end{abstract}


\section{
  Introduction
}\label{sec:01}

The Atacama Large Millimeter/submillimeter Array (ALMA) has been exploring 
astronomical frontiers with unprecedented angular resolutions and sensitivities 
in millimeter/submillimeter 
(mm/submm) waves to observe molecular gas and dust emissions radiated from 
various astronomical phenomena
\citep{Bachiller2008}. 
In theory, ALMA can achieve an angular resolution 
of 15, 9, and 7~mas at observing 
frequencies of 400, 650, and 850~GHz when baselines of up to 16~km 
are available. However, using the longest baselines at these frequencies 
constitutes one of the most challenging observing modes for ALMA.

In order to achieve such high angular resolutions, interferometer phase stabilization of the 
longest baselines is essential. ALMA long baseline and high-frequency (HF) commissioning, 
aimed at developing the high angular resolution capabilities, 
has progressed step-by-step since the construction and commissioning phase. 
The first moderately extended
baseline commissioning test series with respect to the available baselines 
at the time were made in 2010 before starting Cycle~0 early 
science by arranging 600~m baselines between one isolated antenna and an antenna 
cluster distributed in a $100 \times 100$~m area, in order  
to evaluate the system-level phase stabilities in ALMA receiver Bands~3, 6, 7, and 9 
\citep{Matsushita2012}.
In 2012, 2~km baseline experiments were conducted between one isolated antenna 
and an antenna cluster 
distributed in a $500 \times 500$~m area 
\citep{Matsushita2014}. 
The first ALMA Long Baseline Campaign (LBC) was organized in 2013 (LBC-2013) for long baseline 
image capability tests with a maximum of 2.7~km baseline
\citep{Asaki2014,
Matsushita2014,
Richards2014}. 
The second and third ALMA LBCs were organized in 2014 and 2015 (LBC-2014 and LBC-2015), 
respectively, to accomplish the most extended array configuration with 16~km baselines in order 
to make science verification (SV) observations in Bands~3, 6, and 7 
\citep{ALMApartnership2015a,
ALMApartnership2015b,
ALMApartnership2015c,
ALMApartnership2015d}, 
as well as conducting 
user 
observations. 
Phase metrics and phase compensation experiments with baselines up to 16~km 
were also conducted during the LBCs
\citep{Asaki2016,
Hunter2016, 
Matsushita2016}. 
With the success of these campaigns, ALMA has opened 16~km baseline observations 
to the user 
community
at up to Band~6 frequencies 
($\geq$ 1.1~mm wavelength, or $\leq$ 275~GHz) in Cycle~6, 
regularly achieving angular resolutions of 18~mas. 
In 2014, some observations were made in Band 7 on 16~km baselines at angular resolutions of 
$\sim$20~mas in Cycle~3 
\citep[e.g.][]{Andrews2016,
Kervella2016},  
and most recently, Band~7 16~km baseline observations have been 
opened up since Cycle~7.

At ALMA, state-of-the-art technologies are used for stabilizing instrumental phases 
in signal paths required for mm/submm interferometer observations 
\citep{Bryerton2005,
Cliche2006}.
Nevertheless, the phase stability throughout the whole 
observing system cannot 
be well controlled because the largest error is caused by Earth's atmosphere, 
particularly, water vapor in the lower troposphere
\citep{Robson2001}. 
Accurate phase correction, to compensate for the atmospheric phase 
errors, is mandatory to perform  
ALMA long baseline observations at the highest frequencies. 
One of the key techniques 
to correct atmospheric water vapor phase errors is to use  
a water vapor radiometer 
\citep[WVR;][and references therein]{Nikolic2013} 
equipped on each 12~m antenna. 
Since the opacity of the atmospheric water vapor at 183~GHz 
has a strong correlation with the mm/submm 
wave excess path length at the Atacama site, the opacity measured with 
the WVRs can be used to derive precipitable water vapor  
(PWV) content every few seconds and thus derive corrections for the associated 
phase fluctuations.

Phase referencing is another important technique for interferometer phase correction 
\citep[e.g.][]{Beasley1995,Asaki2007}. 
A nearby quasar (QSO) is used as a phase calibrator by being observed
alternately with the science target. 
This is used to calibrate phase 
errors due to instrumental phase offsets and mitigate antenna position
errors, as well as correcting residual atmospheric phase fluctuations
after the WVR phase correction.
Phase referencing using a phase calibrator at the same frequency of the target  
is referred to as 
in-band phase referencing. 
This is the general phase correction technique for all
interferometers and has been adopted at ALMA as a standard phase
correction method, along with WVR phase correction.
The combination of the above two 
techniques is quite successful for the phase correction in ALMA 
\citep{Matsushita2017}.

Note that an extension to phase referencing known as fast switching, is one 
in which a phase calibrator is observed with a cadence of a few tens of seconds. 
If a rapid phase change causes 2$\pi$ phase wrappings between the phase calibrator 
scans, we are unable to track whether the phases are moving positively or negatively 
from the previous scan. The ALMA antennas can quickly change their position 
by several degrees in a few seconds to accommodate 
such a mode, which is most important for HFs  
where the atmospheric fluctuations are most variable.

In the submm wave regime (wavelength $<$1~mm, roughly ALMA Band~7
and higher frequencies), several factors make in-band phase referencing
increasingly difficult. 
Since the atmospheric phase errors are mainly caused by an excess delay change, 
phase corrections must be made more
frequently using the fast switching technique, and the phase
calibrator must be at a smaller separation angle from the target. 
However, as the observing frequency increases, the flux
densities of most QSOs diminish, and the system noise temperature
rises. This indicates that fewer bright-enough phase calibrators are
available close to targets at arbitrary sky positions. 

To alleviate the difficulty, the so-called band-to-band (B2B) phase referencing technique 
is considered to be an alternative phase correction method 
in which a phase calibrator is observed at a lower frequency. 
Since the interferometer phase error is almost proportional to the observing frequency 
(see also  
Section~\ref{sec:02-01}), 
the phase corrections derived from the phase
calibrator at a low frequency (LF) can be multiplied by a ratio of the
HF and LF and applied to the phases of the HF target. 
The B2B phase referencing technique was first demonstrated using 
the Nobeyama Millimeter Array between 148 and 19.5~GHz for a target QSO and 
a reference communication satellite 
\citep{Asaki1998}. 
In B2B phase referencing, it is necessary to remove any instrumental phase 
offset difference between the two frequencies. 
A similar multifrequency phase correction has been made for the 
Combined Array for Research in Millimeter-wave Astronomy 
(CARMA) 
science array 
at 227~GHz using a nearby calibrator's interferometer delay 
measurements
at 30.4~GHz 
obtained 
with a reference 
antenna array and applied to the target using the CARMA pair antenna 
calibration system  
\citep[C-PACS;][]{Perez2010, 
Zauderer2016}. 
In C-PACS, the instrumental phase offset difference was corrected by applying a phase 
difference between the two arrays when they observed the same phase calibrator 
simultaneously. 
Another realization of such a multifrequency phase referencing is made in 
the Korean very long baseline interferometry (VLBI) network, 
which can observe at 22, 43, 86, and 129 GHz simultaneously 
\citep{Dodson2014,
Rioja2014} 
with quasi-optics in the receiver  
\citep{Han2013}.

In order to offer new ALMA image capabilities at the highest angular resolutions, 
implementation of B2B phase referencing is crucial. 
This was the focus of the fourth long baseline capability campaign organized in 2017.
This paper presents results from our feasibility study made during the
High Frequency Long Baseline Campaign 2017 (HF-LBC-2017). We aimed to 
prove that 
ALMA has observation capabilities in Bands 7, 8, 9, and 10 
(285--875~GHz) with up to the longest 16~km baselines using B2B phase referencing.
Section~\ref{sec:02} mentions the basic concept of B2B phase referencing, while 
Section~\ref{sec:03} introduces the strategy of HF-LBC-2017.
The main results of HF-LBC-2017 are presented in Section~\ref{sec:04}. 
The availability of the phase calibrator for B2B and in-band phase referencing 
is discussed in Section~\ref{sec:05}. 
We summarize the overall feasibility study in Section~\ref{sec:06}. 
Note that the details of parts of the experiments are 
described in additional relevant papers 
\citep{Asaki2019, 
Maud2020}

\section{
  Basic concept of B2B phase referencing
}\label{sec:02}

The main goal of interferometer phase correction is 
to remove systematic phase errors due to errors of instrumental 
electrical path length and a priori antenna position and to improve the 
coherence of the visibilities disturbed by atmospheric phase fluctuations. 
Our basic phase correction procedure for HFs on long baselines 
is divided into WVR phase correction and B2B phase referencing. 
The WVR phase correction and its performance are described in previous 
reports
\citep{Matsushita2017,
Maud2017} 
and references therein. 
The effectiveness of phase referencing for ALMA 
was previously investigated for in-band phase referencing in Band~3 
and B2B phase referencing in Band~7 with phase calibrators in Band~3
\citep{Asaki2014,
Asaki2016}. 
Such phase referencing techniques for switching between a 
target and phase calibrator allow very accurate tracking of the 
rapid atmospheric phase fluctuations and can effectively remove 
the phase errors, especially for baselines longer than several 
kilometers. In cases of narrow-bandwidth science observations 
for targeting specific molecular lines, phase calibrators 
are observed in a wider-bandwidth mode in order to obtain higher 
signal-to-noise ratios (S/Ns). 
The combination of the above bandwidth switching and B2B phase referencing 
may provide more flexibility in ALMA HF observations for molecular lines.  
In this section, we describe the basic concept of B2B phase referencing and 
relevant ideas regarding its implementation in HF-LBC-2017.

\subsection{
  B2B phase referencing
}\label{sec:02-01}

\begin{figure}[htbp]
\begin{center}
\begin{tabular}{c}
\includegraphics[width=140mm]{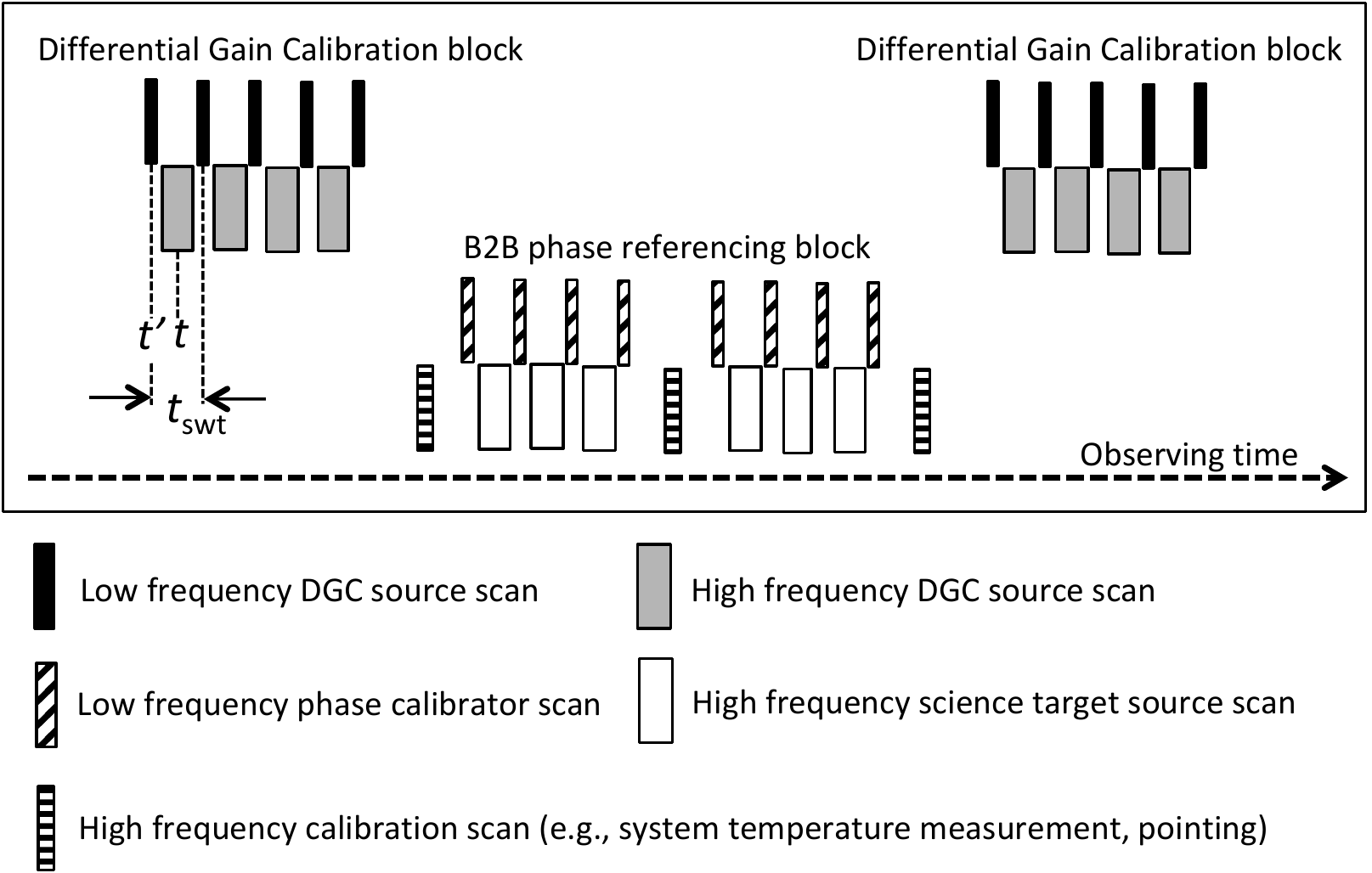}
\end{tabular}
\caption{
Typical sequence of a B2B phase referencing observation. 
}
\label{fig:01}
\end{center}
\end{figure}

Figure~\ref{fig:01} 
shows a typical observation sequence of B2B phase referencing. 
A phase referencing block consists of alternately pointing at a phase calibrator, 
observed at an LF  
$\nu_{\mathrm{_\mathrm{LF}}}$, 
and at a target source, observed at an HF  
$\nu_{\mathrm{_\mathrm{HF}}}$. 
A differential gain calibration (DGC) source 
is observed alternately at both 
frequencies for calibrating the instrumental phase offset difference 
later described in 
Section~\ref{sec:02-02}. 
Other standard calibration scans to measure the system noise temperature, 
as well as bandpass calibration, pointing calibration, and flux calibration, 
are also prepared for the HF target.
We express observed interferometer phases 
$\Phi^{\mathrm{T}}$ 
and 
$\Phi^{\mathrm{C}}$
at $\nu_{\mathrm{_\mathrm{HF}}}$ 
of the 
target and at $\nu_{\mathrm{_\mathrm{LF}}}$
of the phase calibrator, respectively, as follows: 
\\
\begin{eqnarray}
\label{eq:01}
\Phi^{\mathrm{T}}(t) &=& \Phi^{\mathrm{T}}_{\mathrm{total}}(t) - \Phi_{\mathrm{apri}}^{\mathrm{T}}(t) \nonumber \\
&=&
2 \pi \nu_{\mathrm{_\mathrm{HF}}} \left [
    \tau_{\mathrm{trp}}^{\mathrm{T}}(t) 
   + \tau_{\mathrm{bl}}^{\mathrm{T}}(t)
  \right ]  
    + \frac{2 \pi \kappa}{c \nu_{\mathrm{_\mathrm{HF}}}}  \Delta TEC^{\mathrm{T}}(t)  \nonumber \\                                                 
 &+& \Phi_{\mathrm{inst-H}}^{\mathrm{T}}(t) 
    + \Phi_{\mathrm{vis-H}}^{\mathrm{T}}(t)
     + \Phi_{\mathrm{therm-H}}^{\mathrm{T}}(t), \\
\label{eq:02}
\Phi^{\mathrm{C}}(t') &=& 
\Phi^{\mathrm{C}}_{\mathrm{total}}(t') - \Phi_{\mathrm{apri}}^{\mathrm{C}}(t') \nonumber \\
 &=&
2 \pi \nu_{\mathrm{_\mathrm{LF}}} \left [
        \tau_{\mathrm{trp}}^{\mathrm{C}}(t')  
   + \tau_{\mathrm{bl}}^{\mathrm{C}}(t')
  \right ]  
    + \frac{2 \pi \kappa}{c \nu_{\mathrm{_\mathrm{LF}}}}  \Delta TEC^{\mathrm{C}}(t')  \nonumber \\                                                 
 &+& \Phi_{\mathrm{inst-L}}^{\mathrm{C}}(t') 
    + \Phi_{\mathrm{vis-L}}^{\mathrm{C}}(t')
     + \Phi_{\mathrm{therm-L}}^{\mathrm{C}}(t'), 
\end{eqnarray}
\\
where 
\begin{list}{}
{
    \setlength{\itemindent}{0mm}
    \setlength{\parsep}{0mm}
    \setlength{\topsep}{3mm}
}
\item[
$\Phi_{\mathrm{total}}$ ]
               is the total 
               interferometer phase in summation of 
               geometrical delays and all errors; 
\item[
$\Phi_{\mathrm{apri}}$ ] 
              is the a priori phase calculated in the correlator 
              including contributions of the WVR phase correction; 
\item[
$\tau_{\mathrm{trp}}$ ]
              is the atmospheric delay error; 
\item[
$\tau_{\mathrm{bl}}$ ]
              is the delay error due to the baseline vector error coming from 
              the geometrical uncertainties of the antenna 
              positions and uncertainties of the Earth orientation parameters; 
\item[
$\Delta TEC$ ]
              is the spatial difference of the total electron content (TEC) of the ionosphere 
              in the line of sight between two antennas ($\kappa = 40.3$~m$^{3}$~s$^{-2}$);  
\item[
$\Phi_{\mathrm{inst-H}}$ and $\Phi_{\mathrm{inst-L}}$ ]
              are the instrumental phase offsets of frequency standard signals at $\nu_{\mathrm{_\mathrm{HF}}}$ 
              and $\nu_{\mathrm{_\mathrm{LF}}}$, 
              respectively;
\item[
$\Phi^{\mathrm{T}}_{\mathrm{vis-H}}$ and $\Phi^{\mathrm{C}}_{\mathrm{vis-L}}$ ]
              are the visibility phases representing the target source structure at 
              $\nu_{\mathrm{_\mathrm{HF}}}$ and the phase calibrator at $\nu_{\mathrm{_\mathrm{LF}}}$, 
              respectively, 
              with respect to their a priori phase tracking centers; and 
\item[
$\Phi_{\mathrm{therm-H}}$ and $\Phi_{\mathrm{therm-L}}$ ]
              are the thermal noises at $\nu_{\mathrm{_\mathrm{HF}}}$ and $\nu_{\mathrm{_\mathrm{LF}}}$, 
              respectively. 
\end{list}
Superscripts T and C denote target and phase calibrator, 
and $c$ is the speed of light. 
Here the target and phase calibrator are temporally observed in sequence. 
For simplification, let us assume that the target and phase calibrator are both point sources 
at their respective frequencies 
located at their a priori phase tracking centers 
($\Phi_{\mathrm{vis-H}}^{\mathrm{T}} = \Phi_{\mathrm{vis-L}}^{\mathrm{C}} = 0$). 
We denote that 
the mid-time point of the target and phase calibrator scans in one sequence 
occur at $t$ and $t'$, respectively. One important parameter of this sequence is called 
a switching cycle time, which is a length of time denoted $t_{\mathrm{swt}}$. 
This can be understood 
as the interval from the phase calibrator scan midpoint at $t'$ to the next scan 
midpoint of the phase calibrator at $t' + t_{\mathrm{swt}}$. Thus, $t_{\mathrm{swt}}$ 
encompasses the time spent on the calibrator and target and slewing overheads 
twice between the two. 
In order to correct $\Phi^{\mathrm{T}}$, a correcting phase 
$\Phi^{\mathrm{C}}_{\mathrm{cal}}$ at time $t$ is obtained 
by averaging $\Phi^{\mathrm{C}}$ (the temporally closest two phase calibrator 
scans of 
Equation~(\ref{eq:02})) 
and multiplying by an observing frequency ratio 
$R=\nu_{\mathrm{_\mathrm{HF}}}/\nu_{\mathrm{_\mathrm{LF}}}$ 
as follows: 
\\
\begin{eqnarray}
\Phi^{\mathrm{C}}_{\mathrm{cal}}(t)
&=& \frac{\nu_{\mathrm{_\mathrm{HF}}}}{\nu_{\mathrm{_\mathrm{LF}}}} \cdot
                       \frac{\Phi^{\mathrm{C}}(t-t_{\mathrm{swt}}/2)
                      + \Phi^{\mathrm{C}}(t+t_{\mathrm{swt}}/2)}{2} \nonumber \\
&\simeq& 
\label{eq:03}
2 \pi \nu_{\mathrm{_\mathrm{HF}}} \left [
        \overline{\tau_{\mathrm{trp}}^{\mathrm{C}}}(t) 
   + \overline{\tau_{\mathrm{bl}}^{\mathrm{C}}}(t)
    \right ] 
    + R \left [ 
    \frac{2 \pi \kappa}  {c \nu_{\mathrm{_\mathrm{LF}}}} 
         \overline{\Delta TEC^{\mathrm{C}}}(t)                                                 
    + \overline{\Phi_{\mathrm{inst-L}}^{\mathrm{C}}}(t) 
         + \overline{\Phi_{\mathrm{therm-L}}^{\mathrm{C}}}(t) \right ]. 
\end{eqnarray}
\\
This procedure is referred to as frequency phase transfer 
\citep{Rioja2014}. 
The B2B phase referencing is 
carried out by subtracting 
$\Phi^{\mathrm{C}}_{\mathrm{cal}}$ 
of 
Equation~(\ref{eq:03}) 
from Equation~(\ref{eq:01}) as follows:
\\
\begin{eqnarray}
\label{eq:04}
\Phi^{\mathrm{T}}(t)
        - \Phi^{\mathrm{C}}_{\mathrm{cal}}(t)
        &=& 
 2 \pi \nu_{\mathrm{_\mathrm{HF}}} \left [ \tau^{\mathrm{T}}_{\mathrm{trp}}(t) - \overline{\tau^{\mathrm{C}}_{\mathrm{trp}}}(t) \right ] 
       + 2 \pi \nu_{\mathrm{_\mathrm{HF}}} \left [ \tau^{\mathrm{T}}_{\mathrm{bl}}(t) - \overline{\tau^{\mathrm{C}}_{\mathrm{bl}}}(t) \right ]
                             \nonumber \\ 
& + & \left [ \Phi^{\mathrm{T}}_{\mathrm{therm-H}}(t)
                                  - R\ \overline{\Phi^{\mathrm{C}}_{\mathrm{therm-L}}}(t) \right ] \nonumber \\
& + & \frac{2\pi\kappa}{c \nu_{\mathrm{_\mathrm{HF}}}}
     \left[ \Delta TEC^{\mathrm{T}}(t)
           - R^{2}\ \overline{\Delta TEC^{\mathrm{C}}}(t) \right]                                            
                               + \left [ \Phi_{\mathrm{inst-H}}^{\mathrm{T}}(t)
                                  - R\  \overline{ \Phi_{\mathrm{inst-L}}^{\mathrm{C}}}(t) \right ].
\end{eqnarray}
\\

\begin{figure}[htbp]
\begin{center}
\begin{tabular}{c}
\includegraphics[width=140mm]{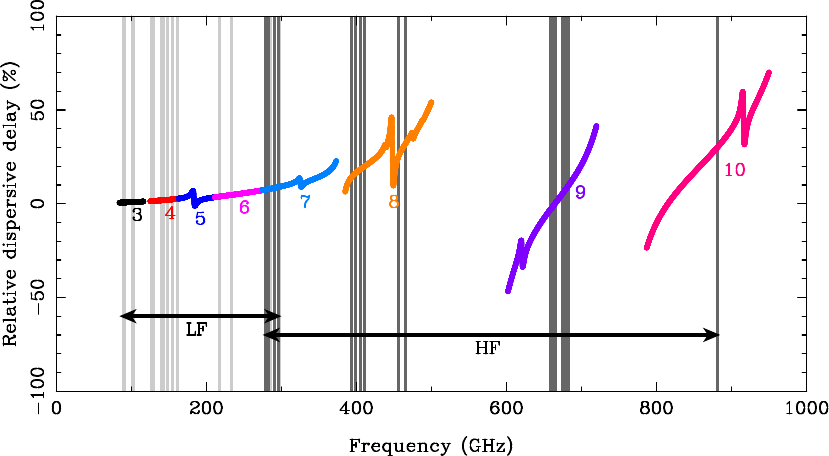}
\end{tabular}
\caption{
Relative dispersive delay of the atmospheric water vapor at the ALMA site (PWV~$=$~1~mm). 
The horizontal axis is the observing frequency, and the vertical axis is the relative dispersive delay compared to 
the nondispersive delay calculated with the ALMA ATM program 
\citep{Nikolic2009}. 
The numbers in the plot represent 
the ALMA receiver bands. 
The light grey stripes in Bands~3--7 represent the LF spectra of 
the phase calibrators in HF-LBC-2017, 
while the dark gray ones in Bands~7--10 represent the 
HF spectra for the targets.
}
\label{fig:02}
\end{center}
\end{figure}

A pervasive concern regarding B2B phase referencing is a dispersive term of mm/submm 
wave delay refraction of the atmospheric water vapor. 
The assumption of the nondispersiveness is likely to be correct except close to strong 
atmospheric water vapor absorption lines
\citep{Pardo2001}.
Figure~\ref{fig:02} 
depicts a relative dispersive delay between the dispersive and nondispersive terms of 
the atmospheric water vapor at the ALMA site (PWV$=$1~mm) calculated with 
the ALMA ATM program  
\citep{Nikolic2009} 
and the ALMA bands. 
Although HF observations are conducted in low-PWV conditions 
(typically, $\leq$1~mm in Band~8), we have to take care of the 
dispersiveness in B2B phase referencing in the submm regime, especially near the 
band edges higher than Band~6, where the relative dispersive term can reach $\sim$50\%.
We discuss our observing frequency selection in HF-LBC-2017, taking into account 
the dispersiveness, in 
Section~\ref{sec:03}. 

Another concern is that ionospheric phase errors have 
an inverse quadratic dependence on the observing frequency and can be expanded using 
the $R$ scaling ratio. 
One of the well-known ionospheric perturbations is nighttime periodic medium-scale traveling 
ionospheric disturbances 
\citep[MSTIDs; e.g.,][]{Otsuka2013} 
with an amplitude of $\sim$1~TECU at most 
(1~TECU$=10^{16}$~electron~m$^{-2}$). 
Typically, MSTIDs 
typically have a horizontal wavelength of a few hundred km, so that the spatial irregularities 
in the longest ALMA baselines do not yield a significant error in the frequency-transferred 
phase. On the other hand, 
it has been reported that  
equatorial plasma bubbles with a horizontal size 
of a few hundred~km are often observed in the early evening to midnight 
above the South America 
\citep[e.g.][]{Takahashi2016,
Takahashi2018}.  
They have a spatial gradient of 0.1~--~0.2~TECU~km$^{-1}$ 
at most, corresponding to a spatial TEC difference of $\sim 2$~TECU at the longest  
ALMA baselines. Let us quantitatively evaluate the frequency-transferred phase error. 
Assuming the sources observed at the elevation angle of $50^{\circ}$, 
%
%
%
the worst case of $\Delta TEC$ at the ALMA site can be $2 \times {\mathrm{sec}}~Z_{\mathrm{i}}$~TECU, 
where $Z_{\mathrm{i}}$ is the zenith angle of the sources for an altitude 
of 300~km (bottom of the ionospheric F-region). 
Considering $\nu_{\mathrm{_\mathrm{HF}}}=873$~GHz and $\nu_{\mathrm{_\mathrm{LF}}}=97$~GHz 
(the largest frequency separation possible for B2B phase referencing at ALMA; also see 
Section~\ref{sec:02-05}), 
the frequency-transferred phase error due to the plasma bubbles in 
Equation~(\ref{eq:04})
is 0.78~rad~TECU$^{-1}$. This leads to a frequency-transferred phase error of $\sim 113^{\circ}$, 
and it becomes one-ninth of this when using another possible $\nu_{\mathrm{_\mathrm{LF}}}$ 
solution of 291~GHz. 
The equatorial plasma bubbles are more frequently observed during southern hemisphere 
summer seasons, so that one may have to consider choosing a smaller $R$ for observations 
conducted before midnight in summer seasons. 
Note that this error term can be mitigated by selecting a nearby DGC source 
later discussed in  
Section~\ref{sec:02-02}.

One note 
regarding B2B phase referencing is the instrumental phase offset difference that 
appears in the last term of 
Equation (\ref{eq:04}). 
The actual stabilities of the instrumental phase in ALMA satisfy the system-level requirements 
\citep{Matsushita2012}, 
so for in-band phase referencing, 
this term cancels out; however, it remains 
as a systematic phase error in B2B phase referencing.
The correction of this phase error is discussed in the next section. 

The term 
$\tau^{\mathrm{T}}_{\mathrm{bl}}(t) - \overline{\tau^{\mathrm{C}}_{\mathrm{bl}}}(t)$ 
is dominated by an inner product of the baseline error vector and the separation angle vector 
between the target and phase calibrator. 
We assess how small the separation angle needs to be in order to mitigate this effect 
in this feasibility study. 

The term 
$\tau^{\mathrm{T}}_{\mathrm{trp}}(t) - \overline{\tau^{\mathrm{C}}_{\mathrm{trp}}}(t)$ 
is randomly variable due to the residual atmospheric phase fluctuations. Here we estimate 
the RMS of 
$\tau^{\mathrm{T}}_{\mathrm{trp}}(t) - \overline{\tau^{\mathrm{C}}_{\mathrm{trp}}}(t)$. 
This is considered to be proportional to the square root of the 
spatial structure function for atmospheric 
phase fluctuations as a function of 
$d + v_{\mathrm{w}} t_{\mathrm{swt}}/2$ 
for baselines whose length is longer than this value, 
where $v_{\mathrm{w}}$ is the velocity in meters per second 
of the atmosphere at the height of the turbulent layer, 
and $d$ is the geometrical distance 
in meters
between the lines of 
sight to the target and the phase calibrator at the altitude 
of the turbulent layer 
\citep{Holdaway2004b}. 
Assuming 
an atmospheric turbulent layer height of 500~m 
and its 
velocity of 6~m~s$^{-1}$ at the ALMA site 
\citep{Robson2001}, 
and that the phase calibrator is horizontally $3^{\circ}$ separated from the target 
at the elevation angle of $50^{\circ}$, 
$d+v_{\mathrm{w}} t_{\mathrm{swt}}/2 = 34 + 3 t_{\mathrm{swt}}$, 
so that the atmospheric phase fluctuation can be greatly reduced by selecting a short switching 
cycle time. 
Regardless of in-band or B2B phase referencing, the switching cycle time must be short enough
\citep{Matsushita2017}
and the separation 
angle should be as small as possible in order to cancel out the 
phase errors from the baseline errors and 
atmospheric phase fluctuations. 

In the case of B2B phase referencing, since 
the thermal noise in the calibrator phase is computationally 
frequency phase-transferred with $R$ 
and applied to the target phase through the phase correction process, 
it is recommended to select a bright calibrator at 
$\nu_{\mathrm{_\mathrm{LF}}}$, which still is much easier than finding a phase calibrator 
at $\nu_{\mathrm{_\mathrm{HF}}}$ 
as later discussed in 
Section~\ref{sec:05}.

\subsection{
  DGC
}\label{sec:02-02}

By applying the frequency-transferred correcting phase as expressed 
in the last term of 
Equation~(\ref{eq:04}), 
B2B phase referencing induces the instrumental phase offset difference. 
The instrumental phase offset difference can be independently measured with 
a cross-band calibration 
\citep{Holdaway2004b} 
in which a 
quasar is observed at 
$\nu_{\mathrm{_\mathrm{HF}}}$ and $\nu_{\mathrm{_\mathrm{LF}}}$ 
in turn to obtain the phase difference. In this paper, this 
cross-band calibration block 
is referred to 
as DGC, such that the observed calibrator for this purpose is 
referred to as a DGC source. 
Figure~\ref{fig:01} 
schematically shows a unit of DGC to obtain a solution, 
which is called the DGC block. 

In the DGC block, the delay error due to the baseline error is canceled out 
by phase referencing between HF and frequency-transferred LF phases 
because 
this delay error is proportional to the inner product of the baseline vector 
error and the separation angle vector, while the separation angle is $0^{\circ}$ in this case. 
Note that there could be a small delay error due to the uncertainty of the a priori source position, 
where the brightness peak for a DGC source may not be exactly consistent between 
$\nu_{\mathrm{_\mathrm{HF}}}$ and $\nu_{\mathrm{_\mathrm{LF}}}$ 
due to core shift properties of active galactic nuclei
\citep{Hada2011}. 
The expected quantities are typically smaller than 
0.1~mas, corresponding to a few tens of femtoseconds for a 16~km baseline, 
so that we can neglect this effect in the 
following discussion. 
The phase difference at time $t$ can be expressed as follows:
\\
\begin{eqnarray}
\label{eq:05}
\Delta \Phi^{\mathrm{DGC}}(t)
& = & 2 \pi \nu_{\mathrm{_\mathrm{HF}}} \left [ 
                        \tau^{\mathrm{DGC}}_{\mathrm{trp}}(t) - \overline{\tau^{\mathrm{DGC}}_{\mathrm{trp}}}(t)
                           \right ] 
                            +  \left [ \Phi^{\mathrm{DGC}}_{\mathrm{therm-H}}(t)
                  - R\ \overline{ \Phi^{\mathrm{DGC}}_{\mathrm{therm-L}}}(t) \right ]  \nonumber \\
& + & \frac{2\pi\kappa}{c \nu_{\mathrm{_\mathrm{HF}}}}
     \left[ \Delta TEC^{\mathrm{DGC}}(t)
           - R^{2}\ \overline{\Delta TEC^{\mathrm{DGC}}}(t) \right ]
 +  \left [ \Phi_{\mathrm{inst-H}}^{\mathrm{DGC}}(t)
                  - R\ \overline{\Phi_{\mathrm{inst-L}}^{\mathrm{DGC}}}(t) \right ]. 
\end{eqnarray}
\\
In 
Equation~(\ref{eq:05}), 
the first and second terms are random variables. 
The RMS of the first term is proportional to the square root of the structure function 
as a function of 
$v_{\mathrm{w}} t_{\mathrm{swt}}/2$ for baselines whose length is longer than this value, 
as discussed in 
Section~\ref{sec:02-01}.  
Therefore, the shorter the switching cycle time is for DGC, the smaller 
the 
atmospheric 
phase noise becomes in 
$\Delta \Phi^{\mathrm{DGC}}$.
The switching cycle time for the frequency switching is of the order of 20~--~30~s. 
In this case, the atmospheric phase fluctuations are tracked exceptionally well and calibrated 
by the frequency phase transfer.
If we take a time average of 
equation~(\ref{eq:05}) 
to suppress the random noise, 
we obtain the following phase error: 
\\
\begin{eqnarray}
\label{eq:06}
\left < \Delta \Phi^{\mathrm{DGC}} (t) \right > &=&
\frac{2\pi\kappa (1 - R^{2})}{c \nu_{\mathrm{_\mathrm{HF}}}}
     \left < \Delta TEC^{\mathrm{DGC}} \right >
+ \left <\Phi_{\mathrm{inst-H}}^{\mathrm{DGC}} - 
                  R\ \overline{ \Phi_{\mathrm{inst-L}}^{\mathrm{DGC}}} \right > . 
\end{eqnarray}
\\
The above phase solution preserves 
the ionospheric phase error directed to the DGC source and 
the instrumental phase offset difference between 
the HF target and LF phase calibrator. 
The first term is negligible for relatively small ionospheric perturbations, such as MSTIDs. 
Even in the case of a relatively large $R$ and an enhanced ionospheric anomaly, 
the frequency-transferred ionospheric phase error 
can be canceled out between the target and DGC source by selecting a nearby DGC source 
\citep{RDodson2009}. 
At last, the DGC solution should be subtracted from 
Equation~(\ref{eq:04}) 
to obtain fully calibrated target visibility data.

\subsection{
  Implementation of B2B phase referencing and DGC
}\label{sec:02-03}

\begin{figure}[htbp]
\begin{center}
\begin{tabular}{c}
\includegraphics[width=160mm]{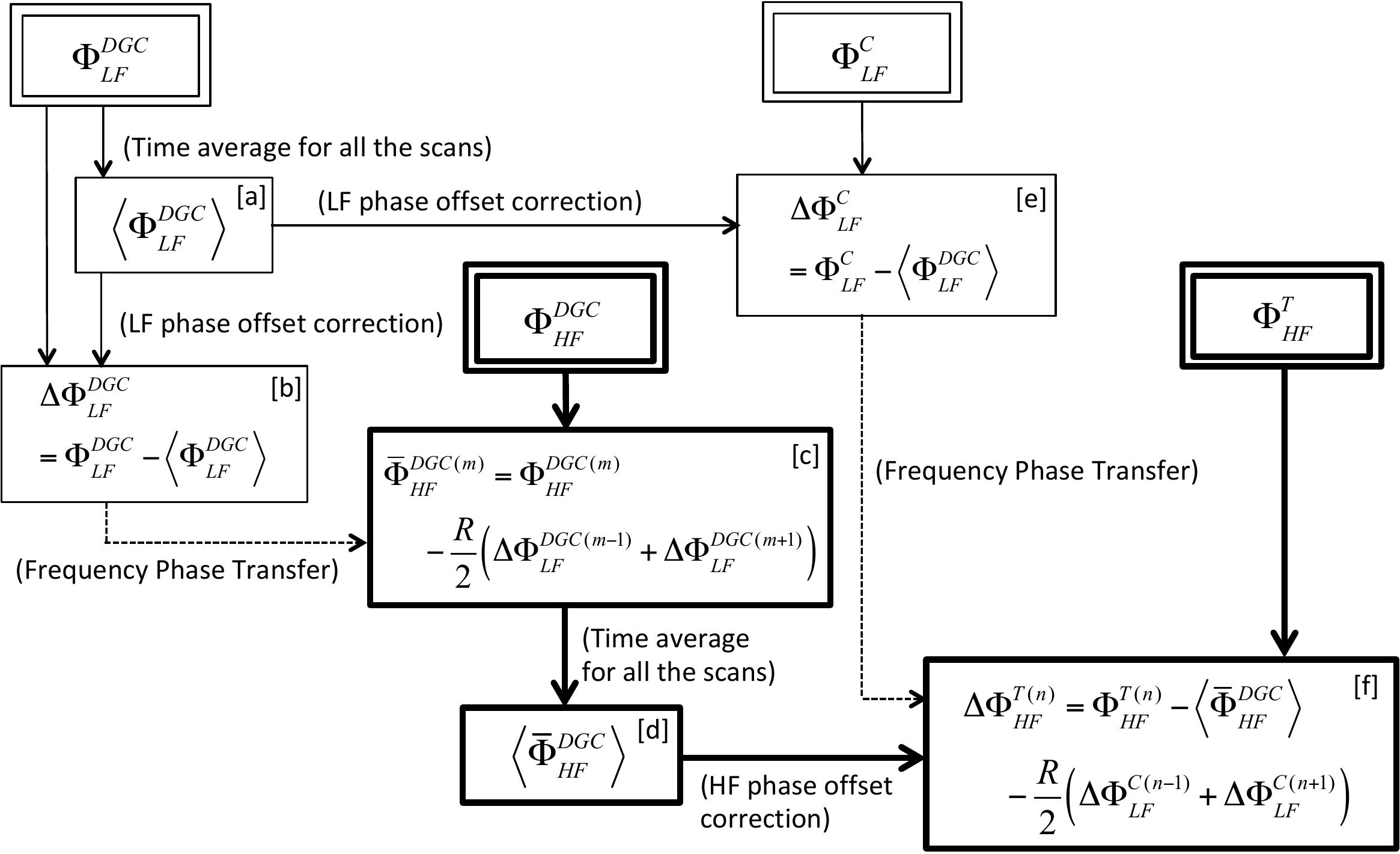}
\end{tabular}
\caption{
Logical workflow for implementation of B2B phase referencing and DGC. here 
$\Phi_{\mathrm{LF}}^{\mathrm{DGC}}$, 
$\Phi_{\mathrm{HF}}^{\mathrm{DGC}}$, 
$\Phi_{\mathrm{LF}}^{\mathrm{C}}$, and 
$\Phi_{\mathrm{HF}}^{\mathrm{T}}$ surrounded by the double squares 
are observed phases of 
a DGC source at $\nu_{\mathrm{_\mathrm{LF}}}$ and $\nu_{\mathrm{_\mathrm{HF}}}$, 
phase calibrator at $\nu_{\mathrm{_\mathrm{LF}}}$, and 
target at $\nu_{\mathrm{_\mathrm{HF}}}$, respectively; 
$R$ is the frequency scaling ratio between $\nu_{\mathrm{_\mathrm{HF}}}$ and 
$\nu_{\mathrm{_\mathrm{LF}}}$; and 
$n$ and $m$ are arbitrary scan numbers. 
The dashed arrows represent frequency phase-transfer. 
The bold squares and arrows represent the HF data. 
}
\label{fig:03}
\end{center}
\end{figure}

In the implementation of B2B phase referencing in the actual data reduction, 
we deal with the instrumental phase offsets at $\nu_{\mathrm{_\mathrm{HF}}}$ and 
$\nu_{\mathrm{_\mathrm{LF}}}$ separately, as depicted in a 
logical workflow in 
Figure~\ref{fig:03}. 
In 
Figure~\ref{fig:03}, 
$\Phi_{\mathrm{LF}}^{\mathrm{DGC}}$, 
$\Phi_{\mathrm{HF}}^{\mathrm{DGC}}$, 
$\Phi_{\mathrm{LF}}^{\mathrm{C}}$, and 
$\Phi_{\mathrm{HF}}^{\mathrm{T}}$ 
are 
observed phases 
of the DGC source at 
$\nu_{\mathrm{_\mathrm{LF}}}$ and 
$\nu_{\mathrm{_\mathrm{HF}}}$, 
phase calibrator at 
$\nu_{\mathrm{_\mathrm{LF}}}$, and 
target at 
$\nu_{\mathrm{_\mathrm{HF}}}$, 
respectively. 
In the workflow, a single interferometer signal output 
is assumed at each of 
$\nu_{\mathrm{_\mathrm{HF}}}$ and $\nu_{\mathrm{_\mathrm{LF}}}$. 
The WVR phase correction and the system noise temperature correction 
have been applied to all of the data before this workflow starts. 
The bandpass calibration and flux scaling using a flux calibrator are 
done only for the HF phase and amplitude.

We derive a time-averaged solution for all of the LF DGC scans 
$\left < \Phi_{\mathrm{LF}}^{\mathrm{DGC}} \right >$ 
(panel~(a) in   
Figure~\ref{fig:03}), 
that represents an LF phase offset for each antenna. 
In the next step, we apply 
$\left < \Phi_{\mathrm{LF}}^{\mathrm{DGC}} \right >$  
to the LF DGC phases 
$\Phi_{\mathrm{LF}}^{\mathrm{DGC}}$ 
and derive short-term phase solutions 
$\Delta\Phi_{\mathrm{LF}}^{\mathrm{DGC}}$  
for atmospheric phase fluctuations alone (free from other offsets) at each LF DGC scan 
(panel~(b)).
These solutions are frequency phase-transferred, and applied to 
the HF DGC phases 
$\Phi_{\mathrm{HF}}^{\mathrm{DGC}}$ 
(panel~(c)). 
This should correct the HF atmospheric fluctuations, so the HF DGC data can then be averaged 
in time and used to derive a HF phase offset   
$\left <\bar{\Phi}_{\mathrm{HF}}^{\mathrm{DGC}} \right >$ 
(panel~(d)).  
We then apply the LF phase offset  
$\left <\Phi^{\mathrm{DGC}}_{\mathrm{LF}} \right >$ 
to the LF phase calibrator phase 
$\Phi^{\mathrm{C}}_{\mathrm{LF}}$ 
to remove the common LF phase offset  
and to obtain the short~term phase solutions $\Delta\Phi^{\mathrm{C}}_{\mathrm{LF}}$ 
containing corrections 
for atmospheric fluctuations at $\nu_{\mathrm{_\mathrm{LF}}}$ 
for each LF phase calibrator scan (panel~(e)).
Finally, the 
HF phase offset correction $\left <\bar \Phi^{\mathrm{DGC}}_{\mathrm{HF}} \right >$ 
is applied to the target phase 
$\Phi^{\mathrm{T}}_{\mathrm{HF}}$, along with the time-dependent phase 
solutions derived from the LF phase calibrator, $\Delta\Phi^{\mathrm{C}}_{\mathrm{LF}}$ (panel~(f)). 
In this paper, we refer to the DGC HF phase offset 
$\left < \bar{\Phi}_{\mathrm{HF}}^{\mathrm{DGC}} \right >$ 
as the DGC solution for the sake of convenience.

\subsection{
  Fast switching
}\label{sec:02-04}

In our feasibility study, we used fast switching with switching cycle times of 
20~--~82~s. 
For almost all of the experiments, 
the scan length per source was 8~s; thus, accounting for a few 
seconds overhead 
for the antenna slew and/or frequency switching, 
the resulting switching cycle time was 20~s, 
considerably faster than the 100~s that 
is currently used for Cycle~7 user observations 
in Band~8 with a 3.6~km maximum baseline ($B_{\mathrm{max}}$) configuration in ALMA. 
We note that the image quality with longer switching cycle times can be investigated by culling 
phase calibrator scans 
\citep{Maud2020}. 

In fast switching, a relative pointing difference of the 12~m antenna between two sources 
separated by less than $2^{\circ}$ is about $0.''6$. 
In B2B phase referencing, 
pointing and focus adjustments between two Bands 
also have to be made by mechanically 
changing the position of the subreflector mounted on a 6-dimensional positioning actuator, as 
well as adjusting the relative pointing offset in the azimuth and elevation axes. 
The pointing offsets of the 12~m antenna between Bands are well maintained with 
an accuracy of $2''$ or higher. 
The relative pointing offset, for instance, between Bands~7 and 9 can be determined with 
a typical deviation of $0.''3$--$0.''5$. 
The 12~m antennas have a field of view of $58''$ and $9''$ in Bands~3 and 9, respectively.

\subsection{
  Harmonic frequency switching
}\label{sec:02-05}

In the case that frequency switching observations like B2B phase referencing 
are made using ALMA without any consideration, a temporal overhead of 
$\sim$20~s takes place for every frequency switching event that requires 
a retuning of the photonic local oscillator (LO) signal 
\citep{Shillue2012}. 
This hinders the fast switching and essentially makes 20~s switching cycle times 
with frequency switching impossible. 
To minimize the overhead time and maximize reliability when using 
B2B phase referencing, 
we switch Bands using a fixed photonic LO frequency 
for both receivers; i.e.,  
the photonic LO is tuned once at the start of an observation 
and not retuned in the repeated frequency switching. Each Band multiplies the photonic LO 
frequency with a small configurable offset of 20~--~45~MHz from an auxiliary oscillator in each 
antenna by a different fixed factor to obtain the actual first LO (LO1) frequency. 
The factor is referred to as a cold multiplier, as it is performed by a multiplier chain in the cold cartridge of the receiver. 
This technique is referred to as harmonic frequency switching 
and can minimize the overhead to $\sim$2~s. 

Each Band can only be used over a particular photonic frequency range, so not all 
receivers available to ALMA can be paired at an arbitrary frequency. The possible band 
combinations in B2B phase referencing are listed in 
Table~\ref{tbl:01}. 
We have to note that 
Bands~1 and 2 have not yet been implemented in ALMA. 
For completeness, Band~1 cannot form a harmonic frequency pair 
with any other Band, 
whereas Band~2 could pair with Bands~6, 8 and 9, the latter being 
most important for HF observations. 
There are some prohibited HF LO1 ranges, 
as listed in 
Table~\ref{tbl:02}, 
for which no matching LF LO1 is available. 
In HF-LBC-2017, we adopted harmonic frequency switching 
for all of the B2B phase referencing experiments.

\section{
  Strategy of HF-LBC-2017
}\label{sec:03}

In HF-LBC-2017, we conducted a series of experiments in order to implement 
B2B phase referencing in ALMA step-by-step in the following four stages:  
(1) confirm that the ALMA observing system correctly 
      executes 
      a B2B phase referencing observation procedure as designed, 
(2) ensure that DGC solutions do not have 
      unexpected 
      temporal instabilities or sky position 
      dependencies, 
(3) quantify the difference between B2B and in-band phase referencing in the image quality, and 
(4) verify that astronomical celestial sources can be imaged with a suitable image 
      fidelity at HFs using long baselines. 
In the campaign, we organized test observations in Bands~7, 8, 9, and 10 (285--875~GHz) with 
the baselines up to 16~km. We conducted two Band~10 experiments, 
one each for stages~2 and 3. Both failed after that 
because the 
observed sources were  
too weak to confirm whether the 
fringes were detected or not. In this paper, the Band~10 results are not mentioned further. 

The stage~1 test began in early 2017 and lasted until June. During this stage, 
we confirmed that the ALMA observing 
system could correctly operate the B2B phase referencing sequence and that the antenna 
and/or frequency fast switching could work without major technical troubles. 
The stage-1 test is not mentioned further in this paper. 

From 2017 March to July, 
we conducted the tests of stages 2 and 3 
tests using the mid-baseline lengths 
($B_{\mathrm{max}}\sim$400~m--4~km) 
to check the DGC solution stability and a basic image 
performance for various separation 
angles and switching cycle times. 
In stage~3, we compared the image quality between B2B and in-band 
phase referencing. 
One of the interesting parts of the study at stage~3 is the investigation 
of imaging performance between in-band and B2B phase referencing 
when the same phase calibrator is used, as well as when more distant 
calibrators are used for in-band phase referencing.

The baselines were longer than 10~km from the middle of July to the end 
of November, 
during which time more stage~3 tests and the stage~4 tests were undertaken. 
For stage~4, using the long baselines ($B_{\mathrm{max}}\sim$14--16~km), 
we conducted high angular resolution imaging 
experiments to demonstrate science-like observations of a QSO 
in Bands~7 and 8 and initiated B2B phase referencing experiments for two complex 
structure sources, HL~Tau and VY~CMa, in Band~9. 
Tables~\ref{tbl:03}--\ref{tbl:05} 
summarize the basic parameters of the experiments performed during 
stages~2, 3, and 4, respectively, together with the information of the ALMA 
execution blocks (EBs).

In selecting the observing frequencies, we prioritized the following two points: 
(1) the harmonic frequency switching is available for a combination of HF and LF 
      to achieve the fast switching, and 
(2) the dispersive term due to the atmospheric water vapor is not very large 
      to adopt a simple frequency scaling ratio 
      ($R=\nu_{\mathrm{_\mathrm{HF}}}/\nu_{\mathrm{_\mathrm{LF}}}$ ).
The selected spectral regions used in HF-LBC-2017 are superimposed in 
Figure~\ref{fig:02}.  
Although spectral regions with a relatively large dispersive term were used in Bands~8 and 10,   
we adopted the simple frequency scaling ratio for B2B phase referencing 
because other comparable delay errors, such as the antenna position errors, do not have 
a frequency dependence. Optimization of the frequency scaling ratio for a given 
frequency combination must be a future subject for B2B phase referencing.

\section{
  Results
}\label{sec:04}

In this section, we discuss the main results from the experiments in 
stages~2--4.
We adopted a consistent switching cycle time of $\sim$20~s, 
typically consisting of 8~s target and phase calibrator scans and 
two 2~s overheads to switch sources and/or frequencies. 
We note that the complex structure source imaging test in stage~4 had longer 
switching cycle times, 
described in 
Sections~\ref{sec:04-03-02} and 
\ref{sec:04-03-03}. 
The WVR phase correction was always applied through HF-LBC-2017. 
Throughout almost all of 
the tests, there were a number of manual flagging commands to be entered 
for these data, mostly for the frequency switching segments, due to the nature of testing 
such an experimental observation mode, though the problem was identified and fixed 
during stage~4. 
For many parts of our data reduction, we made use of the Common Astronomy Software 
Applications (CASA) package 
\citep{McMullin2007}.

\subsection{
  Stage 2: 
  DGC solution stability
}
\label{sec:04-01}

\begin{figure}[htbp]
\begin{center}
\begin{tabular}{c}
\includegraphics[width=60mm]{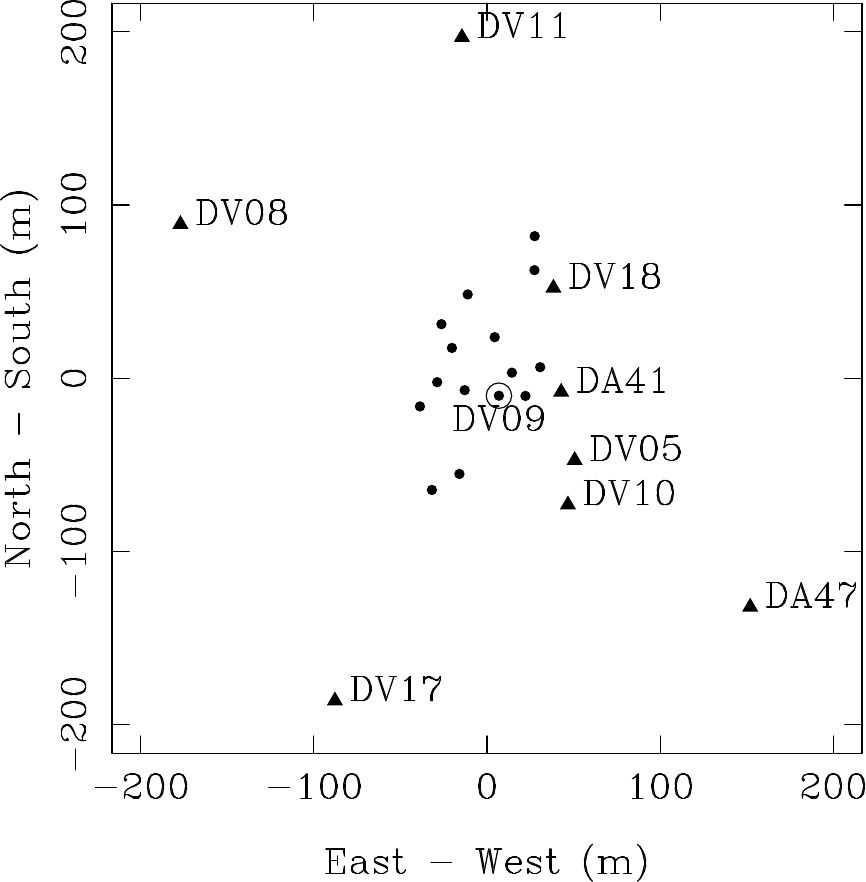}
\includegraphics[width=60mm]{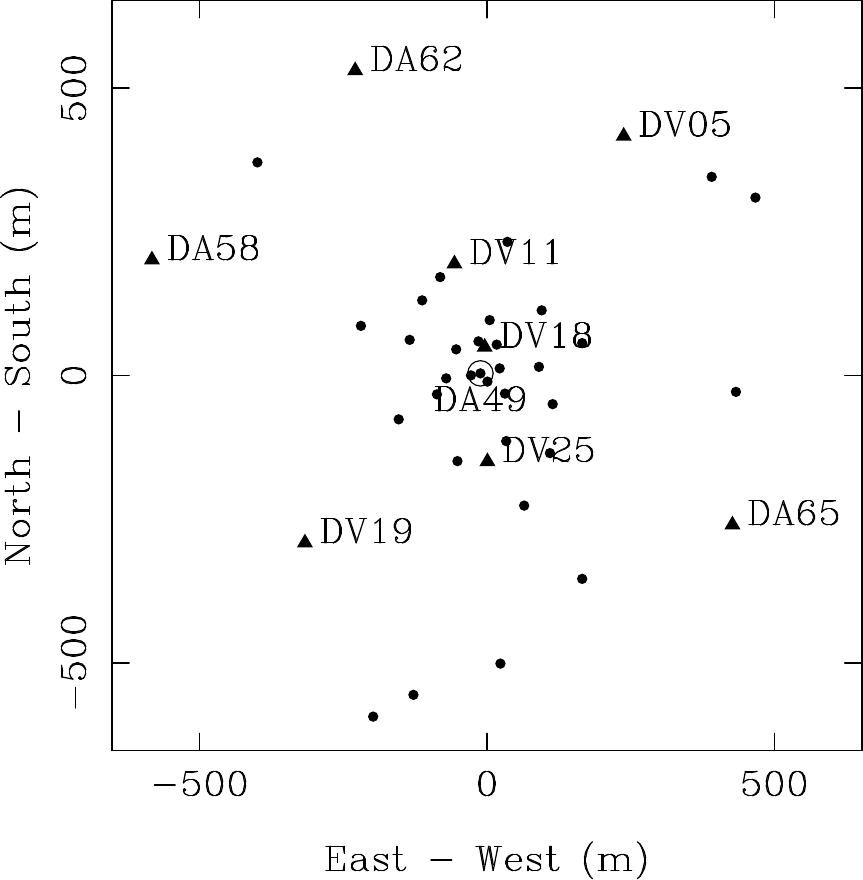}
\includegraphics[width=60mm]{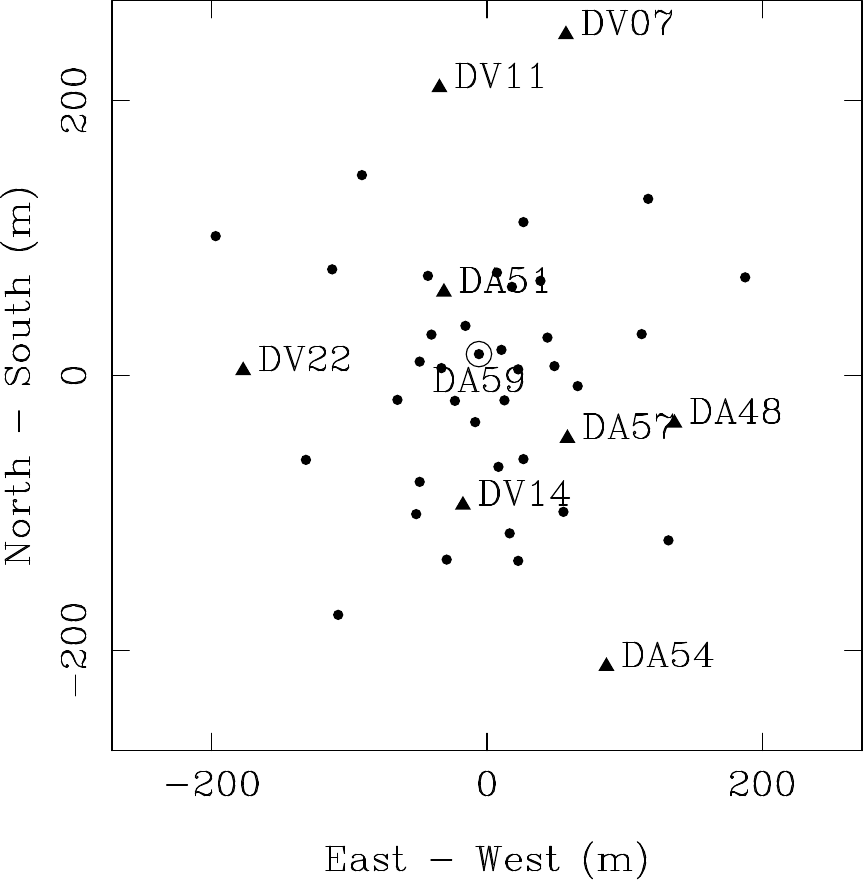}
\end{tabular}
\caption{
Array configurations of the stage~2 
test. 
Left: Band~7--3 experiment on 2017 April~11. 
Filled symbols are antennas used in the experiment. The double circle represents 
the reference antenna (DV09) used for the antenna-based DGC solutions. The DGC solutions are shown in 
Figure~\ref{fig:06} 
for antennas marked with triangles.
Middle: Band~8--4 experiment on 2017 May~4.
Right: Band~9--6 experiment on 2017 April~23. 
The DGC solutions of the Band~8--4 and 9--6 experiments are shown in 
Figure~\ref{fig:07} 
for antennas marked with triangles.
}
\label{fig:04}
\end{center}
\end{figure}

We arranged two sets of experiments in stage~2: 
one set targeting several bright QSOs as DGC 
sources in succession with the sky separations of up to 100$^{\circ}$ to check the 
dependence of the DGC solutions with the sky position, and another targeting a single 
DGC source for over an hour to check the long-term stability. 
In this paper, we report the former stability test results. 
We organized the stage~2 test first from the compact configuration with 
$B_{\mathrm{max}}=$400~m 
in March to $B_{\mathrm{max}}=$3~km in June. 
Figure~\ref{fig:04} 
shows examples of array configurations of the stage~2 test 
in 2017 April and May. 

In the actual ALMA receivers, each Band amplifies two linear polarizations 
($X$ and $Y$) separately, so that each polarized signal has an independent instrumental 
phase offset. The amplified signal is split into four intermediate-frequency signals 
which are called basebands (BBs) with a bandwidth of 2~GHz and an independent 
phase offset from each other. 
The BBs are digitized at the antenna and transferred to the 
correlator. In the correlator, the digitized BBs are filtered out to have a 
user-defined flexible bandwidth, frequency resolution, and polarization pairs 
among $XX$, $XY$, $YY$, and $YX$  
to form a spectral window (SPW) with a uniformly spaced spectral channels. 

\begin{figure}[htbp]
\begin{center}
\begin{tabular}{c}
\includegraphics[width=85mm]{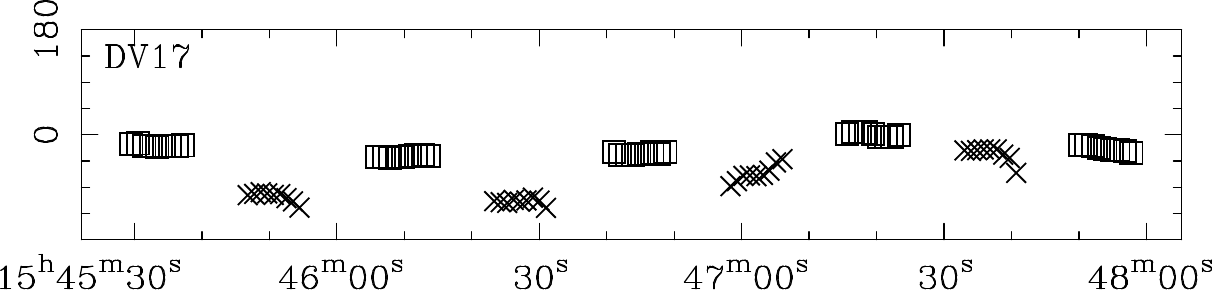} \\ 
\includegraphics[width=85mm]{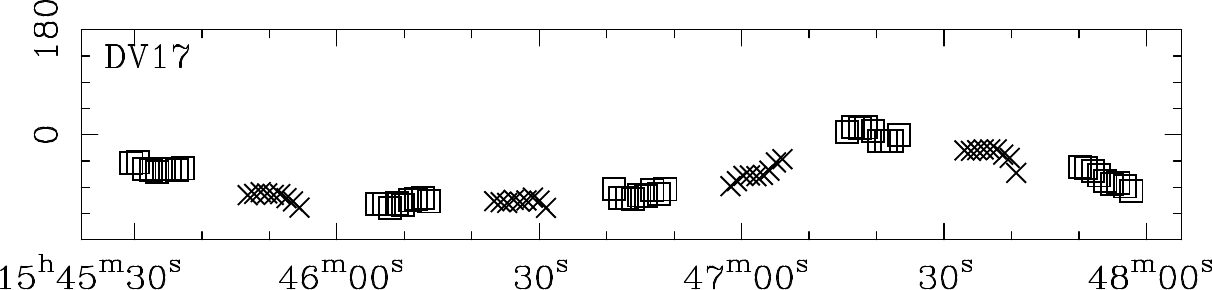} \\ 
\includegraphics[width=85mm]{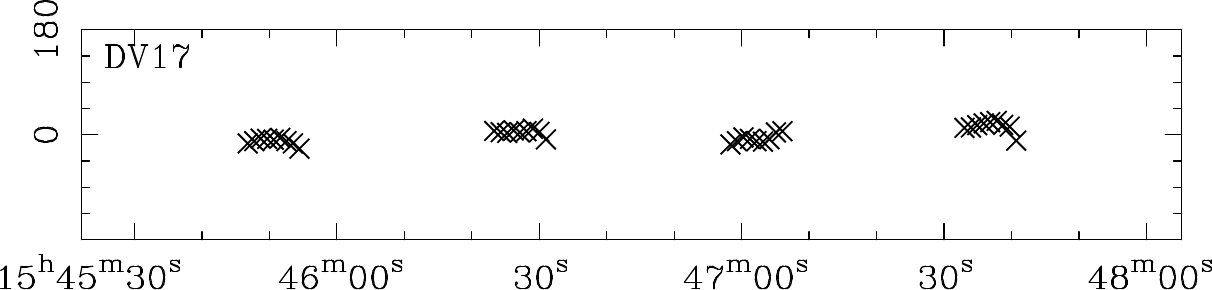}
\end{tabular}
\caption{
Stage~2 DGC stability experiment for Band~7--3 on 2017 April 11.
Top: 
time series of the WVR-corrected antenna-based phase of J2253$+$1608 of the DV17 antenna 
for a single SPW with the $XX$ polarization pair. Each point has a 1~s integration time. 
Crosses and open squares represent the Band~7 and Band~3 interferometer phases, respectively. 
Middle: 
the same as the top panel, but the Band~3 phases are multiplied by the frequency scaling ratio. 
Bottom: 
the Band~7 interferometer phases after correcting with the Band~3 phase time series, 
which are multiplied by the frequency scaling ratio.
}
\label{fig:05}
\end{center}
\end{figure}

Figure~\ref{fig:05} 
shows an example of one DGC block to observe a bright QSO 
(in this case, J2253$+$1608) and the data reduction procedure for the Band combination of 
7 and 3 (Band~7--3) on 2017 April 11, as listed in 
Table~\ref{tbl:03}. 
One DGC block basically consists of four HF and five LF scans. 
The LO1 frequencies in Bands~7 and 3 are 285 and 95~GHz, respectively. 
The top panel shows the WVR-corrected antenna-base phase of a baseline 
between two 12~m antennas (DV17 and DV09 in 
Figure~\ref{fig:04})
for the $XX$ polarization pair. 
Phase offsets for an HF SPW (crosses) and LF SPW (open squares) were 
artificially adjusted to make them align each other (for plotting purposes). 
The middle panel is the same as the top panel, but the Band~3 phases are 
multiplied by the frequency scaling ratio. 
The bottom panel shows the Band~7 phases after correcting with the Band~3 
phase and that the Band~7 phase time variation can be corrected 
using the Band~3 phases that are multiplied by the frequency scaling ratio. 
After the B2B phase referencing correction, the HF 
phase is averaged to one point, so that 
the time interval is $\sim 90$~s for a single DGC solution
in this case. 

\begin{figure}[htbp]
\begin{center}
\begin{tabular}{c}
\includegraphics[width=85mm]{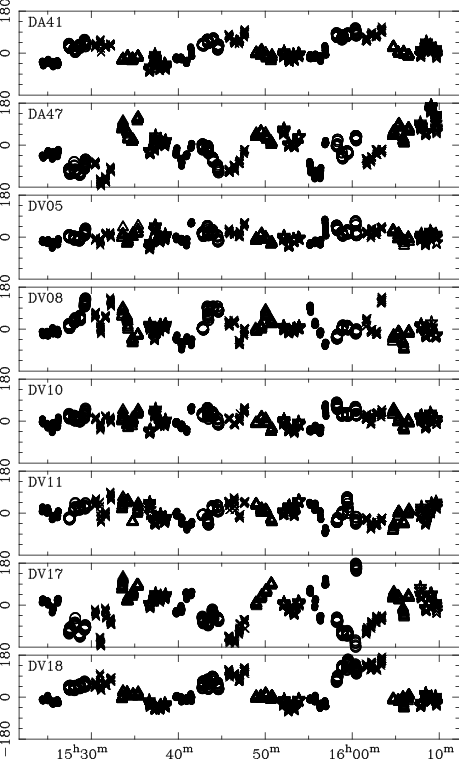}
\hspace{5mm}
\includegraphics[width=85mm]{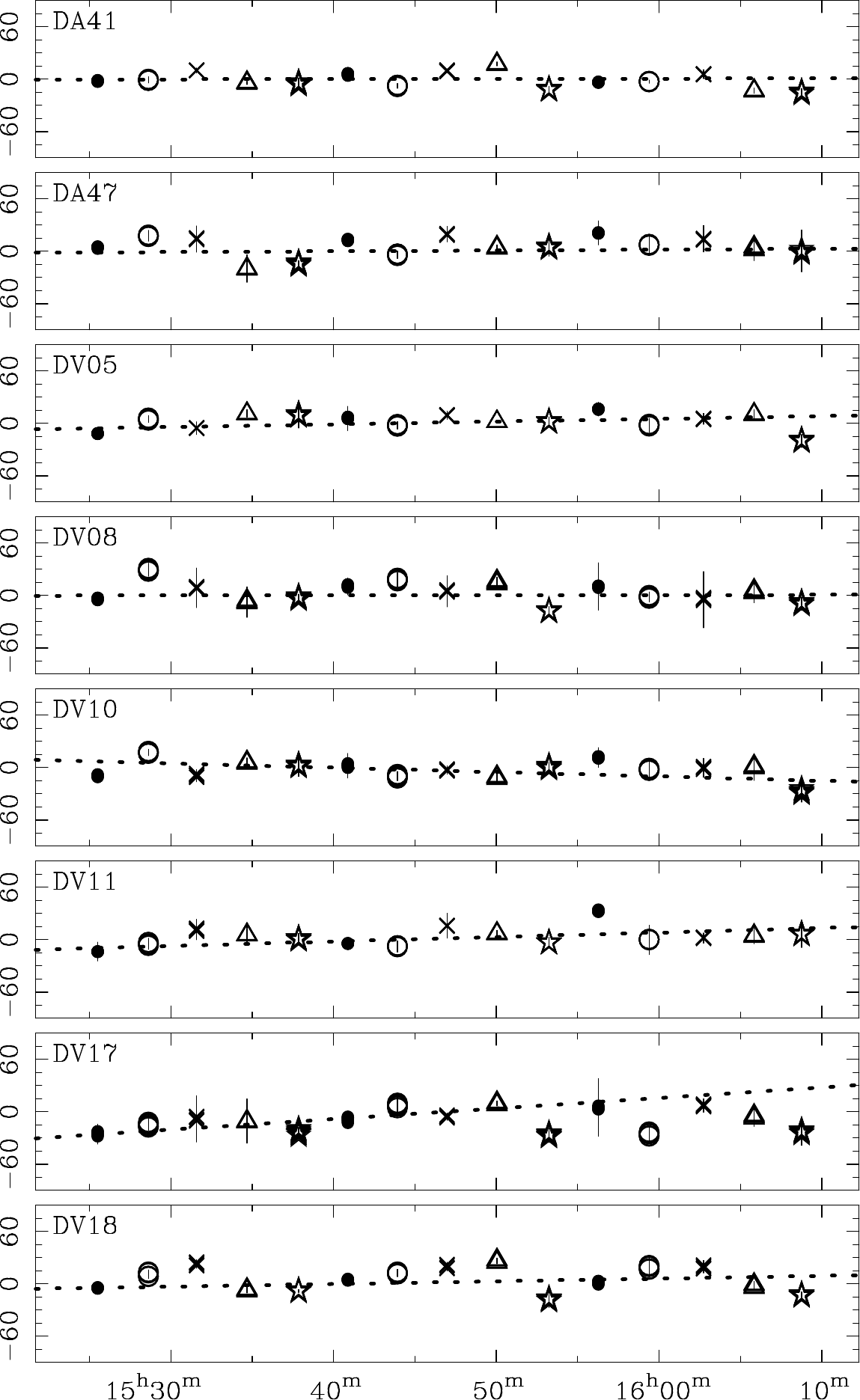}
\end{tabular}
\caption{
Stage~2 DGC stability experiment for Band~7--3 on 2017 April 11.
Left: 
time series of the WVR-corrected 
antenna-based phase 
in Band~7 of J0006$-$0623, J2232$+$1143, J2253$+$1608, J0108$+$0135, and J2348$-$1631 
(filled circles, open circles, crosses, open triangles, and open stars, respectively) after 
subtracting a 
mean phase per SPW per polarization pair ($XX$ and $YY$). 
Each symbol represents an 8~s scan-averaged 
phase for each SPW. 
Right: antenna-based DGC solutions per SPW per polarization pair. 
The time series are shown after subtracting a 
mean 
per SPW per polarization pair. The dotted line in each panel shows a linear 
fit to the phase residuals for each antenna. 
}
\label{fig:06}
\end{center}
\end{figure}

Figure~\ref{fig:06} 
shows the results of the same experiment but for a total of 
five bright QSOs, four SPWs, 
and two polarization pairs ($XX$ and $YY$). 
The left panels plot the WVR-corrected antenna-based phases in Band~7 
with respect to the reference antenna (DV09) before the phase correction, 
whereas the right panels plot the antenna-based DGC solution averaged 
at each DGC block. A simple mean for each SPW and each 
polarization pair was removed. 
In the left panels, there are large time variations, which can be stabilized after 
the phase correction, as shown in the right panels.  
Some antennas still show a common behavior in the corrected phase between 
the SPWs and the polarization pairs in the DGC solution. 
The randomness of the DGC solutions can be due to the residual atmospheric 
phase fluctuations as expressed in the first term of 
Equation~(\ref{eq:05}).  
Considering how B2B phase referencing for a 
target will be implemented, such drifts, if occurring, can be calibrated 
by inserting the interpolated solutions 
determined from multiple DGC blocks in an observation. 
The DGC solution changes in the same way at each antenna for 
all of the SPWs and polarization pairs. 
This indicates that the randomness 
of the DGC solution is attributed not to the thermal noise but rather 
to the atmospheric phase fluctuations and/or 
the LO signal stability difference between the Bands. 

\begin{figure}[htbp]
\begin{center}
\begin{tabular}{c}
\includegraphics[width=85mm]{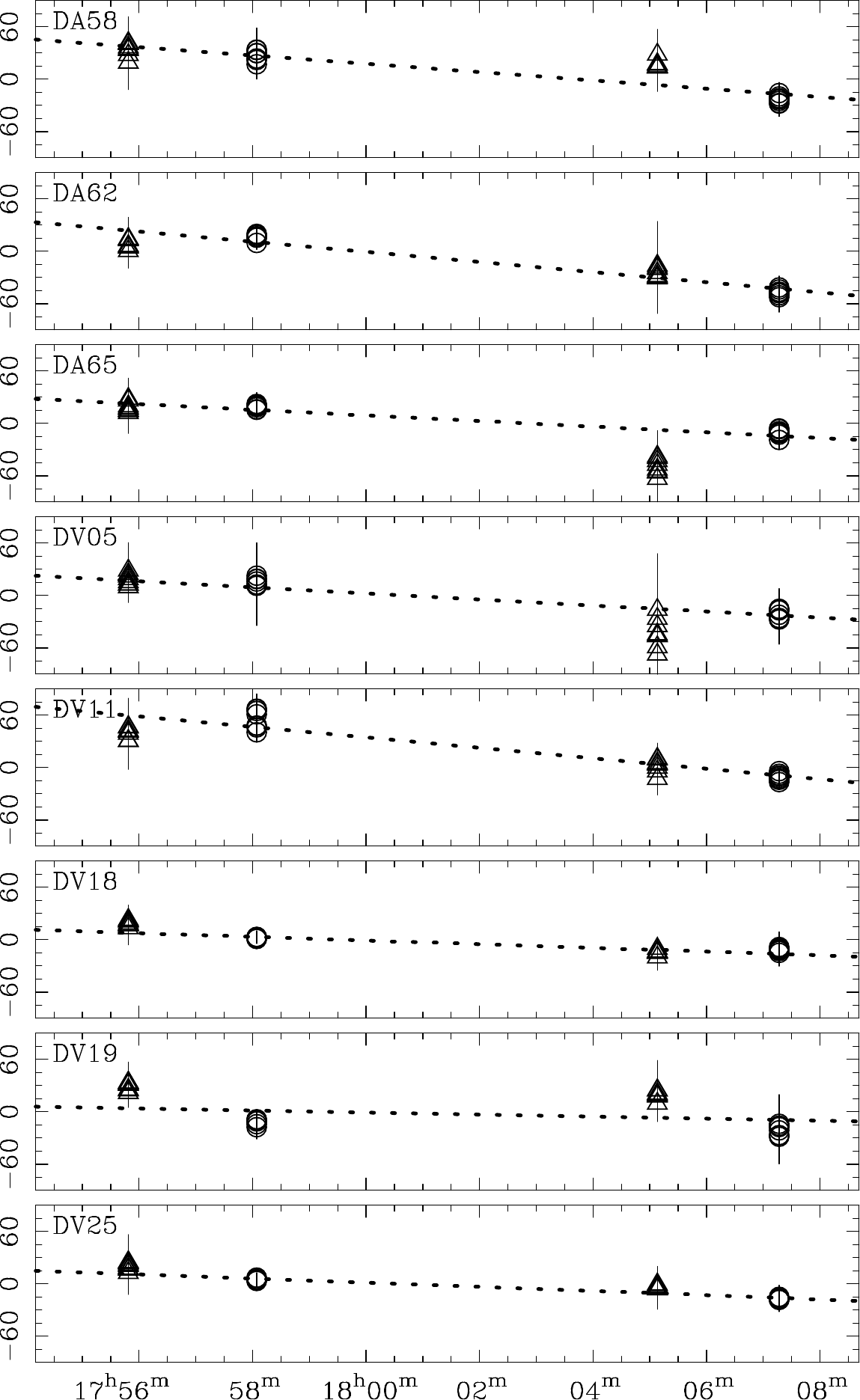}
\hspace{5mm}
\includegraphics[width=85mm]{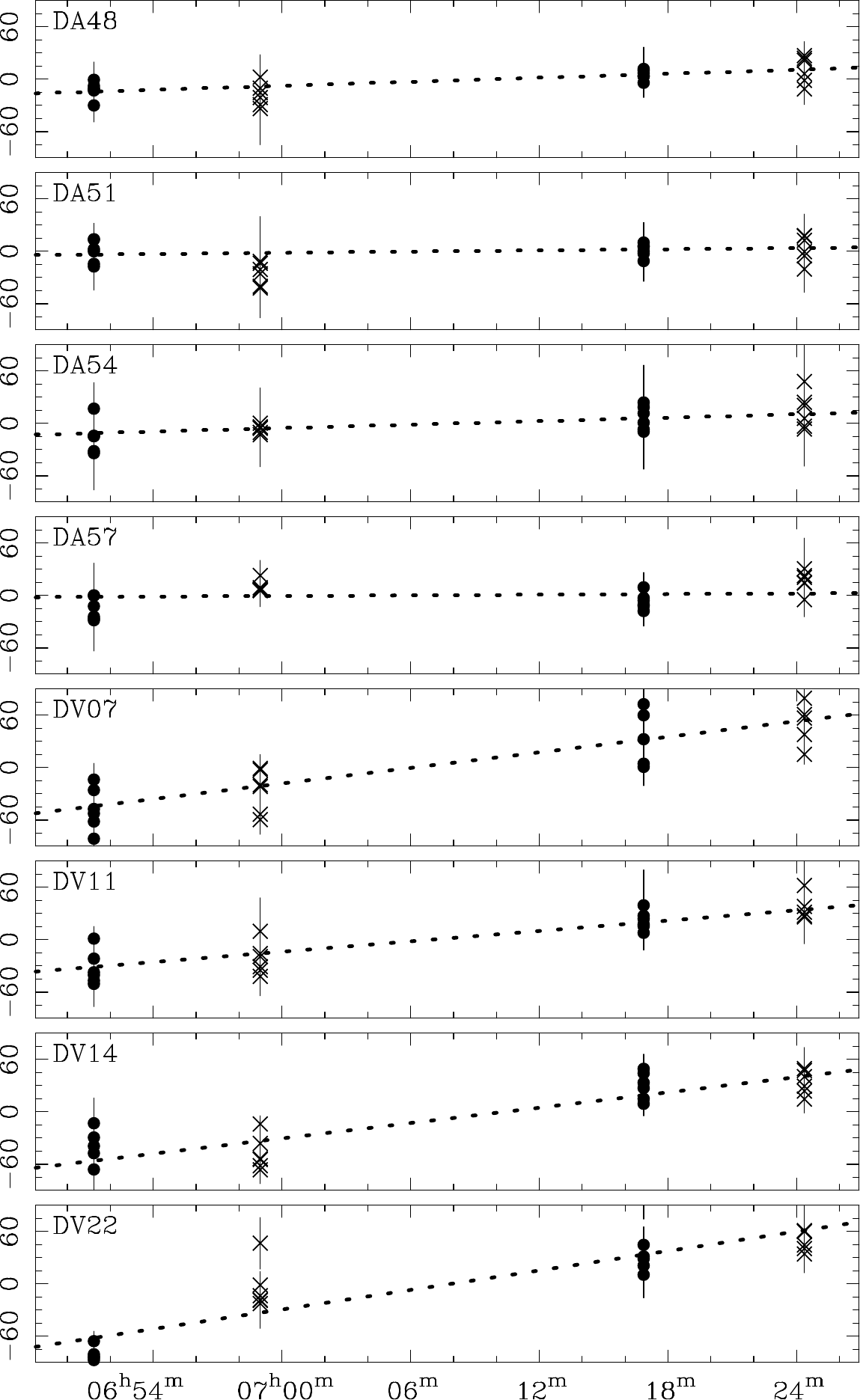}
\end{tabular}
\caption{
Stage~2 DGC stability experiments for Bands~8--4 (left) and 9--6 (right) on 2017 May 4 and April 23, 
respectively. The dotted line in each panel shows a linear 
fit to the phase residuals for each antenna. 
Left: 
in Band~8, two of the 
four observed 
QSOs are plotted 
(J0510$+$1800 and J0522$-$3627; open triangles and open circles, respectively). 
Right: 
in Band~9, two of the five observed QSOs are plotted 
(J1924$-$2914 and J1517$-$2422; filled circles and crosses, respectively). 
}
\label{fig:07}
\end{center}
\end{figure}

Figure~\ref{fig:07} 
is similar to 
Figure~\ref{fig:06}, 
but shows the DGC solutions 
of Band~8--4 (462--154~GHz) and Band~9--6 (675--225~GHz) 
conducted on 2017 May 4 and April 23, respectively. 
In Band~8--4, we observed four bright QSOs but analyzed two of them 
(J0510$+$1800 and J0522$-$3627; open triangles and open circles in the 
left panel, respectively) because the other two 
have antenna shadowing 
effects at around $30^{\circ}$ elevations. 
In Band~9--6, we analyzed two of the 
five observed QSOs 
(J1924$-$2914 and J1517$-$2422; filled circles and crosses, respectively, 
in the right panel) because the other three do not have high enough S/Ns in Band~9. 
In those HF cases, the DGC solutions show not only a 
random phase behavior but also a linear trend more or less 
with approximately a few degree per minute at maximum as represented 
with the dotted lines in the left panels. 
These linear trends  may be caused by instrumental instabilities that are under 
investigation. 

\begin{figure}[htbp]
\begin{center}
\begin{tabular}{c}
\includegraphics[width=90mm]{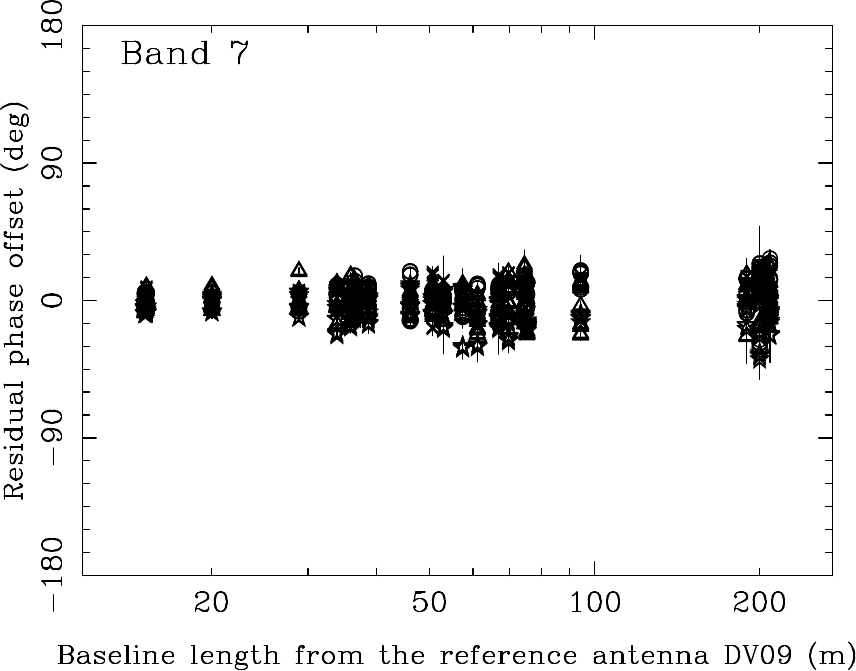} \\
\end{tabular}
\caption{
The DGC solution residuals of the stage~2 
experiment in Band~7--3 (2017 April~11) for all the antennas 
and SPWs after subtracting the linear fit 
as represented by the dotted lines in 
Figure~\ref{fig:06}. 
The horizontal axis is the antenna distance with respect to the reference antenna. 
The vertical axis is the residual DGC solution after subtracting the linear fit. 
}
\label{fig:08}
\end{center}
\end{figure}

Figure~\ref{fig:08} 
shows the 
antenna-based 
DGC solutions 
as a function of baseline length 
to the reference antenna after 
subtracting 
a single linear trend for each antenna. 
The DGC solution in Band~7 has a roughly a standard deviation of $10^{\circ}$--$20^{\circ}$, 
independent of the sky positions of the QSOs. 
We note that 200~m baselines have a larger deviation than 20~m baselines. This is thought 
to be because of the residual atmospheric phase fluctuations. 
As discussed in 
Section~\ref{sec:02-02}, 
the deviation of the DGC solution can increase until a baseline length of 
$v_{\mathrm{w}} t_{\mathrm{swt}}/2$. If we assume $v_{\mathrm{w}}=6$~m~s$^{-1}$, 
this baseline length is expected to 60~m. 

\begin{figure}[htbp]
\begin{center}
\begin{tabular}{c}
\includegraphics[width=90mm]{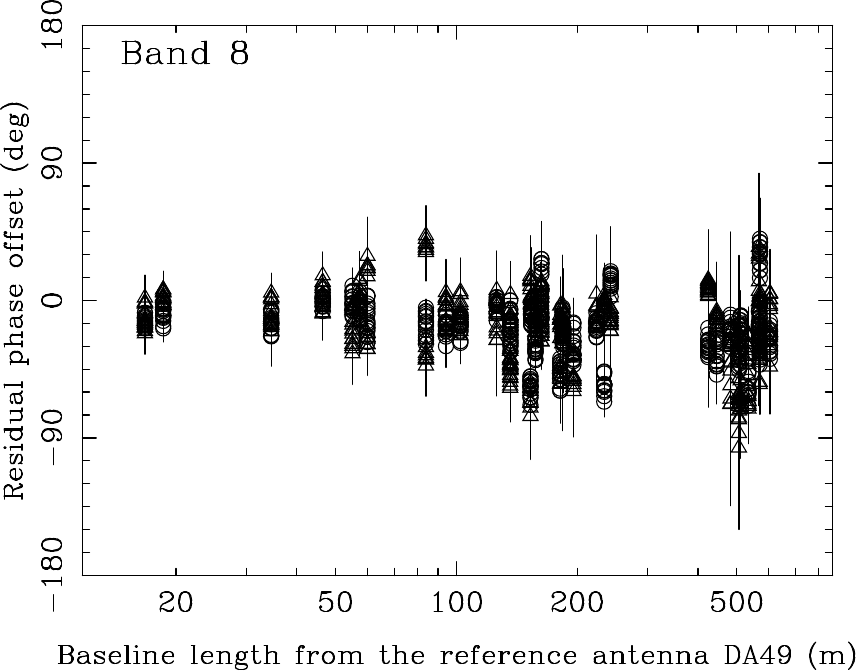}
\\
\includegraphics[width=90mm]{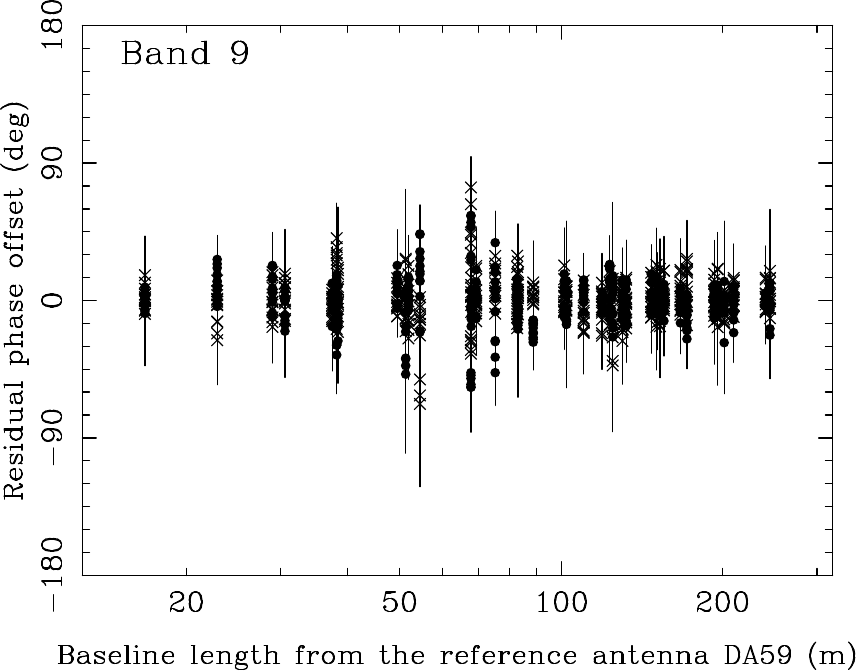} \\
\end{tabular}
\caption{
Same as 
Figure~\ref{fig:08} 
but in Band~8--4 (2017 May 4) and Band~9--6 (2017 April 23) 
in the top and bottom panels, respectively. 
}
\label{fig:09}
\end{center}
\end{figure}

Figure~\ref{fig:09} 
is similar to 
Figure~\ref{fig:08}, 
but in Bands~8--4 and 9--6. 
The DGC solutions at those HFs are also independent of sky position; 
however, the standard 
deviation of the DGC solution is increased to 
$\sim$30$^{\circ}$. 
Comparing with the Band~7--3 result, those higher-frequency experiments show 
higher phase instability mainly because of the 
residual atmospheric phase fluctuations and instrumental phase instabilities 
at the observing frequency. 
Some antennas have a large deviation between the SPWs and/or show 
that the LF and HF phases do not track each other; 
thus, the DGC solution is more variable, 
indicating 
instrumental malfunctions or instabilities, 
generating a more variable DGC solution.
Such issues led to investigation of antenna components 
that were then serviced or replaced. 

From the stage~2 test, we 
found that the DGC is satisfactorily stable, 
i.e. there are no rapid instrumental variations, 
excluding obvious problematic antennas. 
As a whole, the DGC solutions are stable for QSOs 
with positions separated by 100$^{\circ}$ apart on the sky.
For future B2B phase referencing observations, 
the DGC block will be executed two or three times 
in an observation to provide ample calibration of the instrumental phase offset difference 
and linear trend. 

The stabilities of the DGC solutions were also investigated in stages~3 and 4. 
In stage~3, a DGC source was observed at the start 
and end of an observation, separated by around 40--45~minutes; 
in stage~4, a DGC source 
was repeatedly observed every 15~minutes during 1--2 hr.  
In the stage~2 test, 
we reconfirmed that, except for some problematic antennas showing phase drifts, 
the RMS phase noise of the DGC solutions is typically 
$10^{\circ}$--$20^{\circ}$ 
in Band~7, while this can increase to $30^{\circ}$--$40^{\circ}$ in Bands~8 and 9. 
Generally, as part of the later-stage feasibility checks with a baseline out to 16~km, 
the instrumental phase offset difference 
determined from the DGC solution is stable for approximately 1~hr.

\subsection{
  Stage~3: 
  image quality 
comparative test between B2B and in-band phase referencing
}\label{sec:04-02}

Comparative studies of the HF image quality 
between B2B and in-band 
phase referencing falls into two categories: 
(1) the same target and the same phase calibrator are used for both 
      B2B and in-band phase referencing, and 
(2) B2B phase referencing uses a small separation of $1^{\circ}$--$2^{\circ}$, 
      and in-band phase referencing uses a larger separation 
      (typically $3^{\circ}$--$11^{\circ}$).  
The specific goals of this comparative study are summarized as follows: 
(1) confirm that B2B and in-band phase referencing produce the same result when using the 
     same phase calibrator and 
(2) indicate whether B2B phase referencing is an improvement over in-band phase referencing 
     if using a closer phase calibrator, as per the intended use of B2B phase referencing. 
In typical observations, the separation to phase calibrators is usually larger at higher frequencies 
due to the difficulties in finding a bright 
enough source (see 
Section~\ref{sec:05} 
for more details). 
We tried to mimic such a situation in stage~3. 

A series of the stage~3  
experiments started in 2017 July and were performed in a variety of 
weather conditions (atmospheric phase stability) and 
with a range of total number of antennas (typically $> 20$). 
Various harmonic frequency pairs were tested (Bands~7--3, 8--4, 9--4, 9--6, and 10--7).  
A single experiment was formed by the combination of the 
B2B and in-band phase referencing blocks, as well as two DGC blocks 
executed in the following sequence: 
(DGC)---(in-band phase referencing)---(B2B phase referencing)---(in-band phase referencing)---(B2B phase referencing)---(DGC). 
In the above 
sequence, the same target  
was observed for both the B2B and in-band phase referencing blocks, whereas 
different phase calibrators were used 
for each phase referencing block. 
The total length of each observing 
block 
was $\sim8$~minutes, including a system noise temperature measurement 
and pointing calibration. The full run of the 
sequence takes 45--50~minutes. 

The data reduction of the in-band phase referencing was undertaken with 
a standard 
ALMA data reduction procedure, while the B2B phase referencing data reduction procedure 
requires the application of the DGC solutions. 
For the amplitude calibration, we used the DGC source as an HF flux calibrator 
in addition to the 
system noise temperature calibration in both the B2B and in-band phase referencing.
For the imaging, we use a Briggs weighting with a robustness parameter (robust) of 0.5, 
as this is representative of the common robust generally used in ALMA images. 
A fixed number of 50~CLEAN 
iterations were made with a 15~pixel radius 
masking box around the target (located at the phase 
tracking center in the image). 
The cell size is chosen such that 5 (for Bands~7 and 8) 
or 7 (Band~9) pixels cover the synthesized beam major axis.
Due to the nature of the experiment sequence, we can cull phase calibrator scans in order to
mimic longer switching cycle times to investigate 
any potential relationships between 
the image quality, switching cycle time, and weather condition. 

\begin{figure}[htbp]
\begin{center}
\begin{tabular}{c}
\includegraphics[width=150mm]{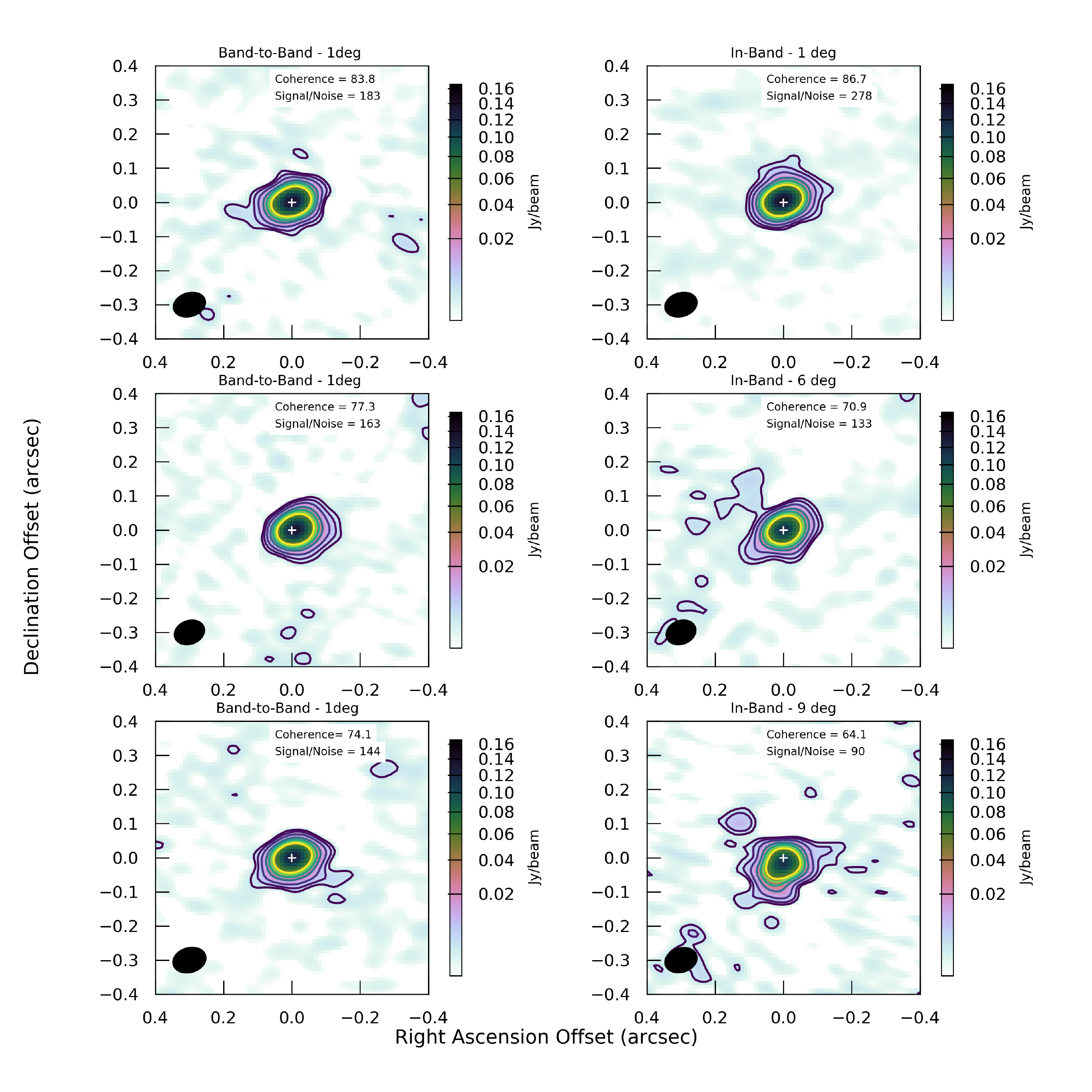}
\end{tabular}
\caption{
Synthesized images of J0633$-$2223 in Band~8 of the comparative 
experiment between B2B and in-band phase referencing on 2017 July 18 
with a $B_{\mathrm{max}}$ of 3.7~km. 
Each horizontal pair represents one experiment. 
Each row shows a consecutive execution pair and thus has similar weather conditions. 
The left panels show the B2B phase referencing images, where J0634$-$2335 at $1.^{\circ}2$ 
was used as the phase calibrator. 
The paired in-band phase referencing images are shown in the right panels 
for the same target 
where 
phase calibrators of 
J0634$-$2335 ($1.^{\circ}2$),  
J0648$-$1744 ($5.^{\circ}8$), and 
J0609$-$1542 ($8.^{\circ}7$) 
were used in the 
top, middle, and bottom panels, 
respectively. 
The images are all scaled to a flux peak of 
0.166~Jy~beam$^{-1}$,
and the colored contours (dark to light) are fixed at 
2.5, 5, 10, 20, 30, 40, and 50~mJy~beam$^{-1}$ 
to highlight the image structure. 
Note that the image coherence is given in percent in the top of each panel. 
}
\label{fig:10}
\end{center}
\end{figure}

The full study of 50~datasets will be detailed in the forthcoming paper 
\citep{Maud2020}. 
In this paper, we detail only one of the Band~8--4 experiments 
that is unique in having four observing sequences run together 
on the same night, comparing four different phase calibrator 
separation angles for the in-band phase referencing.
The LO1 frequencies were 400 and 133~GHz for the 
target and phase calibrator, respectively, in the case of B2B phase referencing. 
Figure~\ref{fig:10} 
shows a subset 
of three out of the four Band~8 images from the consecutive 
experiments on 2017 July 18 targeting QSO J0633$-$2333 
using B2B (left) and in-band (right) phase referencing with 
the switching 
cycle time of 60~s 
obtained 
by culling phase calibrator scans. 
We selected four phase calibrators, 
J0634$-$2335, 
J0620$-$2515, 
J0648$-$1744, and 
J0609$-$1542, located 
at separation angles of 
$1.^{\circ}2$, 
$3.^{\circ}8$, 
$5.^{\circ}8$, and 
$8.^{\circ}7$ 
from the target, respectively. 
The B2B phase referencing was tested only to the closest calibrator, J0634$-$2335, 
while in-band phase referencing was tested to all four. 
The synthesized beam sizes are 80--100~mas. 
The execution pair with the $3.^{\circ}8$ in-band phase 
referencing failed because the sources eventually transited close to 
zenith, 
so that more than 
half of the data had to be flagged out, 
and the resultant image quality was poor. 

In order to check the image quality after phase referencing, 
we performed phase self-calibration 
\citep{Schwab1980} 
with a short solution interval 
(here we could use a solution interval of $\sim 1$~s as the S/N was high) 
and obtained images free from the residual atmospheric and instrumental 
phase errors. 
Here we define the image coherence as the ratio between 
peak flux densities for data with and without self-calibration applied.
The higher the image coherence is, the more effective phase referencing 
we achieve, and the highest image coherence is 1. The top row images of 
Figure~\ref{fig:10} 
indicates that  
the peak flux densities are almost identical for B2B and in-band phase 
referencing---that is, the image coherence is almost unity---and that 
the image structure is pointlike with few defects, although the B2B phase referencing 
image noise is higher. 
The increased image noise is due to inaccuracies in the DGC. 
After self-calibration, the image noise of the B2B and in-band phase referencing 
are equivalent, 
indeed confirming that the residual offsets have been fully corrected.
In the middle and bottom rows of 
Figure~\ref{fig:10}, 
when the in-band 
phase calibrators are located further away, 
the image coherence and image dynamic range begin 
to decrease, and the level of defects increases. 
Since the target image with the $1.^{\circ}2$ 
phase calibrator in B2B phase referencing 
remains largely unchanged, this degradation in the image quality 
is considered to be attributed to the larger separation angles. 
Assuming that the residual RMS phase noise in the visibility has characteristics of a Gaussian 
random noise, the image coherence can be equivalent to a coherence factor 
$\exp(-{{{\sigma_{\phi}}^{2}/{2}}})$, 
where $\sigma_{\phi}$ is the residual RMS phase noise in radians 
\citep{TMS2001}. 
Note, importantly, that the atmospheric phase fluctuations over 
the switching cycle time are low ($<35^{\circ}$), 
and therefore the conditions are stable and should allow all images 
to achieve $>$85\% image coherence. 

\begin{figure}[htbp]
\begin{center}
\begin{tabular}{c}
\includegraphics[width=70mm]{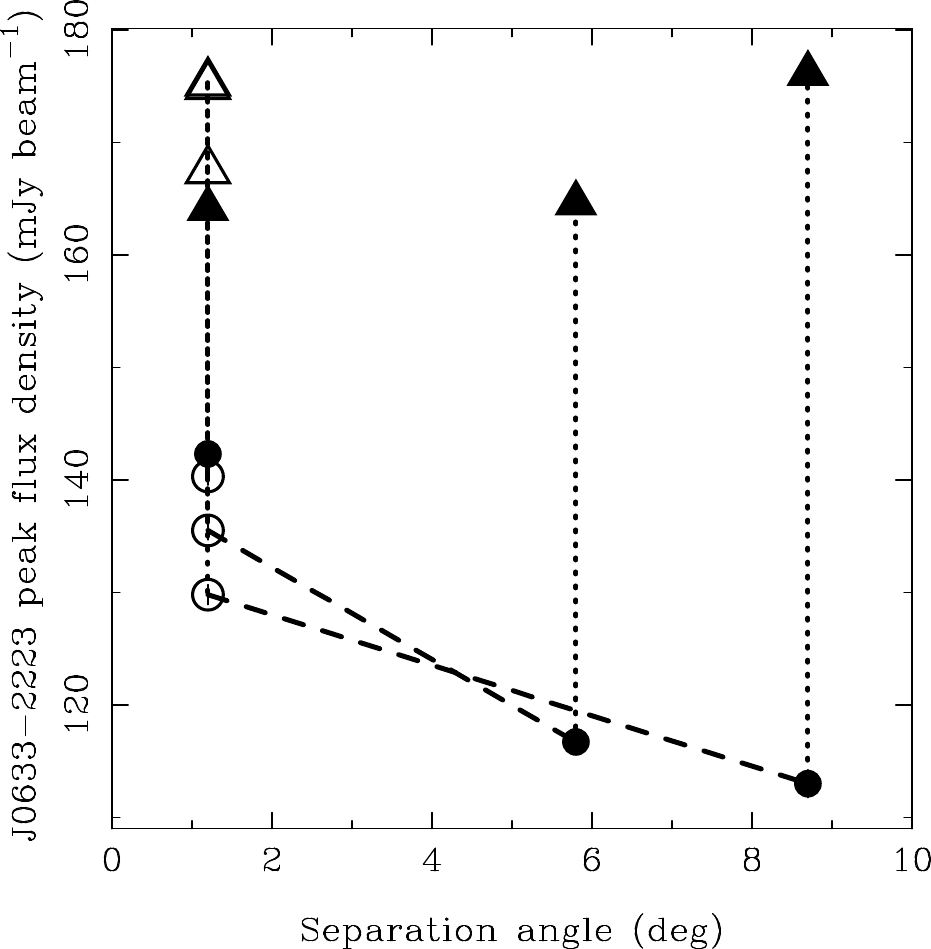}
\hspace{10mm}
\includegraphics[width=70mm]{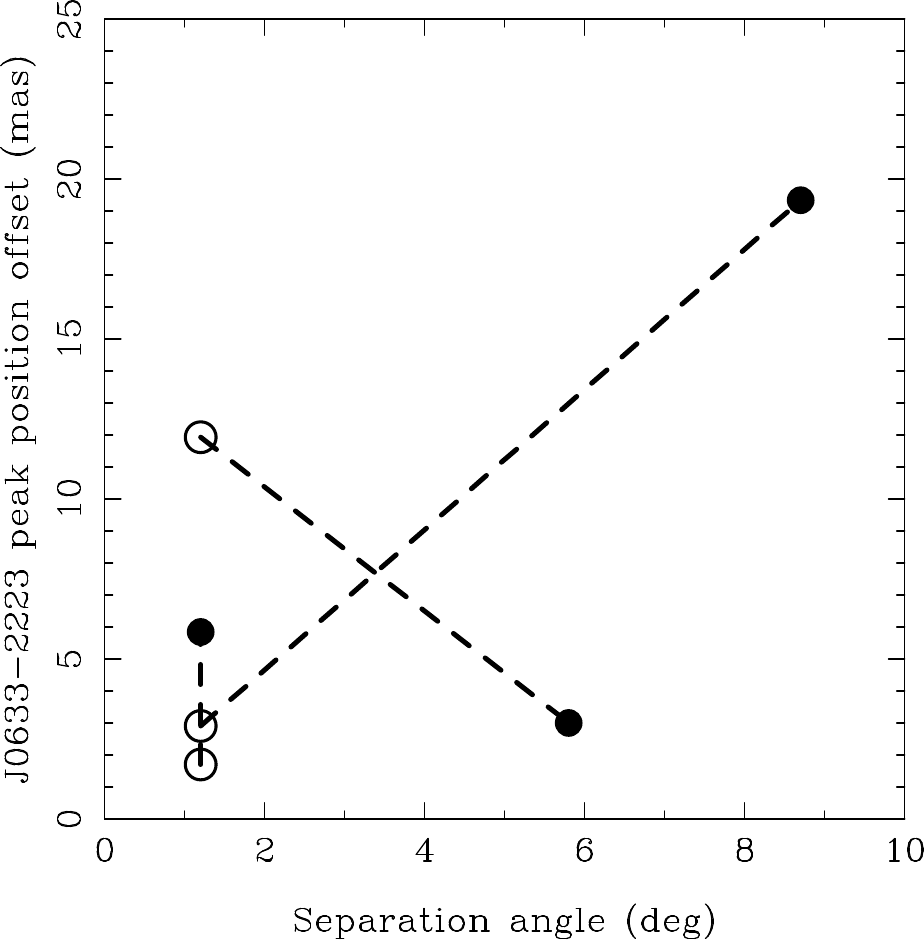}
\end{tabular}
\caption{
Compilation of the stage~3 Band~8--4 comparative study results 
as shown in 
Figure~\ref{fig:10}. 
Left: 
peak flux density of the J0633$-$2223 images. 
The horizontal axis is the separation angle to the phase calibrator. 
The filled and open circles represent the in-band and B2B phase referencing 
images, respectively. 
The triangles connected with the 
dotted lines represent their self-calibrated image. 
The 
{\bf 
dashed 
}
lines indicate 
the consecutive pairs of the B2B and in-band phase referencing 
observations.
Right: peak position offset of the J0633$-$2223 images from 
the a priori phase tracking center. 
}
\label{fig:11}
\end{center}
\end{figure}

Figure~\ref{fig:11} 
compiles the above Band~8--4 experiment. 
The left panel summarizes the peak flux densities of the target images in 
Figure~\ref{fig:10} 
as a function of the separation angle, 
clearly showing 
the decrease in peak flux with increasing separation angle.
Relative to the image obtained for the closest phase calibrator, 
the peak flux density falls 
below 80\% for the images with the $5.^{\circ}8$ and $8.^{\circ}7$ separation 
angles. Relative to the self-calibrated image, the image coherence is 
71\% and 64\% for the $5.^{\circ}8$ and $8.^{\circ}7$ separation angles.

The right panel of 
Figure~\ref{fig:11} 
shows the offset of the image peak positions from 
the a priori phase tracking center of the target. 
The imaging result has a positional uncertainty of 10~--~20\% 
of the synthesized beam size for a separation angle of up to $8.^{\circ}7$, 
although the image with the $8.^{\circ}7$ in-band 
phase calibrator becomes more distorted than the others 
with the closer phase calibrators. Considering the image coherence 
and defects, the phase calibrator in this case should be located closer 
than $5^{\circ}$--$6^{\circ}$. 

Taking into account all the sets of comparative experiments, 
the image coherence of B2B phase referencing is comparable to that of 
in-band phase referencing when using the same close phase calibrator, 
although naturally in-band phase referencing is marginally better considering 
the additional DGC required for B2B phase referencing. On the other hand, 
we found that in-band phase referencing has a noticeable degradation 
of image quality 
in terms of the image coherence and defects with increasing separation angle. 
Specifically, for the longest baseline test data, 
where $\nu_{\mathrm{_\mathrm{HF}}} < $300~GHz and maximal baselines 
were 15~km in stage~3, provided the atmospheric stability is good (over 
the switching cycle time), in order to obtain an image coherence $\geq 70$\% 
the phase calibrator separation angle should be within $\sim 6^{\circ}$ 
Although we have not systematically tested the B2B phase referencing image quality 
for a variety of separation angles, 
the tendency is expected to be the same as the image quality with in-band phase 
referencing if using more distant phase calibrators. 
Since B2B phase referencing requires the additional DGC, 
our findings of the separation angle dependency of in-band phase referencing 
are applicable as an upper limit of the image quality of B2B phase referencing. 

For higher frequencies than 400~GHz, 
although we have small number statistics and the baseline 
lengths are shorter, $< 5$~km, the degradation seems to be respectively worse; 
i.e., it is a function of frequency and baseline length. 
This makes sense as the worse 
calibration causing a reduced coherence are due to delay errors related to underlying 
uncertainties in antenna positions 
\citep{Hunter2016}, 
and are frequency 
dependent when converted to phase. For higher frequencies, it is likely that 
calibrators even closer than $6^{\circ}$ would be 
required for the longer baseline.

\subsection{
  Stage~4: 
  high angular resolution imaging test
}\label{sec:04-03}

In order to investigate the feasibility of high angular resolution imaging with extended 
array configurations at high frequencies, ALMA quasi end-to-end observation experiments 
were arranged 
in stage~4. 
Here we present Band~7--3 and Band~8--4 experiments of a point source (QSO), 
and Band~9--4 experiments of two extended sources (HL~Tau and VY~CMa) that 
have complex structures.
In stage~4, we experimentally applied a $90^{\circ}$ phase switching in the correlator
\citep{TMS2001} 
in Band~9, 
so that the bandwidth of the HL~Tau and VY~CMa observations 
was doubled, compared to that of 
normal science observations in Cycle~5. 

\begin{figure}[htbp]
\begin{center}
\begin{tabular}{c}
\includegraphics[width=80mm]{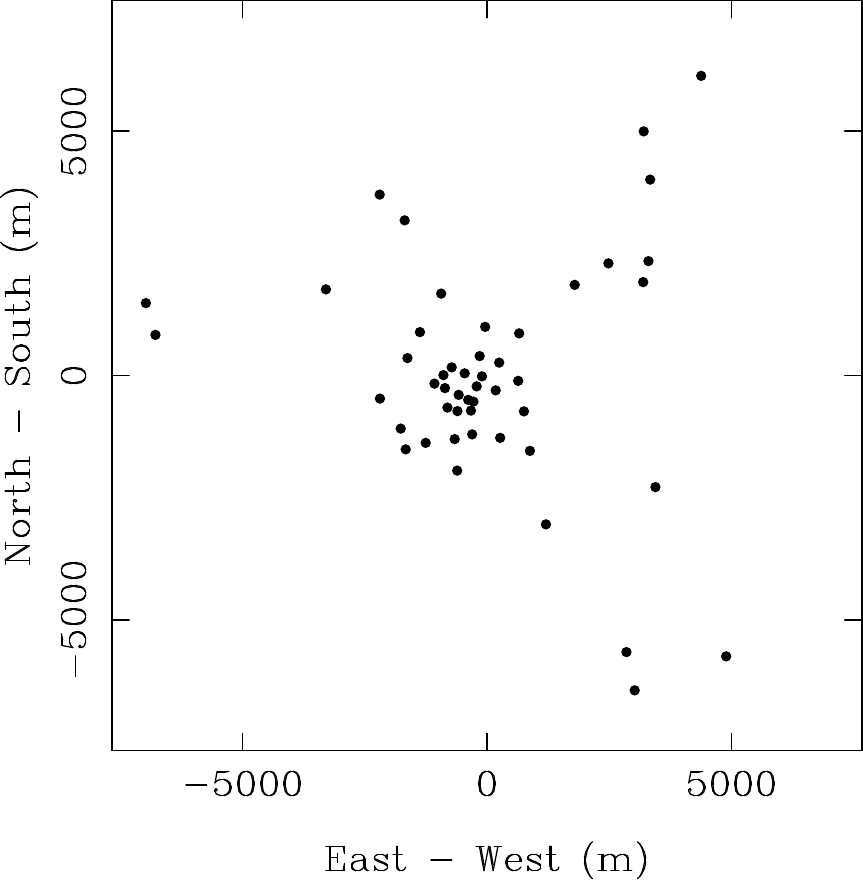}
\end{tabular}
\caption{
Array configuration on 2017 November 3 used for the stage~4 imaging test. 
The black circles represent the 12~m antenna position. 
}
\label{fig:12}
\end{center}
\end{figure}

Figure~\ref{fig:12} shows the array configuration of the stage-4 test on 2017 Nov 3.
The longest and shortest baseline lengths are 13.8~km and 133~m, corresponding to 
angular resolutions of 7~mas and 0.7~arcsec in Band~9, respectively. 
Note, there were only a few baselines shorter than 200~m during the stage-4 
period and thus we could not properly sample structures larger than 0.2~arcsec.
The observed target sources, phase calibrators and DGC sources of the experiments are listed in 
Table~\ref{tbl:05}.

\subsubsection{
  Point-source target: 
  QSO J2228$-$0753 in Bands~7--3 and 8--4
}\label{sec:04-03-01}

\begin{figure}[htbp]
\begin{center}
\begin{tabular}{c}
\includegraphics[height=85mm]{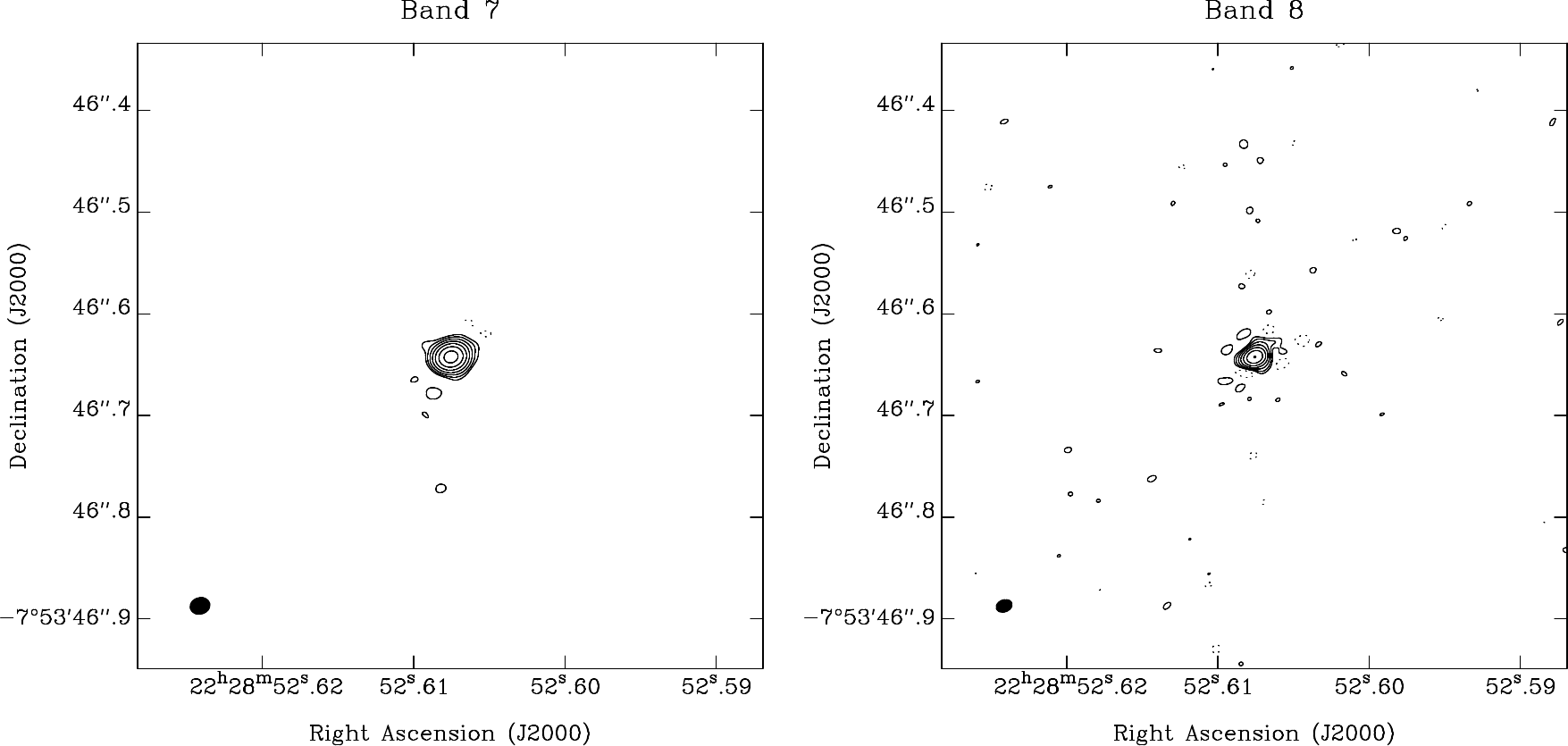}
\end{tabular}
\caption{
Continuum synthesis image of 
{\bf 
QSO J2228$-$0753
}
calibrated with B2B phase referencing.  
The contours are drawn at $-0.5$, 0.5, 1.0, 2.0, 4.0, 8.0, 16.0, and 32.0~mJy beam$^{-1}$ levels. 
The synthesized beam is shown in the bottom left corner of each panel. 
Left: 
Band~7 map using the Band~3 phase calibrator. 
The synthesized beam size with Briggs weighting (robust$=$0.5) is $19 \times 16$~mas. 
The peak flux density and image RMS noise are 
45.32 and 0.10~mJy~beam$^{-1}$, respectively. 
Right: 
Band~8 map using the Band~4 phase calibrator. 
The synthesized beam size is $16 \times 12$~mas. 
The peak flux density and image RMS noise are 
32.35 and 0.16~mJy~beam$^{-1}$, respectively.  
}
\label{fig:13}
\end{center}
\end{figure}

The QSO~J2228$-$0753 was observed as a continuum point-source 
target in Bands~7 and 8, while 
QSO~J2229$-$0832, with a separation angle of $0.^{\circ}7$, was observed as a phase 
calibrator at an LF. We selected a bright QSO, J2253$+$1608, as a DGC source located 
$25^{\circ}$ away from the target. 
The high LO1 frequencies are 289 (Band~7) and 405~(Band~8)~GHz, 
while the corresponding 
low LO1 frequencies are 96~(Band~3) and 135~(Band~4)~GHz, respectively. 
Array configurations with 
$B_{\mathrm{max}}\sim 14-16$~km 
were arranged containing 40--50 12-m antennas. 
The observations were carried out using standard science scheduling blocks 
\citep{Nyman2010}. 
We used a 20~s switching cycle time for B2B phase referencing between 
the target and phase calibrator, 
as well as for the frequency switching cycle on the DGC source. 
The on-source scan length at the HFs was 8~s. 
The left and right panels of 
Figure~\ref{fig:13} 
show the synthesized images in Bands~7 and 8, respectively, 
using a Briggs robust 0.5 weighting in CASA CLEAN 
when combining two EBs taken in the middle of October and the beginning of 
2017 November. 
The achieved beam sizes are  
$19 \times 16$ and $16 \times 12$~mas in Bands~7 and 8, respectively. 
Further details of the observations and data 
analysis results are described in 
another paper
\citep{Asaki2019}. 

We also performed phase self-calibration with a solution interval of the target scan length 
and obtained the images free from the atmospheric and instrumental phase errors in order to 
investigate the image coherence. 
We obtained high image coherences of 94\% and 84\% 
in Bands~7 and 8, respectively, so that B2B phase referencing works effectively 
in those experiments. 
We note that there are cases in imaging where a Briggs robust of $>0.5$ or, 
in extreme cases, a natural weighing with an addition taper are required to notably increase 
the beam size ($>50$\%) and the sensitivity to larger angular scales in order to mitigate resolving 
out a source with considerable extended structures.
In these QSO observations, it does not matter if one adopts 
a robust of 0.5 because 
the observed QSO is a point source even with the above angular resolutions.  
The 94\% image coherence in Band~7 corresponds to an RMS phase noise of 
$21^{\circ}$, 
and the Band~8 image coherence of 84\% is consistent with an RMS phase noise of 
$34^{\circ}$. 

\subsubsection{
  Complex structure target I: HL~Tau in Band~9--4
}\label{sec:04-03-02}


Located in the Taurus molecular cloud,  
HL~Tau is a protoplanetary disk system surrounding a young star at a distance of 
140~pc 
\citep{Rebull2004}. 
This system is very young, and its age is estimated to be less than 1~Myr 
\citep{Beckwith1990,
Robitaille2007}. 
It is the first of two complex structure sources observed to examine whether 
the image quality of B2B phase referencing is as expected in Band~9 when compared 
with the known high-fidelity image from previous ALMA SV data. 
The SV observations of HL~Tau show substructures of bright rings and dark gaps 
in the disk in Bands~3, 6, and 7, 
which strongly indicate the presence of protoplanets 
\citep{ALMApartnership2015b}. 

Two B2B phase referencing EBs were taken on 2017 November~3 with switching 
cycle times of 37 and 26~s, respectively. The scan lengths obtained for 
HL~Tau were 27 and 16~s, while a 6~s scan was obtained on a phase 
calibrator, QSO~J0431$+$1731, 0.$^{\circ}7$ away from HL~Tau. The total observing 
time was 65 and 100~minutes for the first and second executions, respectively. 
Combined, there is 45~minutes on-source time for HL~Tau. The DGC source 
QSO~J0522$-$3627 is 56$^{\circ}$ away from the target and was repeated twice in the 
first EB and three times in the second. The LO1 frequencies for the target and the 
phase calibrator are 671~(Band~9) and 149~(Band~4)~GHz, respectively. 
The correlator was configured to have eight SPWs in Band~9 with a bandwidth of 2~GHz 
each using the $90^{\circ}$ phase switching.
The PWV was 0.53~mm. Due to instrumental instabilities, a number of flags 
were applied so that some of the longest baseline antennas were flagged out; 
thus, we could not achieve the expected $\sim$10~mas angular resolution with the EBs.
We evaluated the image fidelity of our Band~9 data by comparison with the long baseline SV 
in Band~7 in LBC-2014 
(ADS/JAO.ALMA \#2011.0.00015.SV, hereafter LBC-2014-SV). 

\begin{figure}[htbp]
\begin{center}
\begin{tabular}{c}
\includegraphics[width=170mm]{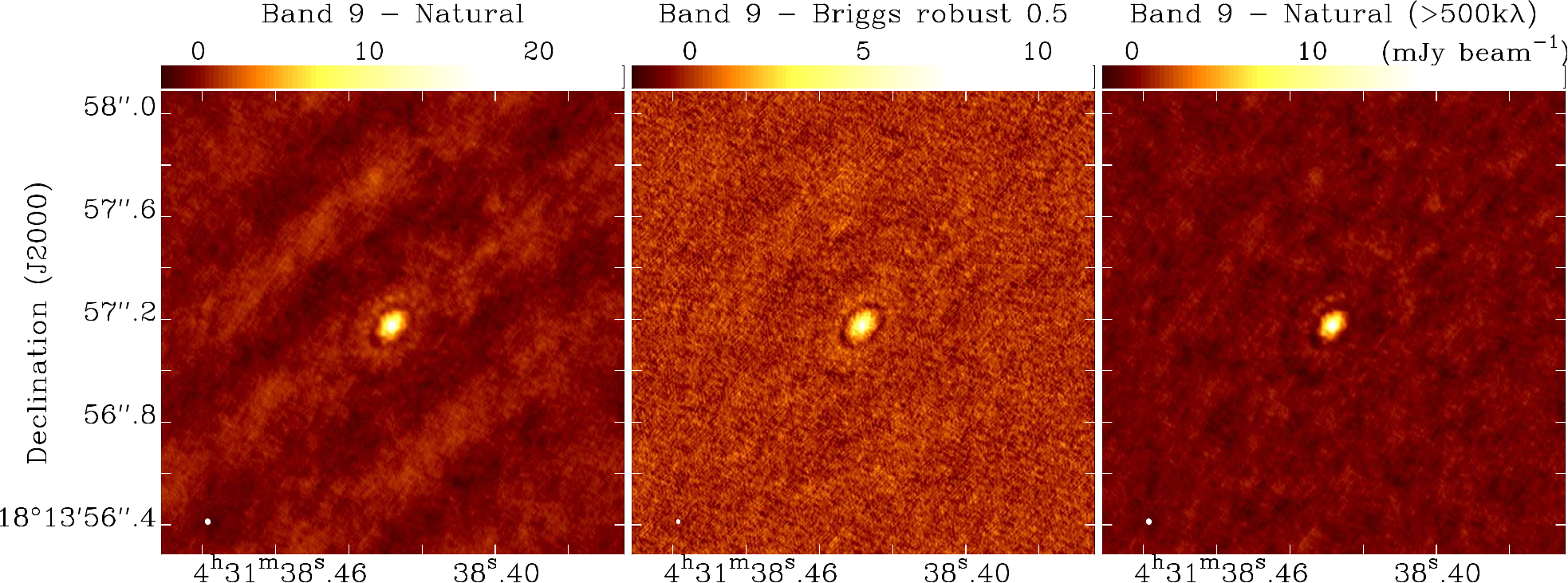}
\\
\\
\includegraphics[width=170mm]{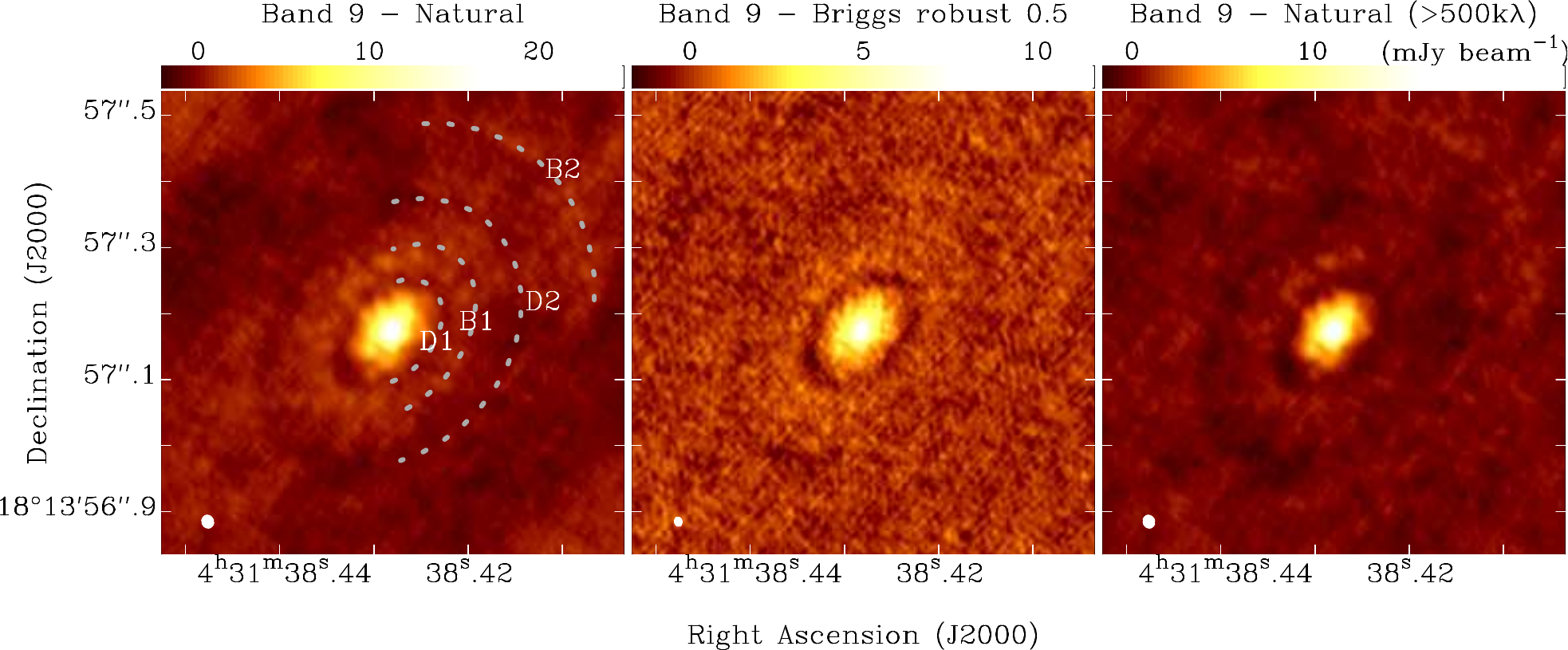}
\end{tabular}
\caption{
Top left: Band 9 continuum map of HL Tau calibrated with B2B phase referencing.  
The resulting beam is 20$\times$18\,mas using a natural weighting. 
Top middle: same as left but where a Briggs robust image weighting
of 0.5 is used so that the resolution is increased to 14$\times$11\,mas. 
Top right: same as left but including baselines $>$500\,k$\lambda$ only. 
The beams are shown in the bottom left corner. 
Bottom row: Enlarged area of the central $700 \times 700$~mas of the top three images. 
The color gradation is modified from a linear scale to highlight the fainter emission. 
}
\label{fig:14}
\end{center}
\end{figure}

Figure~\ref{fig:14} 
presents the HF-LBC-2017 images of HL~Tau in Band~9 after combining the two EBs.  
For the left and middle images, natural and Briggs (robust $=$ 0.5) weightings are used, 
as Briggs weightings with smaller robustness 
parameters form too-sharp beams resolving out 
almost all of the extended emission. The final beams achieved were 20$\times$18 and 
14$\times$11\,mas, respectively. Using the high-S/N bandpass calibrator observation, 
we performed an assessment of the Band~9 RMS phase 
with a time interval of 30~s, slightly longer than the switching cycle time of the first EB. 
The RMS phase was still considerably high around 50$^{\circ}$; thus, the atmosphere 
was at the stability limit for Band~9 when already accounting 
for the very short switching cycle time. 
Under the assumption that phase referencing corrects 
all timescales longer than a half of the switching cycle time, 
we would expect an image coherence of up to 65\%--70\%. 
Using the simple 
assumption that the central peak of emission is completely optically thick 
($\alpha$=2), scaling from the LBC-2014-SV Band~7 data 
(11.56~mJy~beam$^{-1}$, 
\citealp{ALMApartnership2015b}), 
the estimated peak flux density 
at 671~GHz would be expected to be $\sim$44~mJy~beam$^{-1}$. 
The naturally weighted HF-LBC-2017 Band~9 image yields a peak of 
24.9~mJy~beam$^{-1}$, 
suggesting a coherence of around 56\%. We have not, 
however, performed any form of self-calibration and note that the beam is slightly 
smaller compared to the LBC-2014-SV image; hence, this is likely the lower limit 
of the coherence.

In our Band~9 image, a few clear features related to the 
bright and dark lanes 
in the disk can be discerned. Measured along the major axis, 
at a radius of $\sim 0.''09$ (13\,au), we see a strong depression of emission, 
while at $\sim 0.''13$ (19\,au), there is a slight increase in flux. The
features are coincident with D1 and B1, dark and bright features reported by 
\cite{ALMApartnership2015b}. 
The next dark region at $\sim 0.''22$ (31\,au) related to feature D2 is only partially 
visible in the naturally weighted image when contrasted against the somewhat brighter 
feature to the southeast at $\sim 0.''26$ (37\,au) that is an arc shape representing 
the incomplete B2 ring. The Briggs-weighted image resolves out any larger scales.
However, we have to mention that a simple comparison is not exactly fair 
because the total on-source time is 
45\,minutes and we have not performed any self-calibration, whereas in 
\cite{ALMApartnership2015b},  
the images are 
comprised of 10~EBs with 5\,hr on-source time, and 
self-calibration was performed. 
Furthermore, due to the relatively short on-source time and the handful of short baselines, 
the poor ($u$,~$v$) sampling of the largely extended disk structure causes a striped side-lobe 
pattern throughout our Band~9 image at the 50\%--60\% level 
with natural weighting. Spatially the side lobe 
is roughly colocated with the bright large ring B6; however, its structure is far from complete. 
Natural weighting is not the optimal choice; however, the robust 0.5 Briggs weighting
already begins to resolve the disk bright and dark substructures. 
Similarly, the right panels of 
Figure~\ref{fig:14} 
are after reimaging but excluding the short baselines $<$500\,k$\lambda$ 
($\sim$220\,m in Band~9 sensitive to scales $>0.''5$) and using a natural weighting. 
Some of the side lobes are alleviated, but again, the 
disk substructures start to be resolved out.

\begin{figure}[htbp]
\begin{center}
\begin{tabular}{c}
\includegraphics[width=170mm]{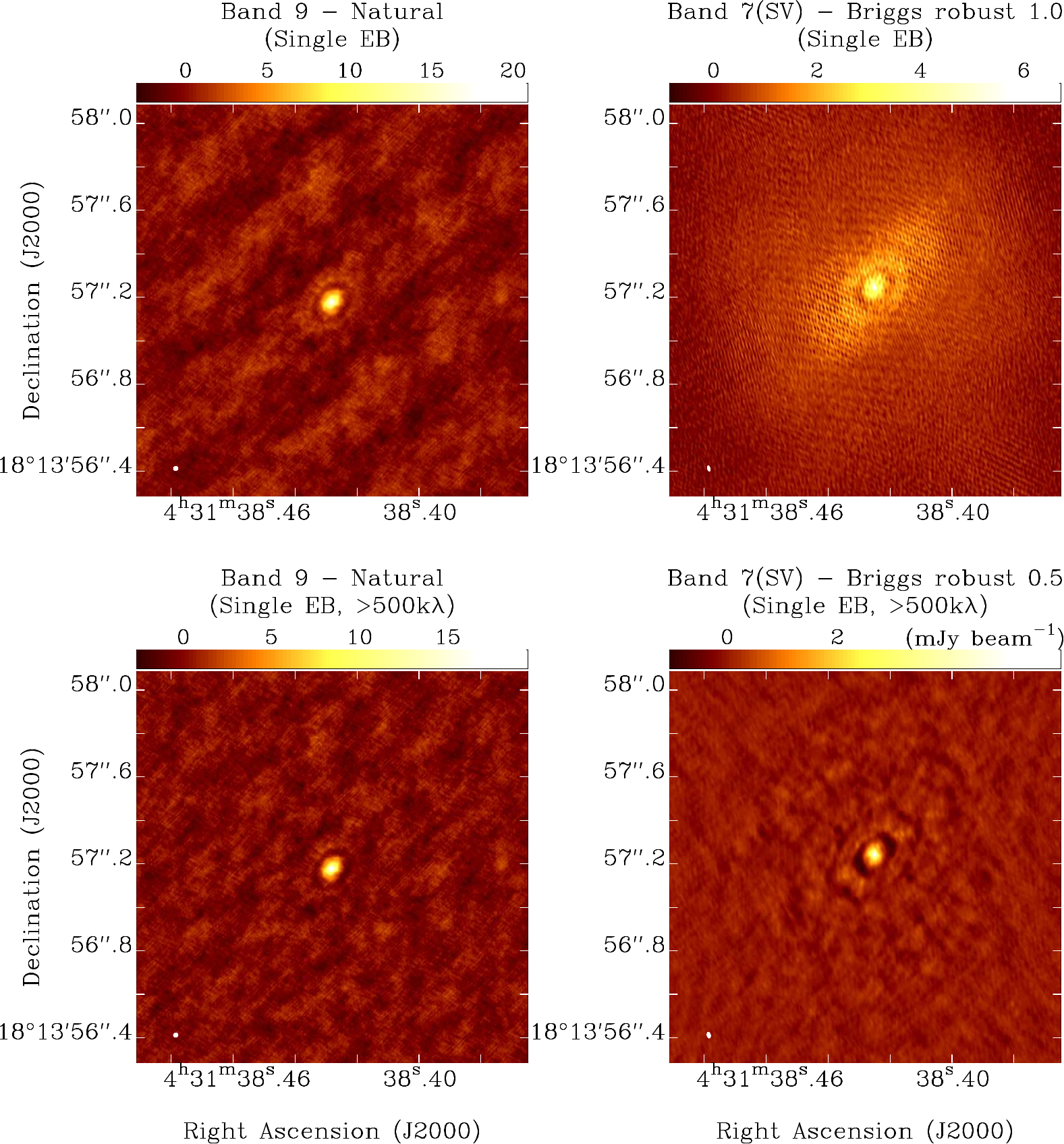}
\end{tabular}
\caption{
Top left: Band~9 continuum map of HL~Tau from one EB only, calibrated with B2B phase referencing. 
Top right: Band~7 continuum map of HL~Tau using only one EB from LBC-2014-SV. 
Bottom left: same as top left except imaged using only baselines $>$500\,k$\lambda$. 
Bottom right: same as top right except imaged using robust 0.5 and only baselines $>$500\,k$\lambda$. 
The beams are shown in the bottom left corner.
}
\label{fig:15}
\end{center}
\end{figure}

To provide a fairer comparison of the image structure, we use only one 
Band~9 HF-LBC-2017 
EB compared with a Band~7 image with a single LBC-2014-SV EB. We used 
the second execution of the HF-LBC-2017 data with the shorter switching cycle 
time and longer on-source time (26~minutes). Comparatively, there was a similar 
on-source time for each single execution of 
the LBC-2014-SV data (30\,minutes), although HF-LBC-2017 has twice the bandwidth using  
the $90^{\circ}$ phase switching. 
The LBC-2014-SV Band~7 image using the 
full ($u$,~$v$) range is shown in the top right panel of 
Figure~\ref{fig:15}. 
A robust of $-1$ was used for the LBC-2014-SV data to push toward higher angular 
resolutions and provide a better match with the HF-LBC-2017 angular resolution. The resultant 
synthesized beam size is 24$\times$10\,mas. The equivalent B2B 
phase referencing image in Band~9 is shown on the 
top left, which has a beam of 
20$\times$18\,mas. In the single EB for the 
LBC-2014-SV data, the ringlike structure in the central $0.''3$--$0.''4$ is reasonably 
evident, whereas in HF-LBC-2017, the image striping dominates the eye, 
again due to using natural weighting to still be sensitive to the 
disk bright and dark substructures. 
For further comparison, when limiting the ($u$,~$v$) range of the LBC-2014-SV 
(and also changing to robust$=$0.5 to keep the beam reasonably matched to 
the B2B phase referencing image) and HF-LBC-2017 data the resultant images are more similar, 
as shown in the bottom left and right panels of 
Figure~\ref{fig:15}, 
respectively, 
because as expected, both resolve out the majority of the extended structure. 
Aside from the unfortunate ($u$,~$v$) sampling, based upon these fairer comparisons, 
our Band~9 image of HL~Tau is 
arguably 
as good as that in Band~7.

\subsubsection{
  Complex structure target II: VY~CMa in Band~9--4
}\label{sec:04-03-03}

Our second high-resolution imaging experiment targeted VY~CMa, 
located at a distance of 1.17~kpc 
\citep{Zhang2012}. 
It is an oxygen-rich red supergiant (RSG) 
with a variable but exceptionally high mass-loss rate. 
The high mass-loss rate (up to $10^{-3}$~$M_{\odot}$~yr$^{-1}$) 
provides bright spectral line and continuum emissions within a few arcseconds, 
making 
it one of the best targets for the image fidelity check by comparison with 
the previous SV in  Band~9 in LBC-2013 with a $B_{\mathrm{max}}$ of 2.7~km 
(ADS/JAO.ALMA~\#2011.0.00011.SV, hereafter LBC-2013-SV) 
not only for the continuum emission but also for the 
658.00655~GHz $v_{2}=1,$~$1_{1,0}-1_{0,1}$ 
H$_{2}$O maser 
\citep{Richards2014}. 

On 2017 November 3,VY~CMa was observed in Band~9 with a switching cycle time 
of 82~s. The scan lengths for VY~CMa and the phase calibrator 
QSO~J0725$-$260, 
at $1.^{\circ}1$ away from VY~CMa, were 62 and 12~s, respectively. 
The total observing time was 47~minutes, 
and the on-source time for VY~CMa was 344~s. 
The PWV was 0.56~mm. 
The LO1 frequencies for the target and the phase calibrator are  
669~(Band~9) and 149~(Band~4)~GHz, respectively. 
The correlator was configured to have eight SPWs 
in Band~9 with a bandwidth of 1.875~GHz each 
using the $90^{\circ}$ phase switching.
One of the SPWs covers the 658~GHz H$_{2}$O maser with 
the frequency resolution of 15.625~kHz. 
A bright QSO, J0522$-$3627, $28^{\circ}$ away from the target, was used as the 
DGC source, as well as for bandpass and flux-scale calibration.

\begin{figure}[htbp]
\begin{center}
\begin{tabular}{c}
\includegraphics[width=180mm]{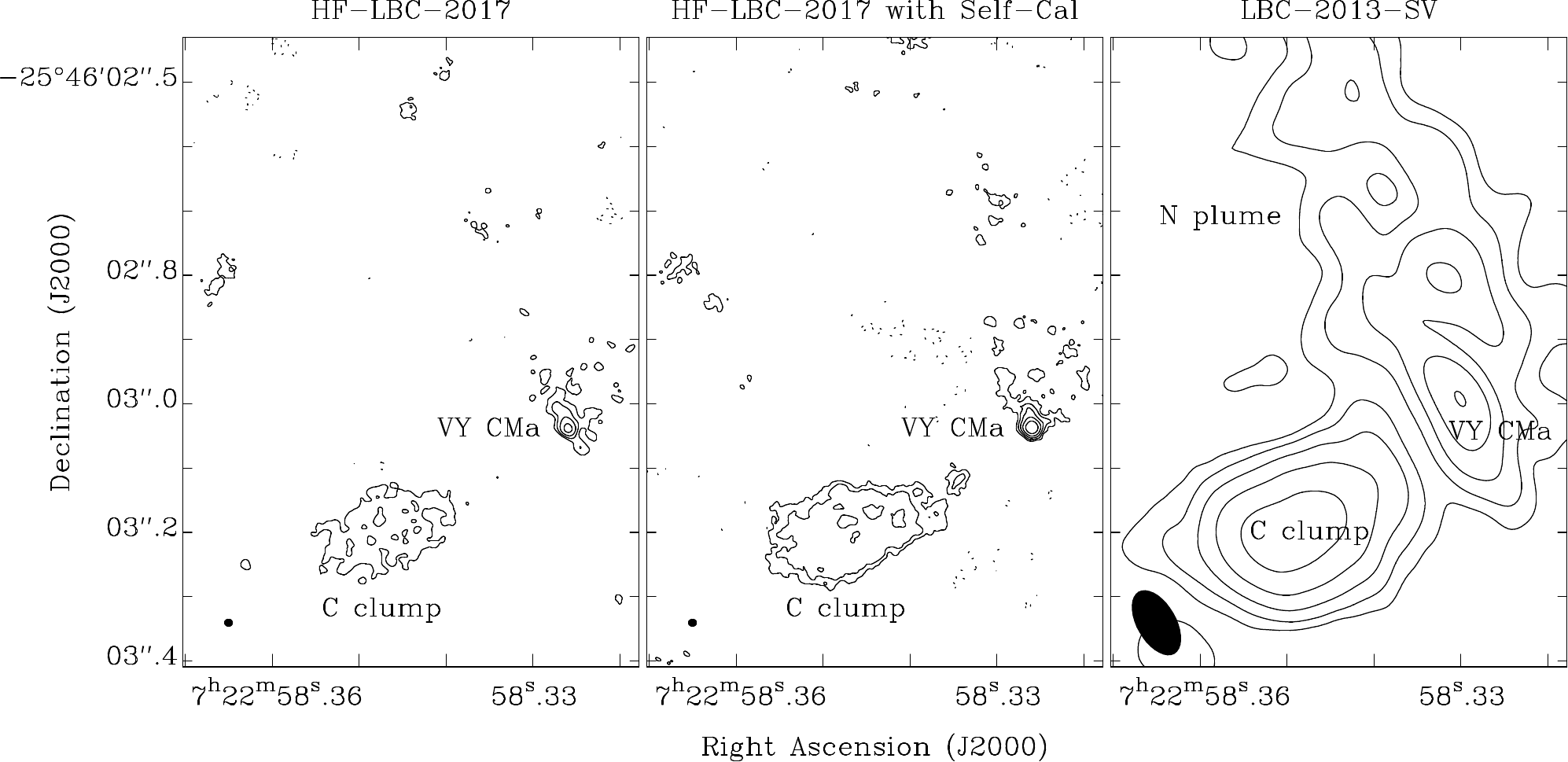}  \\
\end{tabular}
\caption{
Band~9 continuum maps of VY~CMa. 
The synthesized beam is shown in the bottom left corner of each panel. 
The contours are 
drawn
at $-3\sigma$, $3\sigma$, $6\sigma$, 
$12\sigma$, $24\sigma$, and $48\sigma$ levels. 
Left: 
HF-LBC-2017 image 
without self-calibration 
with a longest projected baseline length of 13.8~km. 
The synthesized beam size is $12 \times 11$~mas 
with Briggs weighting (robust$=2$). 
The peak flux density and 
image RMS noise are 41.5 and 1.4~mJy~beam$^{-1}$, respectively.
Middle: 
The same as the left panel but with self-calibration. 
The peak flux density and 
image RMS noise are 135.5 and 1.5~mJy~beam$^{-1}$, respectively.
Right: ALMA SV data taken in 2013 with a $B_{\mathrm{max}}$ of 
2.7~km.
Note that the brightest peak is located not at VY~CMa 
but in the C~clump (316.0~mJy~beam$^{-1}$). 
The synthesized beam size is $110 \times 59$~mas. 
The image RMS noise is 4.1~mJy~beam$^{-1}$. 
}
\label{fig:16}
\end{center}
\end{figure}

The line-free continuum emission image using the averaged eight SPWs 
with Briggs weighting (robust $= 2$) is shown in the left panel of 
Figure~\ref{fig:16}. 
The synthesized beam size is $12 \times 11$~mas, 
and the image RMS noise is 1.5~mJy~beam$^{-1}$. 
In the initial map, there is a bright and compact component (VY~CMa) 
that cannot be resolved even with the longest baselines, 
so that the self-calibration can be applied to the data 
(such a compact component is ideal for performing self-calibration).
In order to obtain the visibility data free from the atmospheric and instrumental phase errors, 
phase and amplitude self-calibrations were performed for the line-free continuum channels 
with the solution interval of 16~s to remove residual phase offsets between 
the SPWs. The self-calibration solutions were then applied to 
all target data, and the bright maser peak, which has a higher S/N, 
was used for further self-calibration, now making and applying 
the solutions with an interval of 4~s to all data. 
Improvements were seen on the continuum peak, which increased 
from 42 to 135 mJy~beam$^{-1}$, 
while the image RMS noise was largely unchanged at 
1.5~mJy~beam$^{-1}$. 
This indicates that stochastic phase errors remained even 
after B2B phase referencing, 
which gave an image coherence loss of $\sim 70$\%. 
The rather long switching cycle time in Band~9 may cause a significant 
coherence loss due to the atmosphere. 
Using the high-S/N bandpass calibrator observation 
(after applying WVR corrections only), we performed an assessment 
of the Band~9 RMS phase with a time interval of 40~s, approximately a half of the 
switching cycle. The RMS phase was extremely high for all of our Band~9 observations 
around $80^{\circ}$, corresponding to a coherence loss of 62\%. 
It is expected that a shorter switching cycle time would have been better 
for the atmospheric phase fluctuation condition, 
although due to the nature of these tests, the observations were performed 
during available time and thus not during the most optimal conditions. 
The final self-calibrated image with the eight SPWs' line-free emission 
is shown in the middle panel of 
Figure~\ref{fig:16}. 

The right panel of 
Figure~\ref{fig:16} 
shows the LBC-2013-SV continuum image for comparison with 
the synthesized beam size of 110$\times$59~mas. 
It is observed in the LBC-2013-SV image that extended, cooler dust from the 
northern plume (N~plume) and eastern clump (C~clump) are prominent, while 
VY~CMa cannot be clearly resolved due to heavy dust obscuration and the 
larger synthesized beam size. 
The diffuse C~clump can still be detected with our high angular resolution. 
This indicates that there are complex substructures in this dust emission, while 
the N~plume is almost resolved out. 
Those resolved emissions probably contribute to the slightly high continuum 
RMS of 1.4~mJy~beam$^{-1}$ while a 0.7--1.0~mJy~beam$^{-1}$ noise level is 
predicted across the line-free continuum, depending on the reference frequency.

Our self-calibrated image shows that VY~CMa can be resolved almost into 
a point source with minor diffuse emission along the same direction as the 
N~Plume. 
\cite{OGorman2015} 
estimated a stellar contribution of 111--124~mJy at around 658~GHz 
based on the analytical stellar properties. The HF-LBC-2017 resolution has 
enabled us to resolve the star with at most 
$\sim10$~--~20\% contribution from dust.  
A resolved two-dimensional Gaussian component fitted to VY~CMa in the 
LBC-2013-SV image (right panel of 
Figure~\ref{fig:16}) 
gave a flux density of 358~mJy in an area of 146$\times$74~mas after deconvolving 
the beam. 
Fitting to a similar sized-aperture for VY~CMa in our image 
gives 394~mJy at the reference frequency of 669~GHz. 
These results are consistent 
within the $\sim 20$\% error for the LBC-2013-SV fluxes.

It is expected that the unresolved point source in HF-LBC-2017 represents 
the RSG photosphere with the diameter of 11.4~mas 
\citep{Wittkowski2012}. 
The peak position of VY~CMa was estimated before the self-calibration 
to be located at 
($\alpha$,~$\delta$)$=$($7^{\mathrm{h}}22^{\mathrm{m}}58^{\mathrm{s}}.3261$, $-25^{\circ}46'03''.038$) (J2000). 
The position is offset by (47,~5)~mas from the position measured from the most astrometrically accurate 
LBC-2013-SV observations at 321~GHz in Band~7 (resolution of 180$\times$90~mas). The LBC-2013-SV 
positional uncertainty is 35~mas, and the shift is likely to be affected also by the contribution of the extended 
dust emission in the larger LBC-2013-SV beam in Band~7. 
As discussed in 
Section~\ref{sec:04-02}, 
the peak offsett is expected to be 10--20\% of the synthesized beam size 
at most using a phase calibrator within 5--6$^{\circ}$. 
We are still investigating this positional offset 
to understand whether it is intrinsic due to the annual parallax and proper motion 
\citep{Zhang2012},
and/or instrumental interferometer phase errors. 

\begin{figure}[htbp]
\begin{center}
\begin{tabular}{c}
\includegraphics[width=160mm]{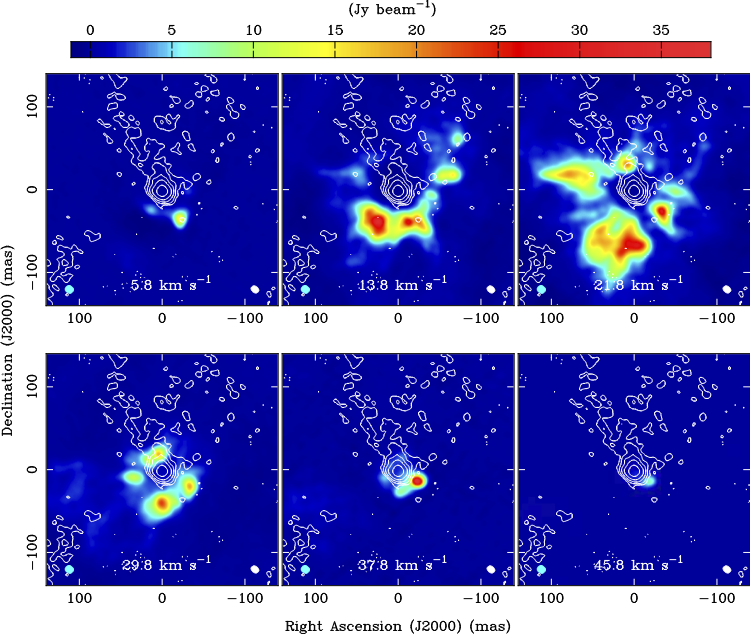}
\end{tabular}
\caption{
The 658~GHz H$_2$O maser cube of VY~CMa (color gradation) 
with Briggs weighting (robust$=0.5$) 
for a $0.''28 \times 0.''28$~region.  
The synthesized beam size is $10 \times 8$~mas, as displayed in the 
bottom left corner by a filled cyan ellipse. 
The map origin is 
centered on 
($\alpha$,~$\delta$)$=$
(07$^{\mathrm{h}}$22$^{\mathrm{m}}$58$^{\mathrm{s}}$.326,~$-25^{\circ}$46$'$03$''$.038) in J2000.
Each map was made by integrating 
the emission over a velocity width of 8~km~s$^{-1}$. 
The radial velocity in LSRK is shown in the bottom in each map. 
The Band~9 continuum emission map with Briggs weighting (robust $= 0$) 
is also displayed with the white contours of 
$2\sigma$, $4\sigma$, $8\sigma$, $16\sigma$, $32\sigma$, and $64\sigma$ 
($1\sigma=1.8$~mJy~beam$^{-1}$), and the synthesized beam is displayed 
in the bottom right corner by a filled white ellipse. 
}
\label{fig:17}
\end{center}
\end{figure}

A bright submm H$_{2}$O maser cube with a velocity resolution 
of 8~km~s$^{-1}$ 
was generated for the 18 averaged frequency channels as shown in 
Figure~\ref{fig:17}. 
For the H$_{2}$O maser cube, we applied Briggs weighting (robust $= 0.5$)  
to achieve a higher angular resolution because astronomical maser emissions are thought to be 
very compact (typically 1~au, or 0.9~mas at a distance of 1.17~kpc), so that the maser emission 
is compact even with a narrower synthesized beam: the resultant 
synthesized beam size is $10 \times 8$~mas. 
For comparison, the continuum emission map is superposed on the H$_{2}$O maser cube. 
The continuum image was made with Briggs weighting (robust $= 0$) to compare relative 
positions between the photosphere emission and maser cloud emission. 
The continuum emission map has a synthesized beam of 
$10 \times 8$~mas and 
the image RMS noise of 1.8~mJy~beam$^{-1}$. 
Since H$_{2}$O masers surrounding evolved stars are highly variable in general, 
we cannot identify which maser emissions correspond to those observed by 
\cite{Richards2014}, 
but it is likely that groups of 
H$_{2}$O 
masers (maser features) 
surround  
the RSG as has been observed in the LB-2013-SV H$_{2}$O maser spot map.

\section{
  Discussions
}\label{sec:05}

We have technically proved the 
effectiveness of B2B phase referencing in HF long baseline observations 
at ALMA, and thus a final question arises: 
is B2B phase referencing more beneficial for selecting a phase calibrator than 
in-band phase 
referencing? 
Since the required flux density for a phase calibrator in B2B phase referencing is a factor of the 
frequency scaling ratio higher than that for in-band phase referencing 
as expressed in 
equation~(\ref{eq:03}), 
the increase in the phase calibrator flux density and the antenna sensitivity at lower 
frequencies may not compensate for the low availability of phase calibrators at higher 
frequencies, especially with the largest frequency scaling ratio 
of Band~10--3 in the harmonic frequency switching. 

To answer this question, we conducted calibrator source counts for B2B and in-band 
phase referencing using the ALMA calibrator source 
catalog.\footnote{https://almascience.nrao.edu/alma-data/calibrator-catalogue}
In the catalogue, 3316 QSOs whose flux densities have been measured with ALMA 
are available. In our analysis, flux density at an arbitrary frequency is evaluated by fitting 
the flux density measurements between Bands~3 and 7 or Bands~3 and 6 
to a power-law function of $\nu^{\alpha}$, 
where $\alpha$ is a spectral index. If a certain calibrator 
has been observed at multiple epochs, 
we refer to the highest value in order to list all possible sources that could be used 
as a phase calibrator. 

The ALMA calibrator sources are not evenly distributed in the sky. For example, 
sources are more crowded along the Galactic plane, especially to the inner Galaxy. 
This is not because the calibrator sources are intrinsically more distributed in the 
Galactic plane but rather because more intensive surveys have been organized based on 
user-proposed targets. 
On the other hand, the flux density limited phase calibrator candidates in our 
investigation are the top 30\% brightest sources in the catalog and are rather 
homogeneously distributed. We conclude that the current investigation is not 
seriously affected by the calibrator distribution bias. 
We constrain the decl. range lower than $+25^{\circ}$. 

The basic parameters to calculate a flux density requirement for a phase calibrator are listed in 
Table~\ref{tbl:06}. 
The system equivalent flux density (SEFD), which is flux density equivalent to the system noise 
temperature of the 12~m antenna 
\citep{TMS2001},
is used for sensitivity calculation. The antennas are assumed to be directed to zenith. 
The available total bandwidth is 7.5~GHz in Bands~3-- 8 assuming that four 1.75~GHz 
SPWs are used, while in Bands~9 and 10, the total bandwidth is 15~GHz by configuring 
eight SPWs 
using the $90^{\circ}$ phase switching in the correlator. 
The phase calibrator scan length is fixed to 
8~s. First, we count only the sources whose flux density is higher than 
the flux density requirement. Second, we obtain a mean solid angle per phase calibrator 
from the above number count. 
Finally, we evaluate a mean angular separation to find a suitable phase calibrator 
in each Band.

The flux density requirement for the phase calibrator and the mean angular 
separation are listed in 
Tables~\ref{tbl:07} 
and 
\ref{tbl:08} for B2B and in-band phase referencing, respectively.  
The phase calibrator availability in B2B phase referencing is superior to in-band phase 
referencing in Bands~8, 9, and 10. In Band~7, on the other hand, 
the availability is almost comparable between B2B and in-band phase referencing, 
so that there may not be a great merit in using B2B phase referencing. 
Band~10 is the most difficult Band in searching for an in-band phase calibrator, 
whose expected mean separation angle is $12.^{\circ}6$.   
By introducing B2B phase referencing, 
one is expected to find a suitable phase calibrator within $4.^{\circ}3$ in Band~3. 
As discussed in 
Section~\ref{sec:04-02}, 
a phase calibrator would be required to be located within 
$\sim 6^{\circ}$, regardless of in-band and B2B phase referencing. 
The present investigation shows that B2B phase referencing is promising 
in searching for a closely located phase calibrator.

\section{
  Summary
}\label{sec:06}

The HF long baseline image capabilities of ALMA were successfully demonstrated by 
HF-LBC-2017 in Bands~7, 8, and 9 with B2B phase referencing when using switching 
cycle times of 20--82~s. The DGC solution shows no significant phase instability 
and is successfully applied to the target phase after B2B phase referencing 
to remove the instrumental phase offset difference. 

We compared the image quality 
between B2B and in-band phase referencing 
by observing QSOs 
with various separation angles and with a range of weather conditions. 
In principle, the closer the separation angle to the phase calibrator, 
the better image quality we can obtain 
in both B2B and in-band phase referencing. 
The image coherence of B2B phase referencing with a 
1--$2^{\circ}$ phase calibrator 
is comparable to that of in-band phase referencing with the same separation angle. 
It is considered that for larger separation angles, the image quality is degraded 
regardless of B2B and in-band phase referencing. 
If we wish to obtain an image coherence of $\geq$70\%, 
the separation angle should be within $\sim 6^{\circ}$.

Our statistical investigation for the phase calibrator availability shows that 
B2B phase referencing is more beneficial for finding a phase calibrator in Bands~8--10 
while there is no apparent merit in Band~7. These results demonstrate that the use of a closer 
phase calibrator together with B2B phase referencing can improve the image quality compared 
with in-band phase referencing in Bands~8--10. 

We have conducted the imaging capability test using B2B phase referencing 
for QSO J2228$-$0753 in Bands~7--3 and 8--4, and for 
of HL~Tau and VY~CMa in Band~9--4. We successfully imaged those 
objects with the highest angular resolutions that can be achieved at each 
frequency with an image fidelity expected from previous ALMA observations.
However, it is apparent for the complex structure sources that the visibility data 
cannot fully recover the extended structure without short ($u$,~$v$) coverage. 
In the case of HL~Tau, comparing to the previous Band~7 SV image, we sampled 
even smaller angular scales with the same baseline lengths in Band~9; thus, 
the image defects due to the undersampling of shorter ($u$,~$v$) coverages 
are more clearly seen. 
Our results highlight that, for high-fidelity images of extended sources, 
it is essential to consider the combination of multiple array configurations 
to acquire shorter ($u$,~$v$) coverages for achieving not only a detectable 
sensitivity but also well-sampled spatial frequency components. 

The basic functionality of B2B phase referencing has been proven in HF-LBC-2017. 
On the other hand, there were several technical problems in the 
experiments, for example, more flagged raw data comparing with the ordinal ALMA observations 
and some problematic antennas showing a relatively large phase drift in the DGC solution in some cases.  
Another important subject not fully investigated for HFs  
is where the dispersive term of the atmospheric water vapor is greater than a few tens of percent.
Solving these issue will be the goal of the next HF observational campaign.

\acknowledgments 
For this research, we used the ALMA data listed in 
Tables~\ref{tbl:03}--\ref{tbl:05}, 
in addition to 
ADS/JAO.ALMA \#2011.0.00015.SV and (HL~Tau in Band~7)
ADS/JAO.ALMA \#2011.0.00011.SV (VY~CMa in Band~9).
This research made use of the online ALMA calibrator catalog 
\\
(https://almascience.nrao.edu/alma-data/calibrator-catalogue).
ALMA is a partnership of ESO (representing its member states), NSF (USA), NINS (Japan), 
together with the NRC (Canada), NSC and ASIAA (Taiwan), and KASI (Republic of Korea), in cooperation 
with the Republic of Chile. The Joint ALMA Observatory is operated by the ESO, AUI/NRAO, 
and NAOJ. 
The authors thank the Joint ALMA Observatory staff in Chile for performing the challenging 
HF-LBC-2017 successfully. 
L.~T.~M. was adopted as a JAO ALMA expert visitor during his stay. 
This work was supported by JSPS KAKENHI grant No. JP16K05306.

\software{
	AATM \citep{Nikolic2009}, 
	CASA \citep{McMullin2007}
}

%

\bibliography{report} 

\begin{thebibliography}{}
\expandafter\ifx\csname natexlab\endcsname\relax\def\natexlab#1{#1}\fi
\providecommand{\url}[1]{\href{#1}{#1}}
\providecommand{\dodoi}[1]{doi:~\href{http://doi.org/#1}{\nolinkurl{#1}}}
\providecommand{\doeprint}[1]{\href{http://ascl.net/#1}{\nolinkurl{http://ascl.net/#1}}}
\providecommand{\doarXiv}[1]{\href{https://arxiv.org/abs/#1}{\nolinkurl{https://arxiv.org/abs/#1}}}

\bibitem[{{ALMA Partnership} {et~al.}(2015{\natexlab{a}}){ALMA Partnership},
  {Fomalont}, {Vlahakis}, {Corder}, {Remijan}, {Barkats}, {Lucas}, {Hunter},
  {Brogan}, {Asaki}, {Matsushita}, {Dent}, {Hills}, {Phillips}, {Richards},
  {Cox}, {Amestica}, {Broguiere}, {Cotton}, {Hales}, {Hiriart}, {Hirota},
  {Hodge}, {Impellizzeri}, {Kern}, {Kneissl}, {Liuzzo}, {Marcelino}, {Marson},
  {Mignano}, {Nakanishi}, {Nikolic}, {Perez}, {P{\'e}rez}, {Toledo}, {Aladro},
  {Butler}, {Cortes}, {Cortes}, {Dhawan}, {Di Francesco}, {Espada}, {Galarza},
  {Garcia-Appadoo}, {Guzman-Ramirez}, {Humphreys}, {Jung}, {Kameno}, {Laing},
  {Leon}, {Mangum}, {Marconi}, {Nagai}, {Nyman}, {Radiszcz}, {Rod{\'o}n},
  {Sawada}, {Takahashi}, {Tilanus}, {van Kempen}, {Vila Vilaro}, {Watson},
  {Wiklind}, {Gueth}, {Tatematsu}, {Wootten}, {Castro-Carrizo}, {Chapillon},
  {Dumas}, {de Gregorio-Monsalvo}, {Francke}, {Gallardo}, {Garcia}, {Gonzalez},
  {Hibbard}, {Hill}, {Kaminski}, {Karim}, {Krips}, {Kurono}, {Lopez}, {Martin},
  {Maud}, {Morales}, {Pietu}, {Plarre}, {Schieven}, {Testi}, {Videla},
  {Villard}, {Whyborn}, {Zwaan}, {Alves}, {Andreani}, {Avison}, {Barta},
  {Bedosti}, {Bendo}, {Bertoldi}, {Bethermin}, {Biggs}, {Boissier}, {Brand},
  {Burkutean}, {Casasola}, {Conway}, {Cortese}, {Dabrowski}, {Davis}, {Diaz
  Trigo}, {Fontani}, {Franco-Hernandez}, {Fuller}, {Galvan Madrid},
  {Giannetti}, {Ginsburg}, {Graves}, {Hatziminaoglou}, {Hogerheijde}, {Jachym},
  {Jimenez Serra}, {Karlicky}, {Klaasen}, {Kraus}, {Kunneriath}, {Lagos},
  {Longmore}, {Leurini}, {Maercker}, {Magnelli}, {Marti Vidal}, {Massardi},
  {Maury}, {Muehle}, {Muller}, {Muxlow}, {O'Gorman}, {Paladino}, {Petry},
  {Pineda}, {Randall}, {Richer}, {Rossetti}, {Rushton}, {Rygl}, {Sanchez
  Monge}, {Schaaf}, {Schilke}, {Stanke}, {Schmalzl}, {Stoehr}, {Urban}, {van
  Kampen}, {Vlemmings}, {Wang}, {Wild}, {Yang}, {Iguchi}, {Hasegawa}, {Saito},
  {Inatani}, {Mizuno}, {Asayama}, {Kosugi}, {Morita}, {Chiba}, {Kawashima},
  {Okumura}, {Ohashi}, {Ogasawara}, {Sakamoto}, {Noguchi}, {Huang}, {Liu},
  {Kemper}, {Koch}, {Chen}, {Chikada}, {Hiramatsu}, {Iono}, {Shimojo},
  {Komugi}, {Kim}, {Lyo}, {Muller}, {Herrera}, {Miura}, {Ueda}, {Chibueze},
  {Su}, {Trejo-Cruz}, {Wang}, {Kiuchi}, {Ukita}, {Sugimoto}, {Kawabe},
  {Hayashi}, {Miyama}, {Ho}, {Kaifu}, {Ishiguro}, {Beasley}, {Bhatnagar},
  {Braatz}, {Brisbin}, {Brunetti}, {Carilli}, {Crossley}, {D'Addario}, {Donovan
  Meyer}, {Emerson}, {Evans}, {Fisher}, {Golap}, {Griffith}, {Hale},
  {Halstead}, {Hardy}, {Hatz}, {Holdaway}, {Indebetouw}, {Jewell}, {Kepley},
  {Kim}, {Lacy}, {Leroy}, {Liszt}, {Lonsdale}, {Matthews}, {McKinnon}, {Mason},
  {Moellenbrock}, {Moullet}, {Myers}, {Ott}, {Peck}, {Pisano}, {Radford},
  {Randolph}, {Rao Venkata}, {Rawlings}, {Rosen}, {Schnee}, {Scott}, {Sharp},
  {Sheth}, {Simon}, {Tsutsumi}, \& {Wood}}]{ALMApartnership2015a}
{ALMA Partnership}, {Fomalont}, E.~B., {Vlahakis}, C., {et~al.}
  2015{\natexlab{a}}, \apjl, 808, L1, \dodoi{10.1088/2041-8205/808/1/L1}

\bibitem[{{ALMA Partnership} {et~al.}(2015{\natexlab{b}}){ALMA Partnership},
  {Brogan}, {P{\'e}rez}, {Hunter}, {Dent}, {Hales}, {Hills}, {Corder},
  {Fomalont}, {Vlahakis}, {Asaki}, {Barkats}, {Hirota}, {Hodge},
  {Impellizzeri}, {Kneissl}, {Liuzzo}, {Lucas}, {Marcelino}, {Matsushita},
  {Nakanishi}, {Phillips}, {Richards}, {Toledo}, {Aladro}, {Broguiere},
  {Cortes}, {Cortes}, {Espada}, {Galarza}, {Garcia-Appadoo}, {Guzman-Ramirez},
  {Humphreys}, {Jung}, {Kameno}, {Laing}, {Leon}, {Marconi}, {Mignano},
  {Nikolic}, {Nyman}, {Radiszcz}, {Remijan}, {Rod{\'o}n}, {Sawada},
  {Takahashi}, {Tilanus}, {Vila Vilaro}, {Watson}, {Wiklind}, {Akiyama},
  {Chapillon}, {de Gregorio-Monsalvo}, {Di Francesco}, {Gueth}, {Kawamura},
  {Lee}, {Nguyen Luong}, {Mangum}, {Pietu}, {Sanhueza}, {Saigo}, {Takakuwa},
  {Ubach}, {van Kempen}, {Wootten}, {Castro-Carrizo}, {Francke}, {Gallardo},
  {Garcia}, {Gonzalez}, {Hill}, {Kaminski}, {Kurono}, {Liu}, {Lopez},
  {Morales}, {Plarre}, {Schieven}, {Testi}, {Videla}, {Villard}, {Andreani},
  {Hibbard}, \& {Tatematsu}}]{ALMApartnership2015b}
{ALMA Partnership}, {Brogan}, C.~L., {P{\'e}rez}, L.~M., {et~al.}
  2015{\natexlab{b}}, \apjl, 808, L3, \dodoi{10.1088/2041-8205/808/1/L3}

\bibitem[{{ALMA Partnership} {et~al.}(2015{\natexlab{c}}){ALMA Partnership},
  {Vlahakis}, {Hunter}, {Hodge}, {P{\'e}rez}, {Andreani}, {Brogan}, {Cox},
  {Martin}, {Zwaan}, {Matsushita}, {Dent}, {Impellizzeri}, {Fomalont}, {Asaki},
  {Barkats}, {Hills}, {Hirota}, {Kneissl}, {Liuzzo}, {Lucas}, {Marcelino},
  {Nakanishi}, {Phillips}, {Richards}, {Toledo}, {Aladro}, {Broguiere},
  {Cortes}, {Cortes}, {Espada}, {Galarza}, {Garcia-Appadoo}, {Guzman-Ramirez},
  {Hales}, {Humphreys}, {Jung}, {Kameno}, {Laing}, {Leon}, {Marconi},
  {Mignano}, {Nikolic}, {Nyman}, {Radiszcz}, {Remijan}, {Rod{\'o}n}, {Sawada},
  {Takahashi}, {Tilanus}, {Vila Vilaro}, {Watson}, {Wiklind}, {Ao}, {Di
  Francesco}, {Hatsukade}, {Hatziminaoglou}, {Mangum}, {Matsuda}, {van Kampen},
  {Wootten}, {de Gregorio-Monsalvo}, {Dumas}, {Francke}, {Gallardo}, {Garcia},
  {Gonzalez}, {Hill}, {Iono}, {Kaminski}, {Karim}, {Krips}, {Kurono},
  {Lonsdale}, {Lopez}, {Morales}, {Plarre}, {Videla}, {Villard}, {Hibbard}, \&
  {Tatematsu}}]{ALMApartnership2015c}
{ALMA Partnership}, {Vlahakis}, C., {Hunter}, T.~R., {et~al.}
  2015{\natexlab{c}}, \apjl, 808, L4, \dodoi{10.1088/2041-8205/808/1/L4}

\bibitem[{{ALMA Partnership} {et~al.}(2015{\natexlab{d}}){ALMA Partnership},
  {Hunter}, {Kneissl}, {Moullet}, {Brogan}, {Fomalont}, {Vlahakis}, {Asaki},
  {Barkats}, {Dent}, {Hills}, {Hirota}, {Hodge}, {Impellizzeri}, {Liuzzo},
  {Lucas}, {Marcelino}, {Matsushita}, {Nakanishi}, {P{\'e}rez}, {Phillips},
  {Richards}, {Toledo}, {Aladro}, {Broguiere}, {Cortes}, {Cortes}, {Espada},
  {Galarza}, {Garcia-Appadoo}, {Guzman-Ramirez}, {Hales}, {Humphreys}, {Jung},
  {Kameno}, {Laing}, {Leon}, {Marconi}, {Mignano}, {Nikolic}, {Nyman},
  {Radiszcz}, {Remijan}, {Rod{\'o}n}, {Sawada}, {Takahashi}, {Tilanus}, {Vila
  Vilaro}, {Watson}, {Wiklind}, {De Gregorio-Monsalvo}, {Di Francesco},
  {Mangum}, {Francke}, {Gallardo}, {Garcia}, {Gonzalez}, {Hill}, {Kaminski},
  {Kurono}, {Lopez}, {Morales}, {Plarre}, {Randall}, {van kempen}, {Videla},
  {Villard}, {Andreani}, {Hibbard}, \& {Tatematsu}}]{ALMApartnership2015d}
{ALMA Partnership}, {Hunter}, T.~R., {Kneissl}, R., {et~al.}
  2015{\natexlab{d}}, \apjl, 808, L2, \dodoi{10.1088/2041-8205/808/1/L2}

\bibitem[{Andrews {et~al.}(2016)Andrews, Wilner, Zhu, Birnstiel, Carpenter,
  P\'{e}rez, Bai, \''{O}berg, Hughes, Isella, \& Ricci}]{Andrews2016}
Andrews, S.~M., Wilner, D.~J., Zhu, Z., {et~al.} 2016, \apjl, 820, L40,
  \dodoi{10.3847/2041-8205/820/2/L40}

\bibitem[{Asaki {et~al.}(2014)Asaki, Matsushita, Kawabe, Fomalont, Barkats, \&
  Corder}]{Asaki2014}
Asaki, Y., Matsushita, S., Kawabe, R., {et~al.} 2014, 2014SPIE 9145, 91454K,
  \dodoi{10.1117/12.2055824}

\bibitem[{Asaki {et~al.}(1998)Asaki, Shibata, Kawabe, Roh, Saito, Morita, \&
  Sasao}]{Asaki1998}
Asaki, Y., Shibata, K.~M., Kawabe, R., {et~al.} 1998, RaSc, 33, 1297,
  \dodoi{10.1029/98RS01607}

\bibitem[{Asaki {et~al.}(2007)Asaki, Sudou, Kono, Doi, Dodson, Pradel,
  Mochizuki, Edwards, Sasao, \& Fomalont}]{Asaki2007}
Asaki, Y., Sudou, H., Kono, Y., {et~al.} 2007, \pasj, 59, 397,
  \dodoi{10.1093/pasj/59.2.397}

\bibitem[{Asaki {et~al.}(2016)Asaki, Matsushita, Fomalont, Corder, Nyman, Dent,
  Philips, Hirota, Takahashi, Vila-Vilaro, Nikolic, Hunter, Remijan, \&
  Vlahakis}]{Asaki2016}
Asaki, Y., Matsushita, S., Fomalont, E.~B., {et~al.} 2016, 2016SPIE 9906,
  99065U, \dodoi{10.1117/12.2232301}

\bibitem[{Asaki {et~al.}(2019)Asaki, Maud, Fomalont, Phillips, Hirota,
  Barcos-Mu{\~{n}}noz, Dent, Takahashi, Corder, Carpenter, \&
  Villard}]{Asaki2019}
Asaki, Y., Maud, L.~T., Fomalont, E.~B., {et~al.} 2019, \apj

\bibitem[{Bachiller \& Cernicharo(2008)}]{Bachiller2008}
Bachiller, R., \& Cernicharo, R.~E. 2008, Science with the Atacama Large
  Millimeter Array: A New Era for Astrophysics,  Berlin Heidelberg New York:
  Springer, Dordrecht

\bibitem[{Beasley \& Conway(1995)}]{Beasley1995}
Beasley, A.~J., \& Conway, J.~E. 1995, 1995ASPC 82, 82, 327

\bibitem[{{Beckwith} {et~al.}(1990){Beckwith}, {Sargent}, {Chini}, \&
  {Guesten}}]{Beckwith1990}
{Beckwith}, S. V.~W., {Sargent}, A.~I., {Chini}, R.~S., \& {Guesten}, R. 1990,
  \aj, 99, 924, \dodoi{10.1086/115385}

\bibitem[{{Bryerton} {et~al.}(2005){Bryerton}, {Saini}, {Morgan}, {Stogoski},
  {Boyd}, \& {Thacker}}]{Bryerton2005}
{Bryerton}, E., {Saini}, K., {Morgan}, M., {et~al.} 2005, in The Joint 30th
  International Conference on Infrared and Millimeter Waves and 13th
  International Conference on Terahertz Electronics, Vol.~1, 72--73

\bibitem[{{Cliche} \& {Shillue}(2006)}]{Cliche2006}
{Cliche}, J.~F., \& {Shillue}, B. 2006, IEEE Control Systems Magazine, 26, 19,
  \dodoi{10.1109/MCS.2006.1580149}

\bibitem[{{Dodson} \& {Rioja}(2009)}]{RDodson2009}
{Dodson}, R., \& {Rioja}, M.~J. 2009, arXiv e-prints, arXiv:0910.1159.
\newblock \doarXiv{0910.1159}

\bibitem[{{Dodson} {et~al.}(2014){Dodson}, {Rioja}, {Jung}, {Sohn}, {Byun},
  {Cho}, {Lee}, {Kim}, {Kim}, {Oh}, {Han}, {Je}, {Chung}, {Wi}, {Kang}, {Lee},
  {Chung}, {Kim}, {Kim}, {Lee}, {Roh}, {Oh}, {Yeom}, {Song}, \&
  {Kang}}]{Dodson2014}
{Dodson}, R., {Rioja}, M.~J., {Jung}, T.-H., {et~al.} 2014, \aj, 148, 97,
  \dodoi{10.1088/0004-6256/148/5/97}

\bibitem[{{Hada} {et~al.}(2011){Hada}, {Doi}, {Kino}, {Nagai}, {Hagiwara}, \&
  {Kawaguchi}}]{Hada2011}
{Hada}, K., {Doi}, A., {Kino}, M., {et~al.} 2011, Natur, 477, 185,
  \dodoi{10.1038/nature10387}

\bibitem[{{Han} {et~al.}(2013){Han}, {Lee}, {Kang}, {Oh}, {Byun}, {Je},
  {Chung}, {Wi}, {Song}, {Kang}, {Lee}, {Kim}, {Sasao}, {Goldsmith}, \&
  {Wylde}}]{Han2013}
{Han}, S.-T., {Lee}, J.-W., {Kang}, J., {et~al.} 2013, \pasp, 125, 539,
  \dodoi{10.1086/671125}

\bibitem[{Holdaway \& D'Addario(2004)}]{Holdaway2004b}
Holdaway, M.~A., \& D'Addario, L. 2004,  ALMA Memo 523.
\newblock \url{http://library.nrao.edu/alma.shtml}

\bibitem[{Hunter {et~al.}(2016)Hunter, Lucas, Brogui\'{e}re, Fomalont, Dent,
  Phillips, Rabanus, \& Vlahakis}]{Hunter2016}
Hunter, T.~R., Lucas, R., Brogui\'{e}re, D., {et~al.} 2016, 2016SPIE 9914,
  99142L, \dodoi{10.1117/12.2232585}

\bibitem[{{Kervella} {et~al.}(2016){Kervella}, {Homan}, {Richards}, {Decin},
  {McDonald}, {Montarg{\`e}s}, \& {Ohnaka}}]{Kervella2016}
{Kervella}, P., {Homan}, W., {Richards}, A.~M.~S., {et~al.} 2016, \aap, 596,
  A92, \dodoi{10.1051/0004-6361/201629877}

\bibitem[{Matsushita {et~al.}(2014)Matsushita, Asaki, Kawabe, Fomalont,
  Barkats, \& Corder}]{Matsushita2014}
Matsushita, S., Asaki, Y., Kawabe, R., {et~al.} 2014, 2014SPIE 9145, 91453I,
  \dodoi{10.1117/12.2056424}

\bibitem[{Matsushita {et~al.}(2012)Matsushita, Morita, Barkarts, Hills,
  Fomalont, \& Nikolic}]{Matsushita2012}
Matsushita, S., Morita, K.-I., Barkarts, D., {et~al.} 2012, 2012SPIE 8444,
  84443E, \dodoi{10.1117/12.925872}

\bibitem[{Matsushita {et~al.}(2016)Matsushita, Asaki, Fomalont, Barkats,
  Corder, Hills, Kawabe, Maud, Morita, Nikolic, Tilanus, \&
  Vlahakis}]{Matsushita2016}
Matsushita, S., Asaki, Y., Fomalont, E.~B., {et~al.} 2016, 2016SPIE 9906,
  99064X, \dodoi{10.1117/12.2231846}

\bibitem[{{Matsushita} {et~al.}(2017){Matsushita}, {Asaki}, {Fomalont},
  {Morita}, {Barkats}, {Hills}, {Kawabe}, {Maud}, {Nikolic}, {Tilanus},
  {Vlahakis}, \& {Whyborn}}]{Matsushita2017}
{Matsushita}, S., {Asaki}, Y., {Fomalont}, E.~B., {et~al.} 2017, \pasp, 129,
  035004, \dodoi{10.1088/1538-3873/aa5787}

\bibitem[{{Maud} {et~al.}(2017){Maud}, {Tilanus}, {van Kempen}, {Hogerheijde},
  {Schmalzl}, {Yoon}, {Contreras}, {Toribio}, {Asaki}, {Dent}, {Fomalont}, \&
  {Matsushita}}]{Maud2017}
{Maud}, L.~T., {Tilanus}, R.~P.~J., {van Kempen}, T.~A., {et~al.} 2017, \aap,
  605, A121, \dodoi{10.1051/0004-6361/201731197}

\bibitem[{{Maud, L. T. et al.}(in preparation)}]{Maud2020}
{Maud, L. T. et al.} in preparation

\bibitem[{McMullin {et~al.}(2007)McMullin, Waters, Schiebel, Young, \&
  Golap}]{McMullin2007}
McMullin, J.~P., Waters, B., Schiebel, D., Young, W., \& Golap, K. 2007,
  2007ASPC 376, 376, 127

\bibitem[{Nikolic(2009)}]{Nikolic2009}
Nikolic, B. 2009,  ALMA Memo 587.
\newblock \url{http://library.nrao.edu/alma.shtml}

\bibitem[{{Nikolic} {et~al.}(2013){Nikolic}, {Bolton}, {Graves}, {Hills}, \&
  {Richer}}]{Nikolic2013}
{Nikolic}, B., {Bolton}, R.~C., {Graves}, S.~F., {Hills}, R.~E., \& {Richer},
  J.~S. 2013, \aap, 552, A104, \dodoi{10.1051/0004-6361/201220987}

\bibitem[{Nyman {et~al.}(2010)Nyman, Andreani, Hibbard, \& Okumura}]{Nyman2010}
Nyman, L.-{\AA}., Andreani, P., Hibbard, J., \& Okumura, S.~K. 2010, 2010SPIE
  7737, 77370G, \dodoi{10.1117/12.858023}

\bibitem[{{O'Gorman} {et~al.}(2015){O'Gorman}, {Vlemmings}, {Richards},
  {Baudry}, {De Beck}, {Decin}, {Harper}, {Humphreys}, {Kervella}, {Khouri}, \&
  {Muller}}]{OGorman2015}
{O'Gorman}, E., {Vlemmings}, W., {Richards}, A.~M.~S., {et~al.} 2015, \aap,
  573, L1, \dodoi{10.1051/0004-6361/201425101}

\bibitem[{Otsuka {et~al.}(2013)Otsuka, Suzuki, Nakagawa, Nishioka, Shiokawa, \&
  Tsugawa}]{Otsuka2013}
Otsuka, Y., Suzuki, K., Nakagawa, S., {et~al.} 2013, AnGeo, 31, 163,
  \dodoi{10.5194/angeo-31-163-2013}

\bibitem[{Pardo {et~al.}(2001)Pardo, Cernicharo, \& Serabyn}]{Pardo2001}
Pardo, J.~R., Cernicharo, J., \& Serabyn, E. 2001, ITAP, 49, 1683,
  \dodoi{10.1109/8.982447}

\bibitem[{{P{\'e}rez} {et~al.}(2010){P{\'e}rez}, {Lamb}, {Woody}, {Carpenter},
  {Zauderer}, {Isella}, {Bock}, {Bolatto}, {Carlstrom}, {Culverhouse}, {Joy},
  {Kwon}, {Leitch}, {Marrone}, {Muchovej}, {Plambeck}, {Scott}, {Teuben}, \&
  {Wright}}]{Perez2010}
{P{\'e}rez}, L.~M., {Lamb}, J.~W., {Woody}, D.~P., {et~al.} 2010, \apj, 724,
  493, \dodoi{10.1088/0004-637X/724/1/493}

\bibitem[{{Rebull} {et~al.}(2004){Rebull}, {Wolff}, \& {Strom}}]{Rebull2004}
{Rebull}, L.~M., {Wolff}, S.~C., \& {Strom}, S.~E. 2004, \aj, 127, 1029,
  \dodoi{10.1086/380931}

\bibitem[{{Richards} {et~al.}(2014){Richards}, {Impellizzeri}, {Humphreys},
  {Vlahakis}, {Vlemmings}, {Baudry}, {De Beck}, {Decin}, {Etoka}, {Gray},
  {Harper}, {Hunter}, {Kervella}, {Kerschbaum}, {McDonald}, {Melnick},
  {Muller}, {Neufeld}, {O'Gorman}, {Parfenov}, {Peck}, {Shinnaga}, {Sobolev},
  {Testi}, {Uscanga}, {Wootten}, {Yates}, \& {Zijlstra}}]{Richards2014}
{Richards}, A.~M.~S., {Impellizzeri}, C.~M.~V., {Humphreys}, E.~M., {et~al.}
  2014, \aap, 572, L9, \dodoi{10.1051/0004-6361/201425024}

\bibitem[{{Rioja} {et~al.}(2014){Rioja}, {Dodson}, {Jung}, {Sohn}, {Byun},
  {Agudo}, {Cho}, {Lee}, {Kim}, {Kim}, {Oh}, {Han}, {Je}, {Chung}, {Wi},
  {Kang}, {Lee}, {Chung}, {Ryoung Kim}, {Kim}, {Lee}, {Roh}, {Oh}, {Yeom},
  {Song}, \& {Kang}}]{Rioja2014}
{Rioja}, M.~J., {Dodson}, R., {Jung}, T., {et~al.} 2014, \aj, 148, 84,
  \dodoi{10.1088/0004-6256/148/5/84}

\bibitem[{{Robitaille} {et~al.}(2007){Robitaille}, {Whitney}, {Indebetouw}, \&
  {Wood}}]{Robitaille2007}
{Robitaille}, T.~P., {Whitney}, B.~A., {Indebetouw}, R., \& {Wood}, K. 2007,
  \apjs, 169, 328, \dodoi{10.1086/512039}

\bibitem[{Robson {et~al.}(2001)Robson, Hills, Richer, Delgado, Nyman,
  Ot\'{a}rola, \& Radford}]{Robson2001}
Robson, Y., Hills, R., Richer, J., {et~al.} 2001,  ALMA Memo 345.
\newblock \url{http://library.nrao.edu/alma.shtml}

\bibitem[{Schwab(1980)}]{Schwab1980}
Schwab, F.~R. 1980, SPIE1980 231, 958828, \dodoi{10.1117/12.958828}

\bibitem[{Shillue {et~al.}(2012)Shillue, Grammer, Jacques, Brito, Meadows,
  Castro, Masui, Treacy, \& Cliche}]{Shillue2012}
Shillue, B., Grammer, W., Jacques, C., {et~al.} 2012, 2012SPIE 8452, 845216,
  \dodoi{10.1117/12.927174}

\bibitem[{Takahashi {et~al.}(2018)Takahashi, Wrasse, Figueiredo, Barros, Abdu,
  Otsuka, \& Shiokawa}]{Takahashi2018}
Takahashi, H., Wrasse, C.~M., Figueiredo, C. A. O.~B., {et~al.} 2018, PEPS, 5,
  32, \dodoi{10.1186/s40645-018-0189-2}

\bibitem[{Takahashi {et~al.}(2016)Takahashi, Wrasse, Denardini, P\'{a}dua,
  de~Paula, Costa, Y., Shiokawa, Galera~Monico, Ivo, \&
  Sant'Anna}]{Takahashi2016}
Takahashi, H., Wrasse, C.~M., Denardini, C.~M., {et~al.} 2016, SpWea, 14, 937,
  \dodoi{10.1002/2016SW001474}

\bibitem[{{Thompson} {et~al.}(2001){Thompson}, {Moran}, \& {Swenson}}]{TMS2001}
{Thompson}, A.~R., {Moran}, J.~M., \& {Swenson}, George~W., J. 2001,
  {Interferometry and Synthesis in Radio Astronomy, 2nd Edition} (A
  Wiley-Interscience Publication, John Wiley \& Sons, Inc.)

\bibitem[{{Wittkowski} {et~al.}(2012){Wittkowski}, {Hauschildt},
  {Arroyo-Torres}, \& {Marcaide}}]{Wittkowski2012}
{Wittkowski}, M., {Hauschildt}, P.~H., {Arroyo-Torres}, B., \& {Marcaide},
  J.~M. 2012, \aap, 540, L12, \dodoi{10.1051/0004-6361/201219126}

\bibitem[{{Zauderer} {et~al.}(2016){Zauderer}, {Bolatto}, {Vogel}, {Carpenter},
  {Per{\'e}z}, {Lamb}, {Woody}, {Bock}, {Carlstrom}, {Culverhouse}, {Curley},
  {Leitch}, {Plambeck}, {Pound}, {Marrone}, {Muchovej}, {Mundy}, {Teng},
  {Teuben}, {Volgenau}, {Wright}, \& {Wu}}]{Zauderer2016}
{Zauderer}, B.~A., {Bolatto}, A.~D., {Vogel}, S.~N., {et~al.} 2016, \aj, 151,
  18, \dodoi{10.3847/0004-6256/151/1/18}

\bibitem[{{Zhang} {et~al.}(2012){Zhang}, {Reid}, {Menten}, \&
  {Zheng}}]{Zhang2012}
{Zhang}, B., {Reid}, M.~J., {Menten}, K.~M., \& {Zheng}, X.~W. 2012, \apj, 744,
  23, \dodoi{10.1088/0004-637X/744/1/23}

\end{thebibliography}
\bibliographystyle{aasjournal}



\clearpage
\newpage

\clearpage
\newpage

\clearpage
\newpage

\clearpage
\newpage

\clearpage
\newpage

\clearpage
\newpage

\clearpage
\newpage

\clearpage
\newpage

\clearpage
\newpage

\clearpage
\newpage

\newpage
\clearpage

\newpage
\clearpage

\clearpage
\newpage

\newpage
\clearpage

\clearpage
\newpage

\clearpage
\newpage

\clearpage
\newpage

\newpage
\clearpage

\begin{table}
\begin{center}
\caption{
  Possible Frequency Combination for the Harmonic Frequency Switching
}\label{tbl:01}
\begin{threeparttable}
\begin{tabular}
{ccccc}
\hline
\hline
HF Band
& (LO1 Frequency Range) 
& LF Band    
& (LO1 Frequency Range)
& LO1 Frequency Ratio \\
\hline
Band~7
& (282.9 -- 324.0~GHz) 
& Band~3
& (94.3 -- 108.0~GHz) 
& 3 \\
\hline
Band~8
& (399.0 -- 465.0~GHz) 
& Band~4   
& (133.0 -- 155.0~GHz) 
&  3 \\
Band~8
& (442.2 -- 492.0~GHz)
& Band~6   
& (221.1 -- 246.0~GHz)
&  2 \\
\hline
Band~9
& (610.2 -- 697.5~GHz) 
& Band~4   
& (135.6 -- 155.0~GHz) 
&  4.5 \\
Band~9
& (663.3 -- 711.9~GHz)
& Band~6   
& (221.1 -- 237.3~GHz)
&  3 \\
\hline
Band~10
& (828.0 -- 942.3~GHz) 
& Band~3   
& (92.0 -- 104.7~GHz) 
&  9 \\
Band~10
& (794.7 -- 913.5~GHz)
& Band~5
& (176.6 -- 203.0~GHz)
& 4.5 
\\
Band~10
& (848.7 -- 942.3~GHz)
& Band~7   
& (282.9 -- 314.1~GHz) 
&  3 \\
\hline
\hline
\end{tabular}
\end{threeparttable}
\end{center}
\end{table}

\clearpage
\newpage

\begin{table}
\begin{center}
\caption{
  Prohibited Frequency Range in the Harmonic Frequency Switching
}\label{tbl:02}
\begin{threeparttable}
\begin{tabular}
{cc}
\hline
\hline
HF Receiver
& LO1 Frequency Range 
 \\
\hline
Band~7
& 324 -- 365~GHz 
\\
Band~8
& 393 -- 399~GHz 
 \\
\hline
\hline
\end{tabular}
\end{threeparttable}
\end{center}
\end{table}

\clearpage
\newpage

\begin{longrotatetable}
\begin{table}
\begin{center}
\caption{
  HF-LBC-2017 
  Stage~2 
  Experiments Reported in This Paper
}\label{tbl:03}
\begin{threeparttable}
\begin{tabular}
{c|c|c|c|c}
\hline
\hline
Date 
& Baseline Range 
& Band\tnote{1} 
& \multicolumn{1}{c|}{DGC Source} 
& EB (uid://A002/)  \\
\hline

& 
&
& \hspace{5mm}J0006$-$0623, J2232$+$1143,\hspace{5mm}~
& 
\\ 
2017 Apr 11 
&15$-$396~m
& 7--3
& 3C454.3, J0108$+$0135, 
& Xbeef47/X20e6
\\ 

& 
&
& J2348$-$1631
&  
\\ 
\hline

& 
&
& J1924$-$2914, J1517$-$2422, 
& 
Xbf894a/X1aab, 
Xbf894a/X1c1e, 
\\
2017 Apr 23
& 15$-$460~m
& 9--6
& J1604$-$4441\tnote{2}, J1751$+$0939\tnote{2},
& 
Xbf894a/X1d58, 
Xbf894a/X1e87, 
\\

& 
&
& J1337$-$1257\tnote{2}
& 
Xbf894a/X1f9a
\\
\hline
\multirow{2}{*}{2017 May 4}
& \multirow{2}{*}{17$-$1110~m}
& \multirow{2}{*}{8--4}
& J0510$+$1800, J0522$-$3627,
& \multirow{2}{*}{Xbff114/X2bc9}
\\

& 
&
& J0006$-$0623\tnote{3}, J0450$-$8101\tnote{3}
& 
\\
\hline
\hline
\end{tabular}
\begin{tablenotes}\footnotesize
\item[1]
HF Band for the target--LF Band for the phase calibrator.
\item[2]
The DGC solutions were not fully analyzed because of an antenna shadowing effect for low elevation angles. 
\item[3]
The DGC solutions were not fully analyzed because the S/Ns were not high enough.
\end{tablenotes}
\end{threeparttable}
\end{center}
\end{table}
\end{longrotatetable}


\clearpage
\newpage

\begin{longrotatetable}
\begin{table}
\begin{center}
\caption{
  HF-LBC-2017 
  Stage~3 
  Experiments Reported in This Paper
}\label{tbl:04}
\begin{threeparttable}
\begin{tabular}
{c|c|c|c|c|c|c}
\hline
\hline
\multicolumn{1}{c|}{Date} 
& \multicolumn{1}{c|}{Baseline Range}
& \multicolumn{1}{c|}{Band} 
& Target 
& Phase Cal. 
& DGC Source 
& EB (uid://A002/Xc23361/) \\
\hline
\multirow{12}{*}{\hfill 2017 July 18 \hfill} 
& \multirow{12}{*}{\hfill 17$-$3697~m \hfill}
& \multirow{8}{*}{\hfill 8--4 \hfill}
& \multirow{4}{*}{\hfill J0633$-$2223 \hfill} 
& \multirow{4}{*}{\hfill J0634$-$2335 \hfill} 
& 
& X4970, X4a9a, 
\\

& 
&
& 
& 
& 
& X4d1e\tnote{2}, X4e7c\tnote{2},  
\\

& 
&
& 
& 
& 
& X5197, X52bb,  
\\

& 
& 
&
& 
& 
& X5550, X56b4
\\
\cline{4-7}

&  
&
&
& 
& \multirow{4}{*}{J0522$-$3627}
& X4865, X4b27, 
\\

&  
& 
&
& 
& 
&  X4bd4, X4f11, 
\\

&  
& 
&
& 
& 
& X5053, X5355, 
\\

&  
& 
&
& 
& 
& X5402, X57c9
\\
\cline{3-7}

&
& \multirow{4}{*}{8--8\tnote{1}}
& \multirow{4}{*}{J0633$-$2223}
& J0634$-$2335 
& 
& X48f1, X49e9 
\\
\cline{5-5}\cline{7-7}

& 
&
& 
&  J0620$-$2515\tnote{2} 
& 
&  X4c81\tnote{2}, X4dbb\tnote{2}
\\
\cline{5-5}\cline{7-7}

& 
&
& 
&  J0648$-$1744 
& 
&  X5105, X5213 
\\
\cline{5-5}\cline{7-7}

& 
&
& 
&  J0609$-$1542
& 
&  X54a3, X55e9
\\
\hline
\hline
\end{tabular}
\begin{tablenotes}\footnotesize
\item[1]
In-band phase referencing.
\item[2]
Executions failed because the target passed close to 
zenith.
\end{tablenotes}
\end{threeparttable}
\end{center}
\end{table}
\end{longrotatetable}

\clearpage
\newpage

\begin{longrotatetable}
\begin{table}
\begin{center}
\caption{
  HF-LBC-2017 
  Stage~4 
  Experiments Reported in This Paper
}\label{tbl:05}
\begin{threeparttable}
\begin{tabular}
{c|c|c|c|c|c|c}
\hline
\hline
\multicolumn{1}{c|}{Date} 
& \multicolumn{1}{c|}{Baseline Range} 
& \multicolumn{1}{c|}{Band} 
& Target 
& Phase Cal. 
& DGC Source 
& EB (uid://A002/) \\
\hline
2017 Oct  9
& 41$-$14969~m
& 8--4
& \multirow{4}{*}{J2228$-$0753}
& \multirow{4}{*}{J2229$-$0832}
& \multirow{4}{*}{J2253$+$1608}
& Xc5802b/X5bb3
\\
\cline{1-3} \cline{7-7}
2017 Oct 10
& 233$-$16196~m
& 7--3
& 
& 
& 
& Xc59134/Xd47
\\
\cline{1-3} \cline{7-7}
2017 Nov 2
& 222$-$13894~m
& 7--3
& 
& 
& 
& Xc65717/X56f
\\
\cline{1-3} \cline{7-7}

& 
& 8--4
& 
& 
& 
& Xc660ef/X8e0
\\
\cline{3-7}
2017 Nov 3
& 230$-$13894~m
& \multirow{2}{*}{\hfill 9--4 \hfill}
& HL~Tau
& \multicolumn{1}{c|}{J0431$+$1731}
& \multirow{2}{*}{\hfill J0522$-$3627 \hfill}
& Xc660ef/X1b89, X660ef/X2046 
\\
\cline{4-5} \cline{7-7}

& 
&
& VY~CMa
& \multicolumn{1}{c|}{J0725$-$2640}
& 
& X660ef/X2d26
\\
\hline
\hline
\end{tabular}
\end{threeparttable}
\end{center}
\end{table}
\end{longrotatetable}

\clearpage
\newpage

\begin{table}
\begin{center}
\caption{
  Assumed Parameters for Phase Calibrator Scan
}\label{tbl:06}
\begin{threeparttable}
\begin{tabular}
{ll}
\hline
\hline
Scan duration & 8~s \\
Antenna number & 43 \\
Required S/N (per scan)  & $20 \times R$ \\
Bandwidth (Bands 3--8)
& 
7.5~GHz (1.875~GHz~$\times$~4~SPWs) 
\\
Bandwidth (Bands 9 and 10) 
& 
15~GHz (1.875~GHz~$\times$~8~SPWs) 
\\
SEFD (Band~3) &  2167~Jy \\
SEFD (Band~4) &  2709~Jy \\
SEFD (Band~5) & 3221~Jy \\
SEFD (Band~6) &  3221~Jy \\
SEFD (Band~7) &  5662~Jy \\
SEFD (Band~8) & 10,932~Jy\\
SEFD (Band~9) & 36,171~Jy \\
SEFD (Band~10) & 75,958~Jy \\
\hline
\hline
\end{tabular}
\end{threeparttable}
\end{center}
\end{table}

\clearpage
\newpage

\begin{table}
\begin{center}
\caption{
  Required Phase Calibrator Flux Density and Mean Separation Angle for B2B Phase Referencing
}\label{tbl:07}
\begin{threeparttable}
\begin{tabular}
{cccc}
\hline
\hline
Target Band 
& Phase Cal. Band
& Required Flux Density 
& Mean Separation \\
(LO1 (GHz)) 
& (LO1 (GHz))
& for Phase Cal. (Jy)
& Angle (deg) \\
\hline
Band~7 (321) 
& Band~3 (107) 
& 0.0957 
& 3.1 \\
\hline
Band~8 (405) 
& Band~4 (135) 
&  0.120 
&  3.5 \\
Band~8 (470)
& Band~6 (235)
&  0.0948
& 3.6 \\
\hline
Band~9 (681) 
& Band~4 (151) 
&  0.180 
& 4.1 \\
Band~9 (681)
& Band~6 (227)
&  0.142 
& 4.1 \\
\hline
Band~10 (873) 
& Band~3 (97) 
&  0.287
& 4.3 \\
Band~10 (873) 
& Band~5 (194) 
& 0.213
& 4.6\\
Band~10 (873)
& Band~7 (291) 
&  0.250
&5.4 \\
\hline
\hline
\end{tabular}
\end{threeparttable}
\end{center}
\end{table}

\clearpage
\newpage

\begin{table}
\begin{center}
\caption{
  Required Phase Calibrator Flux Density and Mean Separation Angle for In-band Phase Referencing
}\label{tbl:08}
\begin{threeparttable}
\begin{tabular}
{cccc}
\hline
\hline
Target Band 
& Phase Cal Band.
& Required Flux Density 
& Mean Separation   \\
(LO1 (GHz)) 
& (LO1 (GHz))
& for Phase Cal. (Jy)
& Angle (deg) 
\\
\hline
Band~7 (321)
& Band~7 (321) 
&  0.0834
&3.7 
\\
Band~8 (470)
& Band~8 (470) 
&  0.161 
&5.0 
\\
Band~9 (681)
& Band~9 (681) 
&  
0.376 
& 
7.8  
\\
Band~10 (873)
& Band~10 (873) 
&  
0.792 
& 
12.6 
\\
\hline
\hline
\end{tabular}
\end{threeparttable}
\end{center}
\end{table}

\end{document}